\PassOptionsToPackage{unicode}{hyperref}
\PassOptionsToPackage{hyphens}{url}
\PassOptionsToPackage{dvipsnames,svgnames,x11names,table}{xcolor}
\documentclass[
  12pt]{article}

\usepackage{amsmath,amssymb}
\usepackage{iftex}
\ifPDFTeX
  \usepackage[T1]{fontenc}
  \usepackage[utf8]{inputenc}
  \usepackage{textcomp} 
\else 
  \usepackage{unicode-math}
  \defaultfontfeatures{Scale=MatchLowercase}
  \defaultfontfeatures[\rmfamily]{Ligatures=TeX,Scale=1}
\fi
\usepackage{lmodern}
\ifPDFTeX\else  
\fi
\IfFileExists{upquote.sty}{\usepackage{upquote}}{}
\IfFileExists{microtype.sty}{
  \usepackage[]{microtype}
  \UseMicrotypeSet[protrusion]{basicmath} 
}{}
\makeatletter
\@ifundefined{KOMAClassName}{
  \IfFileExists{parskip.sty}{%
    \usepackage{parskip}
  }{
    \setlength{\parindent}{0pt}
    \setlength{\parskip}{6pt plus 2pt minus 1pt}}
}{
  \KOMAoptions{parskip=half}}
\makeatother
\usepackage{xcolor}
\makeatletter
\ifx\paragraph\undefined\else
  \let\oldparagraph\paragraph
  \renewcommand{\paragraph}{
    \@ifstar
      \xxxParagraphStar
      \xxxParagraphNoStar
  }
  \newcommand{\xxxParagraphStar}[1]{\oldparagraph*{#1}\mbox{}}
  \newcommand{\xxxParagraphNoStar}[1]{\oldparagraph{#1}\mbox{}}
\fi
\ifx\subparagraph\undefined\else
  \let\oldsubparagraph\subparagraph
  \renewcommand{\subparagraph}{
    \@ifstar
      \xxxSubParagraphStar
      \xxxSubParagraphNoStar
  }
  \newcommand{\xxxSubParagraphStar}[1]{\oldsubparagraph*{#1}\mbox{}}
  \newcommand{\xxxSubParagraphNoStar}[1]{\oldsubparagraph{#1}\mbox{}}
\fi
\makeatother

\usepackage{longtable,booktabs,array}
\usepackage{calc} 
\usepackage{etoolbox}
\makeatletter
\patchcmd\longtable{\par}{\if@noskipsec\mbox{}\fi\par}{}{}
\makeatother
\IfFileExists{footnotehyper.sty}{\usepackage{footnotehyper}}{\usepackage{footnote}}
\makesavenoteenv{longtable}
\usepackage{graphicx}
\makeatletter
\def\maxwidth{\ifdim\Gin@nat@width>\linewidth\linewidth\else\Gin@nat@width\fi}
\def\maxheight{\ifdim\Gin@nat@height>\textheight\textheight\else\Gin@nat@height\fi}
\makeatother
\setkeys{Gin}{width=\maxwidth,keepaspectratio}
\makeatletter
\def\fps@figure{htbp}
\makeatother

\makeatletter
\@ifpackageloaded{caption}{}{\usepackage{caption}}
\AtBeginDocument{%
\ifdefined\contentsname
  \renewcommand*\contentsname{Table of contents}
\else
  \newcommand\contentsname{Table of contents}
\fi
\ifdefined\listfigurename
  \renewcommand*\listfigurename{List of Figures}
\else
  \newcommand\listfigurename{List of Figures}
\fi
\ifdefined\listtablename
  \renewcommand*\listtablename{List of Tables}
\else
  \newcommand\listtablename{List of Tables}
\fi
\ifdefined\figurename
  \renewcommand*\figurename{Figure}
\else
  \newcommand\figurename{Figure}
\fi
\ifdefined\tablename
  \renewcommand*\tablename{Table}
\else
  \newcommand\tablename{Table}
\fi
}
\@ifpackageloaded{float}{}{\usepackage{float}}
\floatstyle{ruled}
\@ifundefined{c@chapter}{\newfloat{codelisting}{h}{lop}}{\newfloat{codelisting}{h}{lop}[chapter]}
\floatname{codelisting}{Listing}

\makeatother
\makeatletter
\@ifpackageloaded{caption}{}{\usepackage{caption}}
\@ifpackageloaded{subcaption}{}{\usepackage{subcaption}}
\makeatother

\ifLuaTeX
  \usepackage{selnolig}  
\fi
\usepackage{amsthm}
\usepackage{comment}
\usepackage{bm}
\usepackage{multirow}
\usepackage{amsfonts}
\usepackage{bbm}
\IfFileExists{bbold.sty}{\usepackage{bbold}}{}
\usepackage{enumitem}
\usepackage{array}
\IfFileExists{cancel.sty}{\usepackage{cancel}}{}
\usepackage{algorithm}
\usepackage{algpseudocode}
\usepackage{threeparttable}
\usepackage{tablefootnote}
\usepackage{adjustbox}
\usepackage{makecell}
\usepackage[]{natbib}
\usepackage{bookmark}

\IfFileExists{xurl.sty}{\usepackage{xurl}}{} 
\hypersetup{
  pdftitle={Bayesian Propensity Score Predictive Inference for Extending Trial Findings to Target Populations},
  pdfauthor={Jungang Zou; Joseph E. Schwartz; Roderick J. A. Little; Nathalie Moise; Qixuan Chen},
  pdfkeywords={Bayesian additive regression trees, Electronic Health Records, Generalizability, Propensity Score, Randomized Trials, Transportability},
  colorlinks=true,
  linkcolor={blue},
  filecolor={Maroon},
  citecolor={Blue},
  urlcolor={Blue},
  pdfcreator={LaTeX via pandoc}}

\newcommand{\anon}{1}

\theoremstyle{plain}

\theoremstyle{definition}

\def\bSig\mathbf{\Sigma}

\newcommand{\bX}{\mathbf{X}}

\newcommand{\cond}{{\, \vert \,}}
\newcommand{\e}{\mathbb{E}}

\let\origtable\table
\renewcommand{\table}[1][]{\origtable[htbp]\spacingset{1}}
\let\origfigure\figure
\renewcommand{\figure}[1][htbp]{\origfigure[#1]\spacingset{1}}
\begin{document}
\def\spacingset#1{\renewcommand{\baselinestretch}%
{#1}\small\normalsize} \spacingset{1}
\if1\anon
{
\title{\textbf{Bayesian Propensity Score Predictive Inference for Extending Trial Findings to Target Populations}}
\vspace{-0.3cm}

\author{
Jungang Zou$^{1}$, Joseph E. Schwartz$^{2,3}$, Roderick J.~A. Little$^{4}$,\\
Nathalie Moise$^{2}$, and Qixuan Chen$^{1,5}$\thanks{Corresponding author.} \\[0.4em]
$^{1}$Department of Biostatistics, Columbia University \\
$^{2}$Center for Behavioral Cardiovascular Health, Columbia University \\
$^{3}$Department of Psychiatry and Behavioral Health, Stony Brook University \\
$^{4}$Department of Biostatistics, University of Michigan \\
$^{5}$Division of Social Solutions and Services Research,\\
Nathan S. Kline Institute for Psychiatric Research
}

\date{}
\maketitle
}
\fi
\if0\anon
{
  \bigskip
  \bigskip
  \bigskip
  \begin{center}
    {\LARGE\bf Bayesian Propensity Score Predictive Inference for Extending Trial Findings to Target Populations}
\end{center}
  \medskip
} \fi
\vspace{-0.5cm}
\begin{abstract}
The CODIACS Vanguard randomized trial compared a depression care intervention with usual care for patients with persistent depressive symptoms following acute coronary syndrome. We assess whether its findings can be extended to broader patient populations represented by two cohorts constructed from the All of Us electronic health records. This extension is challenging when treatment effects vary across covariates whose distributions differ between trial and target populations. We develop Bayesian Propensity Score Predictive Inference (PSPI), an interaction-based outcome modeling framework that combines Bayesian additive regression trees for potentially high-dimensional covariates with natural cubic splines of the sampling propensity score. Two variants include a separate spline in the treatment effect component alone or in both the prognostic and treatment effect components. The spline components allow separate regularization of selection-related variation and thus improve potential outcome estimation within the observed range and allow an explicit extrapolation rule beyond their boundary knots under substantial covariate shift. An inverse Gumbel transformation of the sampling propensity score balances compression near zero and the extent of extrapolation. Simulation studies show that spline-based PSPI models outperform competing methods in the settings studied, achieving lower root mean squared errors and near-nominal coverage, particularly under substantial covariate shift. In the application, PSPI estimates larger intervention benefits in both target populations than the unadjusted trial estimate.
\end{abstract}

\noindent%
\vfill
\newpage
\spacingset{1.8} 

\section{Introduction}
\label{sec:intro}

\subsection{Background} 
\label{sec:objective}
Depression is a major public health burden in the United States, hindering chronic disease management and increasing morbidity and mortality \citep{BUSH2001337}. The burden is particularly pronounced among patients with acute coronary syndrome (ACS), where elevated depressive symptoms are associated with increased risk of ACS recurrence and all-cause mortality \citep{Lichtman2008, Nicholson2006}. Despite the importance of reducing post-ACS depression, routine management of post-ACS depression remains poor \citep{Jha2019}. 

The Comparison of Depression Interventions after Acute Coronary Syndrome (CODIACS) Vanguard trial was a randomized controlled trial evaluating the feasibility, efficacy, and costs of a centralized, stepped, patient preference-based depression care intervention for patients with persistent depressive symptoms following an ACS event  \citep{Davidson2013}. A total of 150 eligible patients were enrolled from five field centers, with 73 randomized to the intervention and 77 to usual care.

The trial showed that depressive symptoms decreased significantly more in the active treatment group than in the usual care group, with evidence of treatment effect heterogeneity across patient subgroups \citep{Davidson2013}. A central scientific question is whether the intervention benefit extends to the broader population of post-ACS patients encountered in routine clinical practice, including patient groups that were underrepresented in the trial. 

The All of Us Research Program is a national electronic health record (EHR) cohort of over one million participants designed to accelerate biomedical research across a diverse U.S. population, with more than 80\% of participants from groups historically underrepresented in biomedical research \citep{AURPI2019}. We therefore define the target population as post-ACS patients with depressive symptoms represented in the All of Us EHR cohort. Compared with the CODIACS trial, this target population differs substantially in characteristics that may modify treatment effects, resulting in a pronounced covariate shift between the trial and target populations. This covariate shift creates the statistical challenge of generalizing treatment effects from randomized trials to real-world populations.

\subsection{Generalizability and Transportability}
Randomized controlled trials (RCTs) provide strong internal validity but often have limited external validity because selective enrollment may result in a trial sample that differs systematically from the target population. This challenge arises when treatment effects are heterogeneous across subgroups and the distributions of treatment effect modifiers differ between the trial and target populations. Such differences occur in both nested designs, where the trial is embedded within the target population, and non-nested designs, where the individuals in the clinical trial do not come from the target population  \citep{Dahabreh2020}. Prior work has shown that bias due to limited external validity can be substantial, potentially even surpassing bias from threats to internal validity \citep{Bell2016}. Although designing more representative trials is the preferred solution, it is often challenging in practice. Consequently, statistical methods are needed to extend trial findings to target populations. Following \cite{Westreich2017}, we use \textit{generalizability} to refer to extending trial findings under nested designs and \textit{transportability} to refer to extending trial findings under non-nested designs. Throughout this paper, we distinguish between the \textit{EHR cohort}, the observed data available for analysis, and the \textit{target population}, which the EHR cohort represents and for which inference is ultimately intended. Figure~\ref{fig:designs} illustrates the sampling structures under both settings; further details on the distinction are formally discussed in Section~\ref{sec:notation}. In this study, the CODIACS Vanguard trial and the All of Us cohort are separately sampled, placing our applied study in the non-nested transportability setting. However, the methods we develop in this work also apply to the nested generalizability setting. We present and evaluate our methods for both settings throughout the paper.

\subsection{Related Work}
Existing methods for generalizability and transportability can be broadly categorized into three classes \citep{Schmid2022, Degtiar2023}: (1) outcome regression, (2) propensity score weighting, and (3) hybrid methods combining both. Outcome regression methods estimate conditional means by modeling outcomes in the trial sample and predicting potential outcomes in the EHR cohort. These models range from simple parametric forms (e.g., linear regression) to flexible machine learning tools such as Bayesian additive regression trees (BART; \citealp{Chipman2010, Kern2017}). Another class centers on the sampling propensity score for trial participation---the probability of being enrolled in the trial given observed covariates. As a balancing score, the propensity score captures all covariate information relevant to participation \citep{ROSENBAUM1983}. Propensity score-based methods aim to align the covariate distribution of the trial sample with that of the EHR cohort \citep{Stuart2011, Tipton2013, OMuircheartaigh2014, CRL2017, Dahabreh2019, Dahabreh2020}. Among these, inverse probability of participation weighting (IPPW; \citealp{Stuart2011, Dahabreh2019,Dahabreh2020}) is the most widely used. However, both outcome regression and propensity score-based methods are vulnerable to model misspecification. Hybrid methods combine both approaches to improve robustness and efficiency. A prominent class is doubly robust estimators, such as the augmented IPPW (A-IPPW; \citealp{Dahabreh2019, Dahabreh2020}) and targeted maximum likelihood estimation \citep{Schmid2022}, which yield consistent estimates if either the outcome or propensity score model is correctly specified. 

Both IPPW and A-IPPW use propensity score weighting, which addresses covariate shift, defined as differences in distributions of covariates between the trial and target populations, by reweighting the trial sample to resemble the target population. However, weighting may become inefficient and unstable because of extreme weights resulting from substantial covariate shift. Moreover, weighting methods primarily balance covariate distributions rather than explicitly modeling treatment effect heterogeneity or extrapolating treatment effects into regions of the covariate space that are poorly represented in the trial. This motivates complementary outcome-modeling approaches that incorporate the sampling propensity score directly into prediction rather than using it solely for weighting.

\subsection{Our Contribution}
Reliable prediction of treatment effects using sampling propensity score-augmented outcome models remains challenging under substantial covariate shift. Nonparametric Bayesian regularized models, such as BART \citep{Hill2011} and Bayesian Causal Forests (BCF; \citealp{Hahn2020}), have demonstrated strong performance for observational causal inference with high-dimensional covariates and complex outcome surfaces. However, in generalizability and transportability settings, these models are trained using trial data but applied to predict outcomes in the EHR cohort. Under substantial covariate shift, outcome models must extrapolate into regions of the covariate space that are poorly represented or absent from the trial, potentially leading to model misspecification, unreliable extrapolation, and biased treatment effect estimates \citep{Oganisian2020}. These challenges are especially pronounced when the shifted covariates also modify treatment effects.

We propose Propensity Score Predictive Inference (PSPI), a Bayesian outcome modeling framework that combines BART with natural cubic splines of the sampling propensity score in both the prognostic and treatment effect components. BART captures nonlinear covariate relationships and interactions, while the separate spline terms explicitly model selection-related variation in outcomes and treatment effects. This construction allows selection-related variation to be regularized separately from the BART components and provides a structured approach to extrapolation beyond the trial sample.

The rationale for this separation is that covariates important for generalizability and transportability may have modest individual predictive signals in the trial data. Their contributions may therefore be attenuated by BART's regularization, even when they modify treatment effects and differ substantially in distribution between the trial and target populations. The sampling propensity score summarizes the covariates' joint relationship with trial participation. Modeling this summary through separate spline terms may help capture selection-related patterns that are underrepresented by BART alone, potentially improving robustness. This motivation is consistent with the finding that including sampling propensity scores in flexible outcome models can improve population mean estimation when important predictors are omitted \citep{Liu2022}.

The spline components also provide an explicit way to model and assess extrapolation. BART models do not naturally extend trends beyond the covariate support of the training data, whereas natural cubic splines impose linear tails beyond their boundary knots on the modeled score scale. Building on the spline-augmented BART strategy of \citet{Nethery2019}, PSPI uses this structure for treatment-effect prediction in generalizability and transportability settings. Separately specifying the score-related components also allows us to modify the assumed prognostic and treatment-effect relationships outside the range represented in the trial, providing a direct basis for sensitivity analyses of target-population inference.

We further introduce an inverse Gumbel transformation of the sampling propensity score before constructing the spline terms. This transformation balances the compression of small scores under the identity transformation against the extended extrapolation distances induced by the logit and complementary log-log transformations. 

Simulation studies show that PSPI can outperform existing methods in both generalizability and transportability settings, particularly under substantial covariate shift. The inverse Gumbel transformation of the sampling propensity score yields lower bias and near-nominal coverage, with substantially narrower credible intervals than alternative transformations. We demonstrate the PSPI framework by extending the treatment effects observed in the CODIACS Vanguard trial to the target population of post-ACS patients with depressive symptoms, represented by the All of Us EHR cohort.

\begin{figure}[!htbp]
  \centering
  \begin{subfigure}[t]{0.32\textwidth}       
    \centering
    \includegraphics[height=5cm,keepaspectratio]{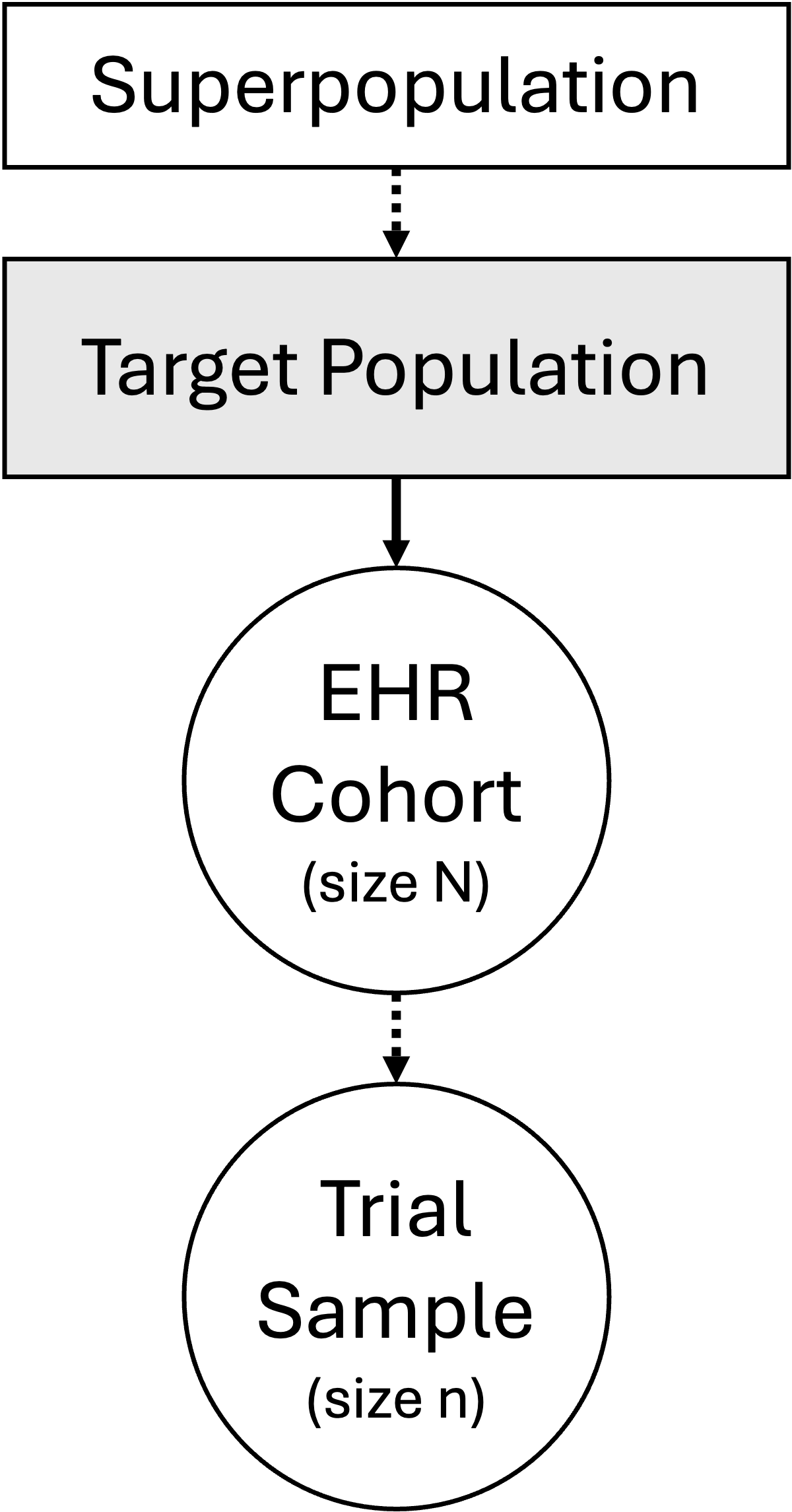}
    \caption{Generalizability}                 
    \label{fig:design-gen}
  \end{subfigure}%
  \hfill
  \begin{subfigure}[t]
  {0.63\textwidth}       
    \centering
    \includegraphics[height=5cm,keepaspectratio]{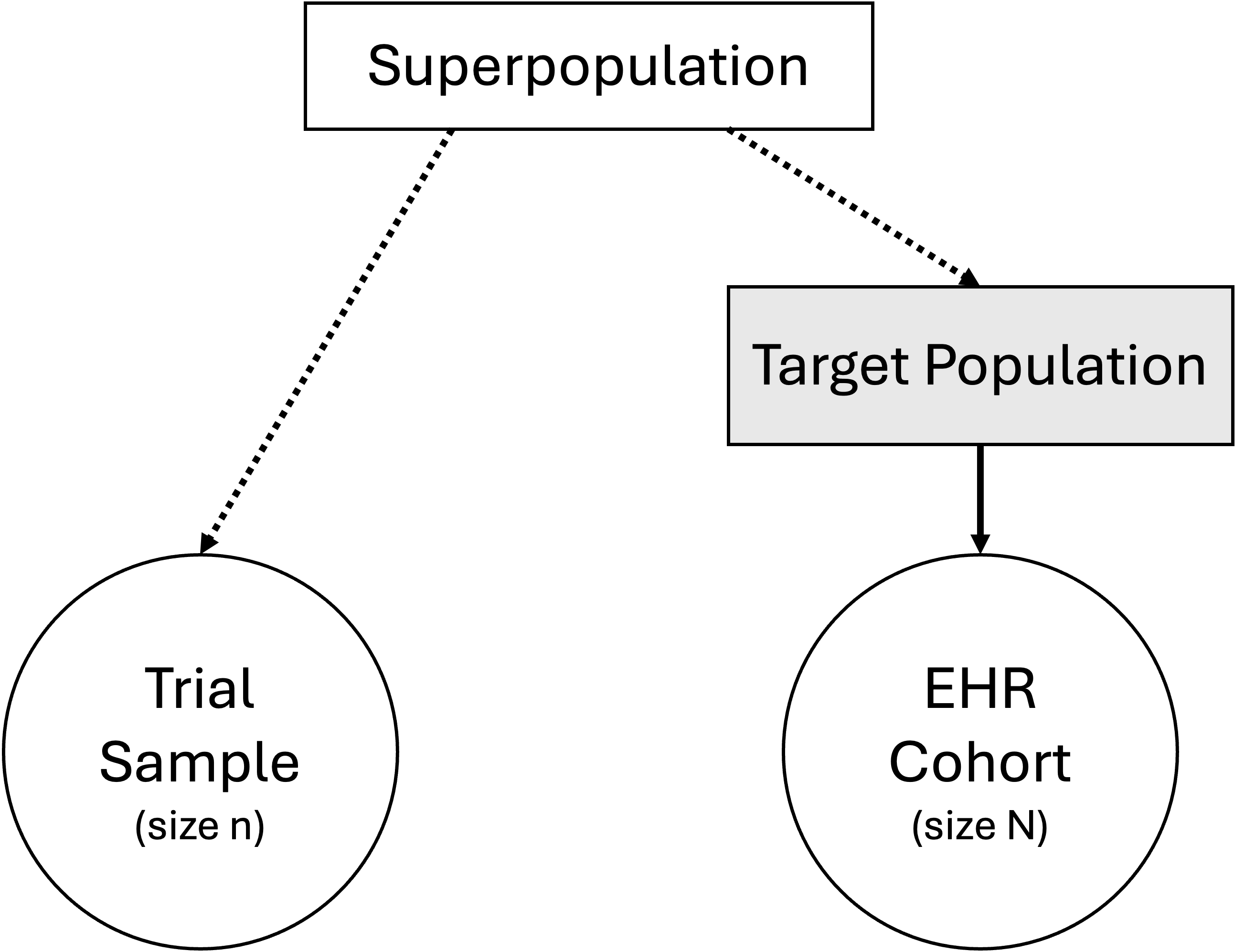}
    \caption{Transportability}                
    \label{fig:design-trans}
  \end{subfigure}
  \caption{Sampling structures under \textbf{(a)} generalizability (nested design) and \textbf{(b)} transportability (non-nested design). Solid arrows: representative sampling. Dotted arrows: selective sampling. Shaded box: target population; circles:
  observed samples.}
  \label{fig:designs}
\end{figure}

\section{Methods}
\label{sec:methods}
\subsection{Notation and Setting}
\label{sec:notation}
Let $\bX$ be baseline covariates, $Y$ a continuous outcome, and $A \in \{0,1\}$ a binary treatment indicator. Let $Y(1)$ and $Y(0)$ denote potential outcomes under intervention and control. We focus on a binary treatment, although our methods can be extended to comparisons of more than two treatments. Consider a superpopulation $\mathbb{P}$ from which individuals' covariates and potential outcomes,  $(X_i, Y_i(0), Y_i(1))$, are  independently and identically distributed. Let $\mathbb{P}_T \subseteq \mathbb{P}$ denote the \textit{target population} to which we seek to extend the trial findings. Although the target population is not observed in its entirety, we assume that an EHR cohort of $N$ individuals is available as a simple random sample from $\mathbb{P}_T$. We use this definition of the EHR cohort in both the nested and non-nested designs.

In a \textit{nested} design (generalizability), a non-random sample of $n$ individuals from the EHR cohort are selected and enrolled in the clinical trial (Figure~\ref{fig:design-gen}). Although the trial sample comes from within the cohort, selective enrollment leaves it unrepresentative of $\mathbb{P}_T$, and trial findings do not directly extend to the target population. In a \textit{non-nested} design (transportability), the $n$ individuals in the clinical trial are not a sample from $\mathbb{P}_T$, but data from the clinical trial are the source of inferences about $\mathbb{P}_T$  (Figure~\ref{fig:design-trans}). In both designs, let $S$ label the observed individuals: $S=1$ for trial participants and $S=0$ for cohort members outside the trial. Treatment is randomized among trial participants with known probability $\Pr(A=1 \cond S=1)$, and the observed outcome is $Y_i = A_i Y_i(1) + (1-A_i) Y_i(0)$. Baseline covariates are observed for all individuals, but treatment and outcome data are observed only for trial participants.

Let $\e$ denote the expectation with respect to $\mathbb{P}$ and $\e_T$ the expectation with respect to $\mathbb{P}_T$. In both designs, the primary estimands are the potential outcome means and the average treatment effect in the target population, $\e_T[Y(a)]$ for $a = 0,1$ and $\e_T[Y(1)-Y(0)]$, so the estimands are unified across the two designs, which differ only in how the trial sample is recruited relative to the EHR cohort. When treatment effects are heterogeneous across $\bX$ and covariate shift exists between trial participants and EHR cohort members outside the trial, $\mathbb{P}(\bX \cond S=1) \neq \mathbb{P}(\bX \cond S=0)$, the average treatment effect among trial participants, $\e[Y(1)-Y(0) \cond S=1]$, generally differs from the target estimand $\e_T[Y(1)-Y(0)]$.

\subsection{Key Assumptions}
\label{sec:assumptions}
Assumptions are needed for identifying the target estimand for generalizability and transportability \citep{Dahabreh2019, Dahabreh2020, Li2022, Li2023, Colnet2024}. We start with three standard assumptions for the internal validity of RCTs: \textbf{A1.\ Consistency}: $Y_i = Y_i(a)$ when $A_i = a$. \textbf{A2.\ Exchangeability over $A$ in the trial}: $(Y(0), Y(1)) \perp A \cond \bX, S = 1$. \textbf{A3.\ Positivity of treatment assignment}: $0 < \Pr(A = a \cond \bX, S = 1) < 1$ for $a = 0, 1$ and all $\bX$ in the support of the trial covariate distribution. In a well-designed RCT, A1--A3 hold by design. To extend inferences to the target population, two additional assumptions are required: \textbf{A4.\ Exchangeability over $S$}: $(Y(0), Y(1)) \perp S \cond \bX$. \textbf{A5.\ Positivity of trial participation}: $0 < \Pr(S = 1 \cond \bX) < 1$ for all $\bX$ in the support of the target covariate distribution. While a weaker condition of A5 suffices for some transportability estimators \citep{Dahabreh2020}, the balanced transformation of the propensity score in the PSPI models requires $\Pr(S = 1 \cond \bX)$, or its estimate, to be supported within $(0,1)$.

Under A1--A5, the potential outcome means are identified in both designs as $\e_T[Y(a)] = \e_T[\e[Y \cond \bX, A=a, S=1]]$, where the inner expectation is the outcome regression among trial participants and the outer expectation is over the covariate distribution of the target population, which the EHR cohort represents by definition. The average treatment effect is identified by the difference $\e_T[Y(1)] - \e_T[Y(0)]$ (proof in Supplementary Materials A.1). 

A4 imposes a strong condition that generally cannot be verified from the observed data, as it requires all covariates associated with potential outcomes to 
be measured. A5 is a population-level condition and does not guarantee adequate covariate overlap in a finite sample. Under substantial covariate shift, some regions of the target covariate distribution may have small but positive probabilities of trial participation and thus be sparsely represented or unrepresented in the observed trial. Consequently, even when A5 holds, estimation may require extrapolation beyond the covariate range observed in the trial and may be sensitive to outcome-model specification \citep{Petersen2012,Tipton2013}. Structural violations of A5, in contrast, require additional identifying assumptions or restriction of the target population. In this study, we assume A1--A5 hold and focus on estimation under limited empirical overlap.

\subsection{Propensity Score Predictive Inference (PSPI)}
\subsubsection{Outcome-Based Estimation and the Target Estimands}
\label{sec:estimation}
The identification results in Section~\ref{sec:assumptions} motivate a Bayesian outcome model strategy. We model the trial outcome as
\begin{equation}
Y_i \cond X_i, A_i, S_i = 1 \;\sim\; \mathcal{N}\big(\mu_{A_i}(X_i),\, \sigma^2\big),
\label{eq:outcome-model}
\end{equation}
where $\mu_a(\cdot)$ is the conditional mean function under treatment $A_i=a$ and $\sigma^2$ is the residual variance. We place priors on $\mu_a$ and $\sigma^2$, obtain their posteriors from the trial data, and for each individual in the EHR cohort draw the imputed potential outcomes $\hat{Y}_i(A_i=a) \cond X_i$ for $a = 0, 1$. 

The estimands in Section~\ref{sec:notation} are averages of the potential outcomes over the target population. Taking the average treatment effect as an example, the target estimand $\e_T[Y(1) - Y(0)]$, referred to as the population average treatment effect (PATE), integrates the treatment effect over the covariate distribution of the target population. However, this covariate distribution is generally unknown in practice. A common strategy in the Bayesian causal inference literature is instead to average over the empirical covariate distribution of the EHR cohort \citep{Li2023b}:
\begin{equation}
\frac{1}{N}\sum_{i} \{\hat{Y}_i(1) - \hat{Y}_i(0)\},
\label{eq:mate}
\end{equation}
which is referred to as the mixed average treatment effect (MATE), a convenient approximation to the PATE. Because the EHR cohort is assumed to be a simple random sample of the target population, we additionally account for uncertainty in the target covariate distribution when estimating the PATE. Specifically, we use a Bayesian bootstrap step \citep{Rubin1981,Li2023b}:
\begin{equation}
\sum_{i} W_i \{\hat{Y}_i(1) - \hat{Y}_i(0)\}, \qquad (W_1, \ldots, W_N) \sim \mathrm{Dirichlet}(1, \ldots, 1).
\label{eq:pate}
\end{equation}

\subsubsection{Outcome Model with the Sampling Propensity Score}
The outcome modeling strategy in Section~\ref{sec:estimation} relies on correct specification of the model fitted to the trial data. Under assumption A4, $Y(a) \perp S \cond \bX$, a correctly specified outcome model in $\bX$, together with the positivity assumptions and the target covariate distribution, is sufficient to identify the target estimands without explicitly modeling trial participation. 
In practice, however, outcome models may be misspecified. To incorporate information about trial participation, we consider the sampling propensity score $\pi(X_i)=\Pr(S_i=1\mid X_i)$, which gives the probability of trial participation conditional on $X_i$, and estimate it as $\hat{\pi}(X_i)$ by regressing $S_i$ on $X_i$. Including $\hat{\pi}(X_i)$ as a predictor may improve predictive performance \citep{Little2004, Liu2022}, particularly when covariates associated with trial participation also predict the outcome but are inadequately represented in the outcome model. We propose the Bayesian Propensity Score Predictive Inference (PSPI) approach, which augments outcome regression with the sampling propensity score and jointly models the trial outcome across treatment arms through an interaction-based formulation:
\begin{equation}
\mu_{A_i}(X_i) = h(X_i, \hat{\pi}(X_i)\cond {\theta_h}) + A_i\cdot s(X_i, \hat{\pi}(X_i)\cond \theta_s)
\label{eq:pspi}
\end{equation}
where $h(\cdot\cond {\theta_h})$ and $s(\cdot \cond {\theta_s})$ are flexible functions with parameters $\theta_h$ and $\theta_s$, respectively. Priors are placed on $\theta_h$ and $\theta_s$. The first component, $h(X_i, \hat{\pi}(X_i)\cond {\theta_h})$, is the prognostic model \citep{Hansen2008}, representing the conditional mean potential outcome under control. The second component, $s(X_i, \hat\pi(X_i)\cond \theta_s)$, captures the treatment effect heterogeneity. Both components include $X_i$ and $\hat\pi(X_i)$ as predictors, because incorporating the sampling propensity score may improve estimation of  both the prognostic and treatment effect functions. 

Since $\hat\pi(X_i)$ is itself a function of $X_i$, including both as predictors in Equation~(\ref{eq:pspi}) may raise identification issues for $\theta_h$ and $\theta_s$. Instead of dropping one covariate in $X_i$ \citep{Zhou2019}, we place regularized priors over $\theta_h$ and $\theta_s$, common in high-dimensional regression. However, this approach raises two concerns. First, treatment effects are typically smaller in scale than the potential outcomes, so a common regularization prior would shrink them disproportionately \citep{Hahn2020}; the interaction-based parametrization in Equation~\eqref{eq:pspi} addresses this by modeling $h(\cdot\cond {\theta_h})$ separately from $s(\cdot\cond {\theta_s})$, allowing distinct regularization priors on the two components. Second, the regularization may attenuate the effect of $\hat\pi(\bX)$. To address this concern, we specify distinct prior structures for $\theta_h$ and $\theta_s$: regularizing priors for the components involving $\bX$ and separate weakly-informative priors for the components involving  $\hat\pi(\bX)$. In Sections~\ref{sec:BART_spline} and \ref{sec:BART_spline_double}, we introduce two PSPI models that implement this split using Bayesian additive regression trees (BART) for the $\bX$ effect and natural cubic splines for the $\hat\pi(\bX)$ effect.

\subsubsection{SplineBART}
\label{sec:BART_spline}
BART, introduced by \citet{Chipman2010}, has gained popularity as a flexible nonparametric method well-suited for modeling complex interactions and nonlinear relationships in high-dimensional settings. Specifically, BART models the regression function as a sum of shallow regression trees that recursively partition the covariate space, with regularization priors on tree structure and leaf values to prevent overfitting. Let $\text{BART}(\cdot)=\sum_{t=1}^{N_{T}}\psi_t(\cdot \cond T_t,B_t)$ be the sum of $N_T$ Bayesian regression trees, where each $\psi_t(\cdot \cond T_t, B_t)$ is a deterministic function representing the $t$-th tree with tree structure parameters $T_t$ and terminal node parameters $B_t$. For brevity, we suppress the tree parameters and denote a BART function by $\text{BART}(\cdot)$.

To estimate the PATE, our first robust PSPI model is \textbf{SplineBART}:
\begin{align}
\label{eq:BART_spline}
\mu_{A_i}(X_i) = \text{BART}_h(X_i, \hat{\tau}_i) + A_i \cdot [\text{BART}_s(X_i) + \text{NS}(\hat{\tau}_i\cond \theta_{s})],
\end{align}
where $\hat{\tau}_i = \phi(\hat{\pi}(X_i))$ is a monotonic transformation on $\hat{\pi}(X_i)$ (the transformation choice will be discussed in Section~\ref{sec:gumbel}), and $\text{NS}(\cdot \cond \theta_{s})$ is a natural cubic spline function with parameters $ \theta_{s} = (\theta_1,  \ldots, \theta_{K-1})^T$ and $K$ pre-specified knots $\xi_1, \xi_2,\ldots,\xi_K$; here, $\xi_1$ and $\xi_K$ are boundary knots, and $\xi_2, \ldots,\xi_{K-1}$ are internal knots \citep{Hastie2009}:
\begin{align*} 
\label{eq:nsb}
& \text{NS}(\hat{\tau}_i \mid \theta_s) = \theta_{1} \hat{\tau}_i + \sum_{k=1}^{K-2} \theta_{k+1} N_k(\hat{\tau}_i) \nonumber \\
& N_k(\hat{\tau}_i) = \frac{(\hat{\tau}_i - \xi_k)^3_+ - (\hat{\tau}_i - \xi_K)^3_+}{\xi_K - \xi_k} - \frac{(\hat{\tau}_i - \xi_{K-1})^3_+ - (\hat{\tau}_i - \xi_K)^3_+}{\xi_K - \xi_{K-1}}, \ k=1,2,\ldots, K-2.
\end{align*} Natural cubic splines provide  extrapolation by enforcing linearity beyond the boundary knots while maintaining flexibility within the boundaries. Here, intercept in $\text{NS}(\hat{\tau}_i \mid \theta_s)$ is removed for identification.

When estimating the PATE using the SplineBART model (\ref{eq:BART_spline}), the prognostic component cancels in predicted treatment contrasts, and the PATE is obtained by averaging the treatment effect component $\text{BART}_s(X_i) + \text{NS}(\hat\tau_i \cond \theta_s)$ over the target covariate distribution. Although the prognostic specification may affect estimation of the treatment effect component through joint model fitting, it does not enter this average directly. Motivated by this distinction, we adopt a more parsimonious prognostic specification by including $\hat{\pi}$ alongside $\bX$ within BART rather than adding a separate spline term. 

Although BART effectively captures complex relationships within the training data, its tree-based predictions do not naturally extend trends beyond the observed covariate range. This limitation is relevant to generalizability and transportability, where covariate shift may lead to some regions of the EHR cohort sparsely represented or unrepresented in the trial. Augmenting BART with a natural cubic spline of the transformed $\hat\pi(\bX)$ not only improves prediction performance within the observed range but also allows an explicit extrapolation rule. Beyond the spline's boundary knots, its contribution extends linearly on the transformed score scale, with slopes determined by the fitted spline at the boundary knots. When multiple covariates exhibit distributional shift, specifying extrapolation relationships directly in the high-dimensional covariate space can be difficult. The sampling propensity score provides a tractable one-dimensional summary that yields balance with respect to trial participation \citep{ROSENBAUM1983}. 

To fit model~\eqref{eq:BART_spline}, we use 200 trees for each of $\text{BART}_h(\cdot)$ and $\text{BART}_s(\cdot)$ and adopt BART prior specifications of \citet{Chipman2010}; additional details are provided in Supplementary Materials A.2. For the error variance $\sigma^2$, we specify a conjugate inverse Gamma prior, $\text{InvGamma}\left(\frac{\upsilon}{2}, \frac{\upsilon\lambda}{2}\right)$, with $\upsilon = 3$ to ensure a weakly informative prior. The scale parameter $\lambda$ is determined using an empirical Bayes approach, matching the residual variance from an initial least squares fit, as described in \citet{Chipman2010}. For $\text{NS}(\cdot \cond \theta_{s})$, we assign a flat prior to $\theta_{1}$, and a multivariate normal prior $\mathcal{MVN}_{K-2}(0, \gamma^2 \mathbb{1})$ for $\theta_{2:K-1}$, where $\gamma^2$ controls global smoothness and $\mathbb{1}$ is an identity matrix. A weakly informative hyperprior $\text{InvGamma}(1, 1)$ is placed on $\gamma^2$. The $K$ fixed knots $\xi_1, \xi_2,\ldots,\xi_K$ are placed at evenly spaced quantiles of $\hat{\tau}$ among intervention-arm  trial participants ($S_i=1$ and $A_i=1$). This choice reflects the fact that the spline appears only in the term multiplied by $A_i$ in Equation~\eqref{eq:BART_spline}, so only intervention arm observations contribute to the estimation of $\theta_s$. To adapt model complexity, $K$ is determined by the sample-size rule $K = \max(2,  (\sum S_iA_i)^{1/3})$ (rounded), so that the spline reduces to linear regression when intervention arm size is very small in trial and becomes more flexible as its size increases. The Gibbs sampler of model~\eqref{eq:BART_spline} is in Web Algorithm 1 in Supplementary Materials A.3. 

\subsubsection{DSplineBART}
\label{sec:BART_spline_double}
Unlike the PATE,  estimation of the mean potential outcomes $\e_T[Y(1)]$ and $\e_T[Y(0)]$ depends directly on the prognostic component, with $\e_T[Y(1)]$ also depending on the treatment effect component. To improve the robustness in estimating the prognostic component, we propose the second PSPI variant that replaces $\text{BART}_h(X_i, \hat{\tau}_i)$ in model (\ref{eq:BART_spline}) with $\text{BART}_h(X_i) + \text{NS}_h(\hat{\tau}_i \cond \theta_h)$. With double splines in the model, we refer to it as \textbf{DSplineBART}:
\begin{equation}
\mu_{A_i}(X_i) =  \text{BART}_h(X_i) + \text{NS}_h(\hat{\tau}_i \cond \theta_{h}) + A_i \cdot [\text{BART}_s(X_i) + \text{NS}_s(\hat{\tau}_i\cond \theta_{s})],
\label{eq:DSplineBART}
\end{equation}
where $\text{NS}_h(\cdot \cond \theta_{h})$ represents another separate natural cubic spline with $H$ pre-specified knots $\xi_{h,1}, \xi_{h,2},\ldots,\xi_{h,H}$ and coefficients $ \theta_{h} = (\theta_{h,1}, \ldots, \theta_{h,H-1})^T$. For the BART components, we place the same priors as SplineBART, each with 200 trees. For $\text{NS}_h(\cdot)$, a flat prior is specified for $\theta_{h,1}$ and multivariate normal prior $\mathcal{MVN}_{H-2}(0, \gamma^2_h\mathbb{1})$ is placed on $\theta_{h,2:(H-1)}$, with $\gamma_h^2\sim\text{InvGamma}(1, 1)$. Since $\text{NS}_h(\cdot)$ is used in estimating potential outcomes for both treatment arms in Equation~\eqref{eq:DSplineBART}, all trial observations inform $\theta_h$. Accordingly, we placed the knots at evenly spaced quantiles of $\hat{\tau}$ across the full trial sample and set $H = \max(2, n^{1/3})$ (rounded). Restricting knot placement to the control arm could require linear extrapolation for intervention-arm $\hat{\tau}$ values outside the control-arm boundary knots.
The Gibbs sampler for the model~\eqref{eq:DSplineBART} is in Web Algorithm 2 in Supplementary Materials A.3.

\subsubsection{Inverse Gumbel Transformation of Sampling Propensity Scores}
\label{sec:gumbel}
In generalizability and transportability analyses, $\pi(X_i)$ denotes the probability of trial participation given $X_i$. When the trial is much smaller than the EHR cohort, estimated sampling propensity scores may be concentrated near zero in the EHR cohort. Trial participants nevertheless tend to have higher scores than EHR participants because individuals with higher participation probabilities are more likely to enter the trial. As a result, the distribution of $\hat \pi(X_i)$ of the trial sample may be shifted toward the upper end of the score distribution of the EHR cohort, leaving many low-score EHR observations sparsely represented or absent from the observed trial. Prediction for these observations may therefore require spline extrapolation toward the lower end of the score distribution. 

For both SplineBART and DSplineBART, the performance of the spline components depends on the scale of $\hat{\pi}(X_i)$ and the choice of monotonic transformation $\phi(\cdot)$. The Identity link $\phi(\hat{\pi}(X_i)) = \hat{\pi}(X_i)$ often suffers from boundary issues, as the narrow region near 0 forces the spline to extrapolate over too little space, leading to undifferentiated and undercovered predictions in extrapolation regions. PENCOMP \citep{Zhou2019} addressed this by using the Logit transformation, $\phi(\hat{\pi}(X_i)) = \log(\frac{\hat{\pi}(X_i)}{1-\hat{\pi}(X_i)})$, mapping $(0, 1)$ to $\mathbb{R}$ and thereby removing boundary bias. However, the Logit transformation is not well suited for generalizability and transportability settings, as it excessively expands values near 0---common in the EHR cohort---into a wide range, while compressing larger values---common in the trial sample---into a narrow band. As a result, the spline is fit on a limited range but must extrapolate across a disproportionately large space, leading to unstable predictions with high variability. To overcome these issues, we exploit the inverse Gumbel function (InvGumbel), \[\phi(\hat{\pi}(X_i)) = -\log(-\log(\hat{\pi}(X_i))),\] which also maps $\hat{\pi}$ from $(0, 1)$ to $\mathbb{R}$, avoiding boundary bias. Compared to the Logit scale, it expands values near 0 more gradually and allocates a larger share of the transformed range to the trial region, thereby improving extrapolation and reducing variability. Additional properties of the InvGumbel function are discussed in Supplementary Materials A.4. We further validate this transformation choice through simulations in Section~\ref{sec:choice_transformation}.

    \section{Simulation Studies}
\label{sec:simulation}
\subsection{Simulation Design}
We conducted simulations to evaluate the performance of PSPI for both generalizability and transportability under various settings. In all settings, simulations were based on a superpopulation of $100{,}000$ individuals, from which 250 replicate datasets were drawn. For each individual in the superpopulation, we generated seven continuous covariates $\bX_{1:7}\sim \mathcal{MVN}_7(\mathbf{0}, \Sigma)$, each with unit variance, and with $\mathrm{cor}(X_1,X_2)=0.2$, $\mathrm{cor}(X_3,X_4)=0.5$, and all other pairs uncorrelated, as well as three independent binary covariates $\bX_{8:10}\sim \mathrm{Bern}(0.5)$. Potential outcomes and the sampling mechanisms depended on $\bX_{1:5}$ and $X_8$, whereas $\bX_{6:7}$ and $\bX_{9:10}$ were included as noise variables. 

We followed the sampling structure shown in Figure~\ref{fig:designs}. For transportability, we first drew a target population of size $10{,}000$ from the superpopulation using the \texttt{sample()} function in \texttt{R} without replacement and with \texttt{prob} parameter specified as $\text{invlogit}(-0.3X_1)$. We then drew a simple random sample from this target population as the EHR cohort. Separately, to mimic the motivating data application, we drew a trial sample of size $n=150$ from the superpopulation using the same \texttt{sample()} function without replacement, with the \texttt{prob} argument given by different participation mechanisms below. We present four settings of transportability and their results in the main text, while the other settings of transportability and the corresponding generalizability settings and results are reported in Supplementary Materials B.2.

\noindent \textbf{SS1. Linear setting with varying EHR cohort size}: We considered a linear model for potential outcomes: $Y(a)\sim\mathcal{N}(2X_{1} -1.5 X_{2} + 0.5 X_{3} + X_{4} + X_{8} + \mathbb{I}(a = 1)[2 + 3 X_{1}+2X_{2} + X_{3} + X_{4} + 3 X_{8}],1)$, for $a=0,1$. The EHR cohort size was set to $N=500$, $1,000$, or $2,000$. For the trial sample, the \texttt{prob} argument in \texttt{sample()} was specified as $\mathrm{invlogit}(-1.41 + 1.5 X_{1} - 0.7 X_{2}  + 0.5 X_{5}+ X_{4} - 2X_{8})$. Treatment was randomized among trial participants with $\Pr(A=1)=0.5$.

\noindent \textbf{SS2. Linear settings with substantial covariate shift}: We used the same treatment assignment and outcome models as in SS1 with $N=500$ or $1,000$. Trial participation was modeled through the \texttt{prob} argument: $\text{invlogit}(-2.38 + 1.8X_1 - 1.5X_2^2  - 0.8X_5X_8 - 6 (X_4 - 0.5)^2)$. Individuals with extreme values of $X_{4}$ were much less likely to be selected into the trial sample, resulting in a substantial covariate shift in $X_{4}$ between the trial sample and the EHR cohort.

\noindent \textbf{SS3. Nonlinear settings}: Potential outcomes were generated nonlinearly as $Y(a)\sim\mathcal{N}(2X_{1} + X_{2} +3 X_{3} + 2X_{4} +3X_{8} + 2 X_{1}X_{3}X_{8}+0.8X_{2}X_{4}+0.5X_{1}^2 + \mathbb{I}(a = 1)[2 + 3 X_{1}+X_{2} - X_{3} - 2X_{4} + 3 X_{8} + 1.5X_{1}X_{4} + 0.8 X_{3}^2 + 0.5X_{2}X_{8} + 0.3X_{4}^3X_{8} + \text{invlogit}(X_{1}X_{2})],1)$, for $a=0,1$. We considered nonlinear participation models: $\text{invlogit}(-2.5 + 4 X_{1} - 1.5 X_{2}^2 + 1.2X_{4} -2.4X_8 + 0.6X_{5}+0.6X_{1}X_4)$. The EHR cohort size was $N=500$ or $1,000$. Treatments were randomly assigned with $\Pr(A=1)=0.5$.

\noindent \textbf{SS4. Nonlinear settings with substantial covariate shift}: We used the same EHR cohort size, treatment assignment, and outcome models as in SS3. Trial participation was modeled as: $\text{invlogit}(-2 + 8X_1 - 5X_2^2 -0.8X_5X_8 - 5 (X_4 - 0.5)^2)$. Individuals with extreme values of $X_{2}$ or $X_4$, or small values of $X_1$ were less likely to be selected, resulting in a strong covariate shift between the trial sample and the EHR cohort.

We compared SplineBART and DSplineBART with several alternative methods for estimating $\e_T[Y(1) - Y(0)]$, $\e_T[Y(1)]$, and $\e_T[Y(0)]$.
$\bullet$~\textbf{Trial-only}: the mean estimator within the trial sample.
$\bullet$~\textbf{IPPW} \citep{Dahabreh2019}: the inverse probability of participation weighting estimator, with $\hat{\pi}$ fit by a probit BART model on all covariates; this estimated sampling propensity score is used by every method requiring $\hat{\pi}$.
$\bullet$~\textbf{A-IPPW} \citep{Dahabreh2019}: a doubly robust augmented IPPW estimator, combining participation weighting with outcome regression using two separate arm-specific BART models to predict the potential outcomes.
$\bullet$~\textbf{BART}: an outcome regression applying equation~(\ref{eq:outcome-model}) with $\mu_{A_i}=\text{BART}(X_i, A_i)$ \citep{Kern2017}.
$\bullet$~\textbf{BART-PS}: BART with the propensity score added, $\mu_{A_i}=\text{BART}(X_i, A_i, \hat{\pi}_i)$.
$\bullet$~\textbf{InterBART}: a PSPI approach without $\hat{\pi}$, $\mu_{A_i}=\text{BART}_h(X_i)+A_i\cdot\text{BART}_s(X_i)$.
$\bullet$~\textbf{InterBART-PS}: InterBART extended by including $\hat{\pi}$ in both terms, $\mu_{A_i}=\text{BART}_h(X_i, \hat{\pi}_i)+A_i\cdot\text{BART}_s(X_i, \hat{\pi}_i)$.
$\bullet$~\textbf{FullBART}: a PSPI variant of SplineBART replacing the spline with a separate BART, $\mu_{A_i}=\text{BART}_h(X_i, \hat{\pi}_i)+A_i\cdot[\text{BART}_s(X_i)+\text{BART}_{\pi}(\hat{\pi}_i)]$, under the identifiability constraint on the trial sample: $\frac{1}{n}\sum_{i=1}^n\text{BART}_s(X_i)=0$.
$\bullet$~\textbf{True model}: a benchmark using the Bayesian outcome regression with the correctly specified outcome models. 

For outcome regression methods that include the sampling propensity score as a predictor, we used the posterior mean of the propensity score in the regression \citep{Wang2025}. For IPPW and A-IPPW, we obtained posterior draws of  $\hat{\pi}$ and conditional mean outcomes $\hat{Y}(a)$ from separate BART models. We computed IPPW estimates using each posterior draw of $\hat{\pi}$ and A-IPPW using paired draws of $\hat{\pi}$ and $\hat{Y}(a)$. Point estimates were obtained by averaging these draw-specific estimates. Following the approach of \citet{Antonelli2022}, we estimated the variance of each estimator by summing the bootstrap variance of its posterior-averaged estimate with its posterior variance evaluated on the original data. The propensity-score model was fitted once, and its posterior draws were reused across bootstrap samples. 

For all the comparison methods, we estimated the MATE and corresponding mean potential outcomes by averaging over the EHR cohort, as in Equation (\ref{eq:mate}), and compared these estimates with true quantities averaged over the same individuals. For SplineBART, DSplineBART, and the true model, we additionally estimated the PATE and population mean potential outcomes using the Bayesian bootstrap approach in Equation (\ref{eq:pate}) and compared these estimates with the corresponding target-population quantities. All Bayesian models were fit using 5{,}000 burn-in iterations followed by 5{,}000 posterior draws. Point estimates were obtained as posterior means, with 95\% equal-tailed probability intervals. Additional details on each comparison method are provided in Supplementary Materials B.1.

Performance was evaluated based on average bias (Bias) and root mean squared error (RMSE), defined as $\text{Bias} = \frac{1}{250}\sum_{r=1}^{250}\text{bias}_r$ and $\text{RMSE} = \sqrt{\frac{1}{250}\sum_{r=1}^{250}\text{bias}_r^2}$, where $\text{bias}_r$ denotes the difference between the estimated and true values of the estimand in the $r$-th replicate ($r=1,\ldots,250$). We also reported the empirical coverage rate and the average width of the 95\% credible intervals. For readability, all performance metrics were multiplied by 100.

\subsection{Simulation Results}

\begin{table}[t]
\caption{Comparison of bias (×100), RMSE (×100), and coverage rate (×100) and average width (×100) of 95\% confidence/credible interval in simulation scenarios SS1 and SS2.}
\label{tab:s12}
\resizebox{\textwidth}{!}{%
\begin{tabular}{lcccccccccccccc}
\toprule
\multicolumn{15}{c}{\textbf{SS1. Linear, $N = 500$, $n=150$}} \\
\hline
\multirow{2}{*}{\textbf{Estimators}} & \multicolumn{4}{c}{\textbf{$\e_T[Y(1)-Y(0)]$}} & & \multicolumn{4}{c}{\textbf{$\e_T[Y(1)]$}} & & \multicolumn{4}{c}{\textbf{$\e_T[Y(0)]$}} \\ \cline{2-5} \cline{7-10} \cline{12-15}
  & \textbf{Bias} & \textbf{RMSE} & \textbf{Coverage} & \textbf{Width} & & \textbf{Bias} & \textbf{RMSE}  & \textbf{Coverage} & \textbf{Width} & & \textbf{Bias} & \textbf{RMSE}  & \textbf{Coverage} & \textbf{Width} \\
 \hline
\textbf{Existing methods} & & & & & & & & & & & & & &\\
\ \ Trial-only & 218 & 231 & 18 & 289 & & 449 & 455 & 0 & 265 & & 231 & 234 & 0 & 116\\
\ \ IPPW & 71 & 147 & \bf 95 & 648 & & 180 & 210 & 74 & 608 & & 109 & 121 & 54 & 264\\
\ \ A-IPPW & 37 & \bf 67 & 86 & 201 & & 103 & 112 & 37 & 177 & & 65 & 74 & 39 & 116\\
\ \ BART & 92 & 104 & 44 & 173 & & 98 & 107 & 26 & 137 & & 6 & 35 & 93 & 132\\
\ \ BART-PS & 95 & 107 & 50 & 185 & & 110 & 121 & 28 & 160 & & 15 & 42 & 93 & 157\\
\textbf{Interaction-based outcome models} & & & & & & & & & & & & & &\\
\ \ InterBART & 93 & 101 & 45 & 171 & & 102 & 109 & 25 & 144 & & 9 & 33 & 95 & 131\\
\ \ InterBART-PS & 104 & 111 & 37 & 173 & & 116 & 124 & 24 & 166 & & 12 & 40 & 93 & 155\\
\ \ FullBART & 84 & 95 & 58 & 188 & & 108 & 118 & 34 & 171 & & 24 & 45 & 93 & 158\\
\ \ SplineBART & \bf -8 & \bf 66 & \bf 96 & 271 & & 41 & 70 & 90 & 223 & & 49 & 62 & 80 & 164\\
\ \ DSplineBART & \bf 4 & \bf 65 & \bf 95 & 271 & & \bf 18 & \bf 56 & \bf 93 & 214 & & 13 & 46 & \bf 94 & 187\\
\textbf{True model} & -1 & 29 & 93 & 107 & & -1 & 23 & 95 & 88 & & 1 & 16 & 95 & 60\\
\midrule
\textbf{With Bayesian bootstrap} & & & & & & & & & & & & & &\\
\ \ SplineBART & \bf -11 & \bf 69 & \bf 96 & 279 & & 39 & 73 & 89 & 243 & & 49 & 64 & 84 & 172\\
\ \ DSplineBART & \bf 2 & \bf 68 & \bf 96 & 279 & & \bf 15 & \bf 61 & \bf 95 & 235 & & 14 & 48 & \bf 94 & 195\\
\ \ True model  & -4 & 34 & 96 & 136 & & -3 & 34 & 97 & 142 & & 1 & 19 & 95 & 78\\
\midrule
\midrule
\multicolumn{15}{c}{\textbf{SS2. Linear, strong covariate shift, $N = 500$, $n=150$}} \\
\hline
\multirow{2}{*}{\textbf{Estimators}} & \multicolumn{4}{c}{\textbf{$\e_T[Y(1)-Y(0)]$}} & & \multicolumn{4}{c}{\textbf{$\e_T[Y(1)]$}} & & \multicolumn{4}{c}{\textbf{$\e_T[Y(0)]$}} \\ \cline{2-5} \cline{7-10} \cline{12-15}
  & \textbf{Bias} & \textbf{RMSE} & \textbf{Coverage} & \textbf{Width} & & \textbf{Bias} & \textbf{RMSE}  & \textbf{Coverage} & \textbf{Width} & & \textbf{Bias} & \textbf{RMSE}  & \textbf{Coverage} & \textbf{Width} \\
 \hline
\textbf{Existing methods} & & & & & & & & & & & & & &\\
\ \ Trial-only & 502 & 506 & 0 & 247 & & 806 & 809 & 0 & 228 & & 305 & 306 & 0 & 98\\
\ \ IPPW & 241 & 283 & 60 & 619 & & 390 & 409 & 25 & 582 & & 150 & 160 & 32 & 251\\
\ \ A-IPPW & 146 & 165 & 37 & 231 & & 255 & 262 & 1 & 194 & & 109 & 115 & 16 & 135\\
\ \ BART & 233 & 239 & 0 & 185 & & 251 & 258 & 0 & 172 & & 18 & 45 & 95 & 168\\
\ \ BART-PS & 211 & 219 & 9 & 215 & & 235 & 246 & 4 & 215 & & 25 & 57 & 95 & 213\\
\textbf{Interaction-based outcome models} & & & & & & & & & & & & & &\\
\ \ InterBART & 237 & 243 & 1 & 195 & & 264 & 270 & 0 & 192 & & 27 & 48 & 93 & 177\\
\ \ InterBART-PS & 251 & 257 & 0 & 197 & & 265 & 273 & 0 & 231 & & 13 & 50 & 97 & 220\\
\ \ FullBART & 211 & 218 & 5 & 226 & & 248 & 257 & 2 & 239 & & 37 & 60 & 94 & 225\\
\ \ SplineBART & \bf -29 & \bf 124 & \bf 95 & 491 & & 58 & 127 & 90 & 428 & & 87 & 99 & 72 & 232\\
\ \ DSplineBART & \bf -14 & \bf 126 & \bf 96 & 503 & & \bf 33 & \bf 120 & \bf 93 & 432 & & 47 & 90 & \bf 96 & 374\\
\textbf{True model} & 2 & 32 & 93 & 126 & & 1 & 31 & 93 & 116 & & -1 & 12 & 94 & 48\\
\midrule
\textbf{With Bayesian bootstrap} & & & & & & & & & & & & & &\\
\ \ SplineBART & \bf -32 & \bf 124 & \bf 95 & 494 & & 55 & 126 & 91 & 435 & & 87 & 98 & 73 & 236\\
\ \ DSplineBART & \bf -16 & \bf 127 &\bf  96 & 507 & & \bf 30 & \bf 120 & \bf 93 & 440 & & 46 & 90 & \bf 96 & 377\\
\ \ True model & -1 & 38 & 95 & 152 & & -2 & 41 & 94 & 161 & & -1 & 16 & 96 & 70\\
\bottomrule
\end{tabular}
}
\end{table}

\begin{table}[t]
\caption{Comparison of bias (×100), RMSE (×100), and coverage rate (×100) and average width (×100) of 95\% confidence/credible interval in simulation scenarios SS3 and SS4.}
\label{tab:s34}
\resizebox{\textwidth}{!}{%
\begin{tabular}{lcccccccccccccc}
\toprule
\multicolumn{15}{c}{\textbf{SS3. Nonlinear, $N = 500$, $n=150$}} \\
\hline
\multirow{2}{*}{\textbf{Estimators}} & \multicolumn{4}{c}{\textbf{$\e_T[Y(1)-Y(0)]$}} & & \multicolumn{4}{c}{\textbf{$\e_T[Y(1)]$}} & & \multicolumn{4}{c}{\textbf{$\e_T[Y(0)]$}} \\ \cline{2-5} \cline{7-10} \cline{12-15}
  & \textbf{Bias} & \textbf{RMSE} & \textbf{Coverage} & \textbf{Width} & & \textbf{Bias} & \textbf{RMSE}  & \textbf{Coverage} & \textbf{Width} & & \textbf{Bias} & \textbf{RMSE}  & \textbf{Coverage} & \textbf{Width} \\
 \hline
\textbf{Existing methods} & & & & & & & & & & & & & &\\
\ \ Trial-only & 340 & 363 & 21 & 479 & & 830 & 837 & 0 & 388 & & 490 & 495 & 0 & 284\\
\ \ IPPW & 166 & 284 & \bf 91 & 969 & & 433 & 464 & 42 & 816 & & 268 & 298 & 48 & 585\\
\ \ A-IPPW & 99 & \bf 146 & 87 & 438 & & 238 & 253 & 30 & 371 & & 138 & 149 & 46 & 266\\
\ \ BART & 155 & 176 & 49 & 304 & & 181 & 202 & 36 & 284 & & 26 & 73 & 93 & 281\\
\ \ BART-PS & 157 & 177 & 57 & 332 & & 232 & 249 & 28 & 340 & & 75 & 104 & 86 & 336\\
\textbf{Interaction-based outcome models} & & & & & & & & & & & & & &\\
\ \ InterBART & 185 & 195 & 24 & 281 & & 201 & 216 & 29 & 300 & & 16 & 66 & 98 & 289\\
\ \ InterBART-PS & 204 & 213 & 14 & 280 & & 257 & 270 & 20 & 358 & & 53 & 89 & 94 & 344\\
\ \ FullBART & 152 & 169 & 64 & 347 & & 241 & 255 & 30 & 373 & & 88 & 113 & 88 & 362\\
\ \ SplineBART & \bf -46 & 156 & \bf 97 & 677 & & 87 & 161 & \bf 93 & 590 & & 133 & 150 & 72 & 371\\
\ \ DSplineBART & \bf -20 & \bf 146 & \bf 98 & 683 & & \bf 32 & \bf 136 & \bf 96 & 560 & & 52 & 112 & \bf 98 & 507\\
\textbf{True model} & 0 & 58 & 92 & 213 & & -2 & 62 & 95 & 238 & & -2 & 53 & 96 & 210\\
\midrule
\textbf{With Bayesian bootstrap} & & & & & & & & & & & & & &\\
\ \ SplineBART & \bf -48 & 157 & \bf 98 & 680 & & 83 & 161 & \bf 92 & 605 & & 130 & 147 & 78 & 387\\
\ \ DSplineBART & \bf -21 & \bf 148 & \bf 98 & 686 & & \bf 28 & \bf 139 & \bf 96 & 577 & & 49 & 110 & \bf 98 & 520\\
\ \ True model & -2 & 62 & 93 & 233 & & -6 & 69 & 96 & 273 & & -4 & 58 & 95 & 231\\
\midrule
\midrule
\multicolumn{15}{c}{\textbf{SS4. Nonlinear, strong covariate shift, $N = 500$, $n=150$}} \\
\hline
\multirow{2}{*}{\textbf{Estimators}} & \multicolumn{4}{c}{\textbf{$\e_T[Y(1)-Y(0)]$}} & & \multicolumn{4}{c}{\textbf{$\e_T[Y(1)]$}} & & \multicolumn{4}{c}{\textbf{$\e_T[Y(0)]$}} \\ \cline{2-5} \cline{7-10} \cline{12-15}
  & \textbf{Bias} & \textbf{RMSE} & \textbf{Coverage} & \textbf{Width} & & \textbf{Bias} & \textbf{RMSE}  & \textbf{Coverage} & \textbf{Width} & & \textbf{Bias} & \textbf{RMSE}  & \textbf{Coverage} & \textbf{Width} \\
 \hline
\textbf{Existing methods} & & & & & & & & & & & & & &\\
\ \ Trial-only & 403 & 417 & 4 & 419 & & 876 & 880 & 0 & 336 & & 472 & 477 & 0 & 253\\
\ \ IPPW & 261 & 307 & 66 & 704 & & 607 & 619 & 7 & 619 & & 346 & 355 & 11 & 420\\
\ \ A-IPPW & 167 & 185 & 64 & 386 & & 334 & 341 & 0 & 320 & & 166 & 174 & 23 & 249\\
\ \ BART & 226 & 234 & 6 & 246 & & 303 & 313 & 1 & 253 & & 77 & 100 & 78 & 255\\
\ \ BART-PS & 221 & 229 & 13 & 276 & & 316 & 326 & 3 & 320 & & 95 & 119 & 81 & 321\\
\textbf{Interaction-based outcome models} & & & & & & & & & & & & & &\\
\ \ InterBART & 241 & 245 & 1 & 238 & & 314 & 322 & 1 & 279 & & 74 & 96 & 85 & 273\\
\ \ InterBART-PS & 252 & 256 & 1 & 241 & & 335 & 343 & 1 & 349 & & 83 & 108 & 88 & 339\\
\ \ FullBART & 204 & 211 & 22 & 318 & & 317 & 326 & 5 & 366 & & 113 & 131 & 80 & 357\\
\ \ SplineBART & \bf -24 & \bf 164 & \bf 98 & 809 & & 130 & 203 & \bf 94 & 726 & & 154 & 167 & 64 & 361\\
\ \ DSplineBART & \bf 1 & \bf 163 & \bf 99 & 833 & & \bf 88 & \bf 180 & \bf 96 & 721 & & 88 & 147 & \bf 98 & 653\\
\textbf{True model} & 4 & 55 & 96 & 218 & & -1 & 70 & 96 & 277 & & -6 & 64 & 96 & 251\\
\midrule
\textbf{With Bayesian bootstrap} & & & & & & & & & & & & & &\\
\ \ SplineBART & \bf -24 & \bf 163 & \bf 99 & 811 & & 128 & 201 & \bf 93 & 737 & & 152 & 165 & 69 & 375\\
\ \ DSplineBART & \bf 1 & \bf 162 & \bf 99 & 835 & & \bf 86 & \bf 179 & \bf 95 & 733 & & 85 & 148 & \bf 97 & 662\\
\ \ True model & 5 & 59 & 96 & 238 & & -3 & 75 & 96 & 310 & & -8 & 68 & 96 & 270\\

\bottomrule
\end{tabular}
}
\end{table}

We present results with $N=500$ for Scenarios SS1 and SS2 in Table~\ref{tab:s12}, and SS3 and SS4 in Table~\ref{tab:s34}, respectively. Additional results are summarized in the Supplementary Materials B.2. In Scenario SS1 with $N=500$, the true model achieved the most consistent point estimates and nominal-level coverage with the narrowest intervals, indicating that correctly specified outcome models are sufficient for both the MATE and the PATE. The trial-only and IPPW estimators exhibited the largest RMSE and interval widths when estimating both $\e_T[Y(1) - Y(0)]$ and $\e_T[Y(a)]$. For $\e_T[Y(1) - Y(0)]$, SplineBART and DSplineBART achieved lower bias and comparable RMSE to A-IPPW, while substantially outperforming the other methods. Only IPPW, SplineBART, and DSplineBART attained near-nominal coverage, with IPPW having wider intervals. FullBART outperformed both InterBART and InterBART-PS, demonstrating the benefit of modeling $\hat{\pi}$ with a separate component. When estimating $\e_T[Y(0)]$, DSplineBART outperformed SplineBART, while BART, BART-PS, InterBART, InterBART-PS, and FullBART also performed well. However, most methods failed to estimate $\e_T[Y(1)]$ consistently, with DSplineBART being the only estimator that achieved both good point estimates and near-nominal coverage, demonstrating its added robustness. The PATE estimates of the two spline-based models  using Bayesian bootstrap were similar to their MATE estimates. When the EHR cohort size increases to $1,000$ and $2,000$ (Web Table 1), the overall patterns were largely unchanged across most models, with SplineBART and DSplineBART exhibiting narrower 95\% credible intervals. 

In Scenario SS2 with strong covariate shift, the spline-based models using $\hat{\pi}$ showed stable extrapolation. SplineBART and DSplineBART achieved much smaller bias and RMSE than all other methods, with coverage near nominal, though with wider intervals in estimating both $\e_T[Y(1) - Y(0)]$ and $\e_T[Y(a)]$. All other methods performed poorly. As in Scenarios SS1, DSplineBART outperformed SplineBART for estimating $\e_T[Y(1)]$ and $\e_T[Y(0)]$. These findings also held for $N=1,000$, with slightly narrower intervals. While FullBART outperformed InterBART and InterBART-PS in SS1, the advantage of modeling $\hat{\pi}_i$ using a separate BART function disappeared, given that BART has limited extrapolation ability under strong covariate shift.

In scenarios SS3 and SS4, where outcome models were nonlinear, the results were generally consistent with those from the linear settings. Although the intervals for SplineBART and DSplineBART widened further, both spline-based PSPI methods maintained near-nominal coverage even under covariate shift when estimating $\e_T[Y(1) - Y(0)]$ and $\e_T[Y(1)]$, whereas most other models showed severe undercoverage. These findings suggest that the spline-based PSPI models provide robust extrapolation performance even under nonlinear settings. 

We also considered unequal treatment allocation, with $\Pr(A=1)=0.3$ or $0.7$. The findings were similar (Web Table 3), suggesting that the spline-based PSPI methods are robust to these levels of treatment imbalance. Similar advantages were observed in the generalizability simulations (Web Tables 4--6). Together, these results support the use of SplineBART and DSplineBART for PATE estimation, and DSplineBART for estimating mean potential outcomes, in generalizability and transportability analyses, particularly under substantial covariate shift.

\subsection{Choices of Transformations on Estimated Propensity Scores}
\label{sec:choice_transformation}

To assess the impact of different transformations of $\hat{\pi}$ in the spline-based PSPI models, we conducted additional simulations to compare the performance of DSplineBART under four transformations: Identity ($\phi(\hat{\pi}) = \hat{\pi}$), Logit ($\phi(\hat{\pi}) = \log[\frac{\hat{\pi}}{1 - \hat{\pi}}]$), Complementary log-log (Cloglog; $\phi(\hat{\pi}) = \log[-\log(1 - \hat{\pi})]$), and InvGumbel ($\phi(\hat{\pi}) = -\log[-\log(\hat{\pi})]$). The results are summarized in Web Tables 7 and 8 of the Supplementary Materials B.3.

The InvGumbel transformation achieved the lowest bias and a lower or comparable RMSE to the Identity transformation for estimating $\e_T[Y(1) - Y(0)]$. Although the Identity transformation produced narrowest intervals, it suffered from undercoverage. In contrast, InvGumbel attained near-nominal coverage with substantially narrower intervals than the Logit and Cloglog transformations, and only slightly wider than Identity. Similar patterns were observed for $\e_T[Y(a)]$. 

\begin{figure}[t]
\caption{Histogram of $\phi(\hat{\pi})$ and posterior estimates of the treatment-effect spline $\text{NS}_s(\phi(\hat{\pi}_i)\cond\theta_s)$ from DSplineBART in a simulation replicate under Scenario SS1 ($N=500$). The upper panel displays histograms of $\hat{\pi}$ after transformations in the EHR cohort (above x-axis) and the trial (below x-axis). The lower panel shows the posterior means and 95\% credible intervals for $\text{NS}_s(\phi(\hat{\pi}_i)\cond \theta_s)$ by different values of $\phi(\hat{\pi}_i)$. Solid, dashed, and light-grey lines denote the spline over the common-support, out-of-trial extrapolation, and trial-specific regions, respectively (the trial-specific region is excluded from estimands); shaded bands are 95\% credible intervals, darker over the common-support region.}
\begin{center}
\includegraphics[width=\textwidth]{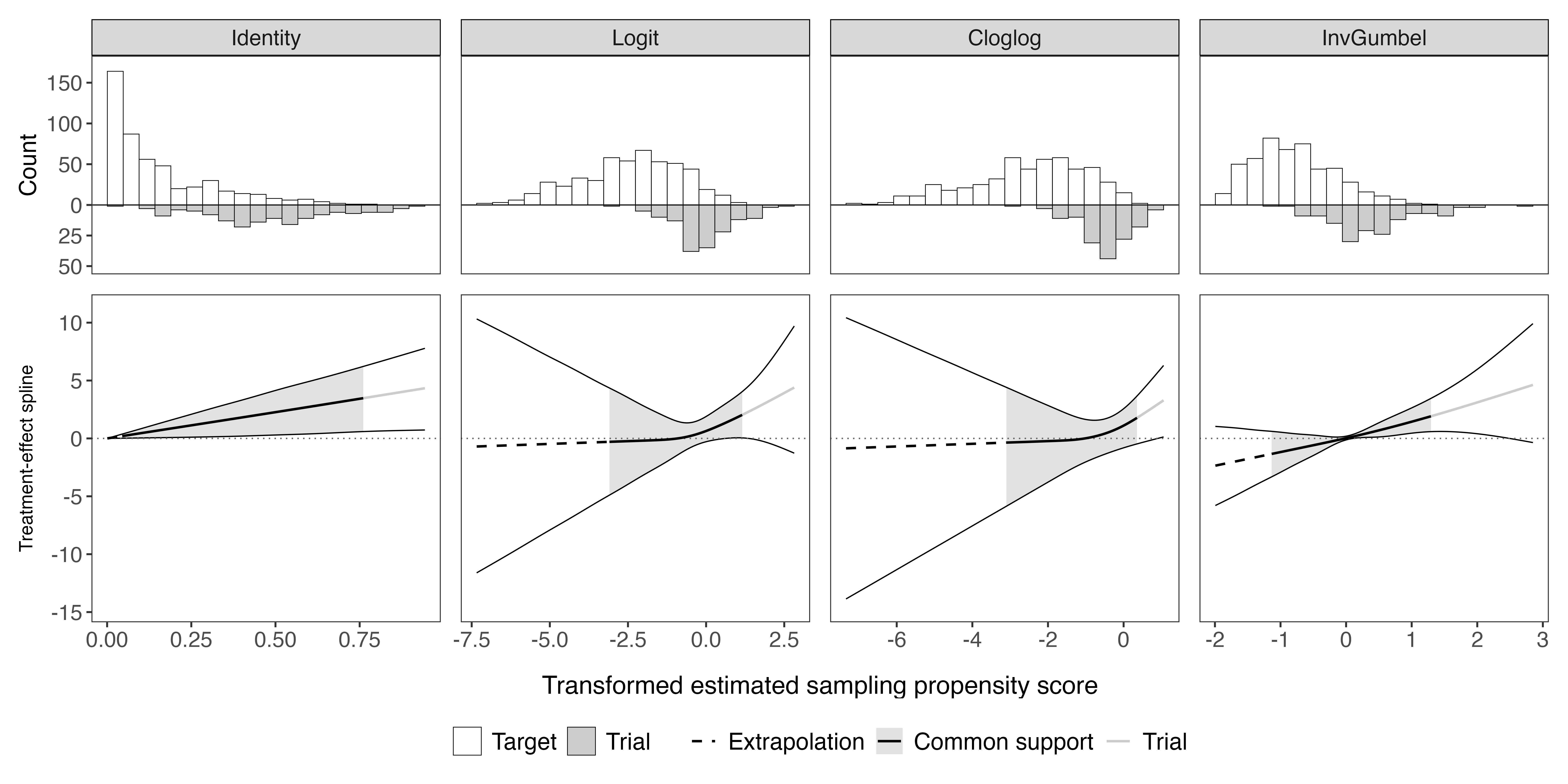}
\end{center}
\label{fig:transformation}
\end{figure}

To further illustrate these results, we examined one replicate of simulation from Scenario SS1 with $N=500$. Figure~\ref{fig:transformation} displays the histograms of $\phi(\hat{\pi})$ and the corresponding posterior means and 95\% credible intervals from DSplineBART for estimating $\e_T[Y(1) - Y(0)]$. The upper panel shows the distributions of $\hat{\pi}$ under each transformation in the EHR cohort and the trial sample, while the lower panel presents estimates of $\text{NS}_s(\phi(\hat{\pi}_i)\cond \theta_s)$. Among the four transformations, the Identity transformation showed the closest visual alignment between the trial and the EHR cohort score distribution, and most values of $\hat{\pi}$ were concentrated near zero. This concentration compressed the score scale, over which the spline extrapolated toward lower values of $\hat{\pi}$, yielding limited posterior uncertainty in predicted potential outcomes and contributing to undercoverage of intervals for the PATE and mean potential outcomes. Conversely, the Logit and Cloglog transformations pushed trial observations toward the right tail, forcing the spline to extrapolate excessively at lower values of $\hat{\pi}$ and producing unstable estimates with overly wide intervals. In contrast, the InvGumbel transformation provided a balanced mapping between these extremes---offering proper spread for spline fitting within the trial sample while enabling moderate, stable extrapolation in the EHR cohort. This balance mitigated both boundary compression from the Identity link and excessive extrapolation from the Logit and Cloglog links, resulting in more consistent point estimates and well-calibrated credible intervals.

\section{Application to Transport the CODIACS Vanguard Trial to All of Us EHR Data}
\label{sec:case}

\subsection{Comparison of trial to EHR cohorts}

As discussed in Section~\ref{sec:objective}, our objective is to transport the treatment effect of the CODIACS Vanguard intervention to a real-world  population of post-ACS patients with elevated depressive symptoms, as represented by cohorts constructed from the All of Us EHR data. The All of Us Research Program has recruited adults across the United States through participating health care provider organizations and direct volunteer enrollment since 2018 \citep{AURPI2019}. We used the All of Us Registered Tier Dataset v9 to construct the EHR cohorts. We consider two EHR cohorts, differing in the timing of depression onset after ACS. Sample $T_1$ ($N=389$) represents patients with depression onset 2--6 months after ACS, matching the trial's enrollment window. Sample $T_2$ ($N=956$) represents a broader population with depression onset at least 2 months after ACS, with no upper time limit. Detailed inclusion and exclusion criteria are provided in Supplementary Materials C.1.

\begin{table}[h]
\centering
\caption{Summary statistics for the EHR cohorts and the CODIACS Vanguard trial. EHR cohorts are post-ACS patients with elevated depressive symptoms in the All of Us EHR data, defined by two time windows of depression onset after acute coronary syndrome (ACS): $T_1$ (2--6 months) and $T_2$ (2+ months). Continuous variables are reported as mean [min--max] and categorical variables as count (\%). The last two columns report mean (SD) of BDI score change from baseline to 6-month follow-up by subgroup within each trial arm.}
\label{tab:summary}
\begin{adjustbox}{width=\linewidth,center}
\begin{threeparttable}
\begin{tabular}{lcccccc}
\toprule
\multirow{3}{*}{\textbf{Variables}}
  & \multicolumn{2}{c}{\textbf{Target EHR cohort}}
  & \multicolumn{2}{c}{\textbf{Trial Sample}}
  & \multicolumn{2}{c}{\textbf{Outcome in Trial}} \\
\cmidrule(lr){2-3} \cmidrule(lr){4-5} \cmidrule(lr){6-7}
  & $T_1$ & $T_2$
  & Intervention & Control
  & Intervention & Control \\
  & (size: $389$) & (size: $956$)
  & (size: $73$) & (size: $77$)
  & (size: $73$) & (size: $77$) \\
\midrule
\textbf{Age}, mean [min--max]
  & 63.5 [36--85] & 63.8 [36--86]
  & 59.2 [36--83] & 60.0 [38--82]
  & --- & --- \\
\textbf{Sex} & & & & & & \\
\ \ Male
  & 155 (40.4\%) & 402 (42.6\%)
  & 43 (58.9\%) & 44 (57.1\%)
  & $-$7.4 (8.5) & $-$6.8 (6.2) \\
\ \ Female
  & 229 (59.6\%) & 542 (57.4\%)
  & 30 (41.1\%) & 33 (42.9\%)
  & $-$9.8 (7.5) & $-$5.2 (7.6) \\
\textbf{Race} & & & & & & \\
\ \ Black
  & 54 (14.2\%) & 141 (15.0\%)
  & 23 (31.5\%) & 27 (35.1\%)
  & $-$4.8 (7.6) & $-$5.8 (7.1) \\
\ \ Non-Black
  & 326 (85.8\%) & 796 (85.0\%)
  & 50 (68.5\%) & 50 (64.9\%)
  & $-$10.0 (7.9) & $-$6.3 (6.7) \\
\textbf{Ethnicity} & & & & & & \\
\ \ Hispanic
  & 60 (16.0\%) & 134 (14.5\%)
  & 16 (22.2\%) & 13 (16.9\%)
  & $-$6.8 (6.1) & $-$6.4 (5.3) \\
\ \ Non-Hispanic
  & 315 (84.0\%) & 791 (85.5\%)
  & 56 (77.8\%) & 64 (83.1\%)
  & $-$8.8 (8.7) & $-$6.1 (7.2) \\
\textbf{ACS Type} & & & & & & \\
\ \ NSTEMI
  & 269 (69.2\%) & 655 (68.5\%)
  & 19 (26.0\%) & 21 (27.3\%)
  & $-$9.2 (8.7) & $-$5.8 (8.1) \\
\ \ STEMI
  & 58 (14.9\%) & 143 (15.0\%)
  & 19 (26.0\%) & 21 (27.3\%)
  & $-$8.8 (6.7) & $-$7.2 (6.0) \\
\ \ Unstable angina
  & 62 (15.9\%) & 158 (16.5\%)
  & 35 (47.9\%) & 35 (45.5\%)
  & $-$7.7 (8.7) & $-$5.6 (6.7) \\
\textbf{Currently taking} & & & & & & \\
\textbf{antidepressant} & & & & & & \\
\ \ Yes
  & 153 (39.3\%) & 364 (38.1\%)
  & 25 (34.2\%) & 24 (32.4\%)
  & $-$7.8 (8.0) & $-$6.4 (5.8) \\
\ \ No
  & 236 (60.7\%) & 592 (61.9\%)
  & 48 (65.8\%) & 50 (67.6\%)
  & $-$8.6 (8.3) & $-$6.3 (7.1) \\
\bottomrule
\end{tabular}

\begin{tablenotes}[flushleft]\footnotesize
\item STEMI: ST-segment elevated myocardial infarction.
\item NSTEMI: Non–ST-segment elevated myocardial infarction, or bundle branch block/uncertain type of myocardial infarction.
\end{tablenotes}
\end{threeparttable}
\end{adjustbox}
\end{table}

Table~\ref{tab:summary} compares patient characteristics across the two EHR cohorts and the trial sample, including age, sex, race, ethnicity, ACS type, and current antidepressant use. It also reports the mean and standard deviation of the primary trial outcome ---the change in Beck Depression Inventory (BDI) score from baseline to six months after randomization ---within each subgroup in the trial. The BDI is a 21-item, self-report questionnaire that measures the severity of depression symptoms. Each question has four choices scored from 0 to 3, making the total score range from 0 to 63. Higher scores indicate more severe symptoms. Compared with the EHR cohorts, the trial included a smaller proportion of women, who showed a larger observed treatment effect than men; a larger proportion of Black patients, who showed a smaller observed treatment effect than non-Black patients; and a smaller proportion of patients with NSTEMI, who showed the largest observed treatment effect among the ACS subtypes. These differences in covariate distributions, together with the apparent treatment effect heterogeneity, suggest that the trial-only treatment effect may differ from those in the target populations represented by the EHR cohorts. In addition, the EHR cohorts include patients slightly older than those included in the trial, requiring extrapolation beyond the trial sample's upper age limit.

\subsection{Transportability analyses}
To transport treatment effects from the CODIACS Vanguard trial to the EHR-defined populations, we applied the methods from the Simulation Section, with the same set of hyperparameters. Missing covariates in EHR and trial were assumed missing at random (MAR) and imputed by multiple imputation by chained equations \citep{VanBuuren2011} to generate 20 imputed datasets. For Bayesian outcome models, we pooled posterior draws across the 20 multiply-imputed datasets to compute posterior estimates, following the recommendations of \citet{gelman2013}. For the trial-only, IPPW, and A-IPPW  estimators, we applied Rubin's Rules \citep{Rubin1987} to yield pooled estimates and corresponding 95\% confidence intervals. The pooled estimates for the average treatment effects and means of potential outcome are reported in Figure~\ref{fig:gen_results}.

\subsection{Main results}

The trial-only analysis, calculated as the unadjusted difference in mean outcomes between the two trial arms, estimated a 2.2-point greater reduction in BDI score for the intervention group compared with the control group. IPPW and A-IPPW estimated larger effects, with 95\% confidence intervals of A-IPPW excluding zero. IPPW estimated reductions of 3.7 points in $T_1$ (95\% CI: $[-7.8,0.3]$) and 3.6 points in $T_2$ ($[-7.7,0.5]$), while A-IPPW estimated larger reductions of 5.0 points in $T_1$ ($[-9.0,-1.1]$) and 4.7 points in $T_2$ ($[-8.8,-0.7]$). In contrast, BART and BART-PS yielded smaller reductions of approximately 2.6--2.7 points, close to the trial-only estimate, with credible intervals including zero. The interaction-based models estimated larger reductions, ranging from 4.3 to 5.1 points in $T_1$ and from 4.2 to 4.9 points in $T_2$. In $T_1$, all PSPI models except FullBART had credible intervals excluding zero. DSplineBART estimated the largest reduction ($-5.1$; $[-9.8,-0.4]$), followed by SplineBART ($-4.9$; $[-9.5,-0.3]$). The same pattern was observed in $T_2$, where all PSPI models except FullBART yielded credible intervals excluding zero. DSplineBART again estimated the largest reduction among the PSPI models ($-4.9$; $[-9.4,-0.3]$). Overall, SplineBART and DSplineBART produced the largest estimated reductions, consistent with A-IPPW. The Bayesian-bootstrap analyses yielded nearly identical point estimates with slightly wider credible intervals.

\begin{figure}[t]
  \centering
\caption{Point estimates and 95\% probability intervals for average treatment effects and mean potential outcomes in the CODIACS Vanguard trial application, as changes in BDI-score from baseline to six months. The top and bottom rows correspond to EHR cohorts $T_1$ and $T_2$, respectively. SplineBART-Bootstrap and DSplineBART-Bootstrap estimate the PATE and population mean potential outcomes for the corresponding target populations, while the remaining estimators report the MATE and mean potential outcomes averaged over each EHR cohort. Numbers below the intervals show point estimates.}\vspace{-0.3cm}
  \label{fig:gen_results}
  \includegraphics[width=\textwidth]{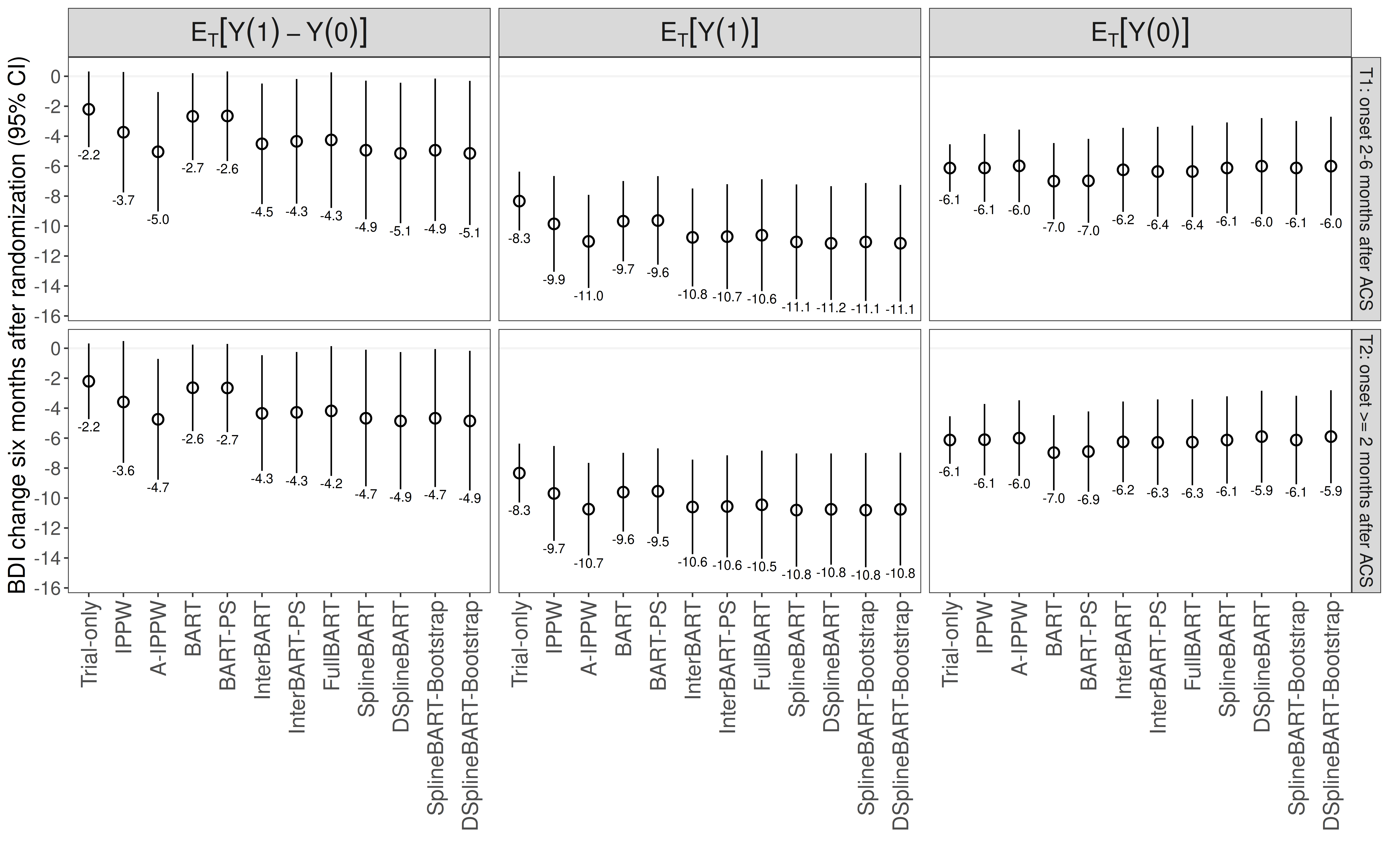}\vspace{-0.5cm}
\end{figure}

For the treated potential outcome mean, the trial-only analysis estimated a mean reduction of 8.3 points in BDI score at 6 months (95\% CI: $[-10.3, -6.4]$). SplineBART and DSplineBART estimated larger reductions, with 11.1 to 11.2 points in $T_1$ and 10.8 points in $T_2$. For the control potential outcome mean, the trial-only analysis estimated a mean reduction of 6.1 points (95\% CI: $[-7.7, -4.5]$), and all interaction-based models gave estimates close to it in both samples (5.9 to 6.4 points). The larger transported effects of two spline-based PSPI models therefore came almost entirely from the treated potential outcome, while the untreated mean was more similar across methods.

\begin{figure}[t]
\caption{Comparison of the estimated sampling propensity-score between the trial sample and the two EHR cohorts and posterior treatment-effect spline estimates using SplineBART and DSplineBART. Top: histograms of the InvGumbel-transformed sampling propensity score in the target (white, above the axis) and the trial (grey, below), with error bars of $\pm 1$ standard deviation across imputations. Bottom: posterior means and 95\% credible intervals of the treatment-effect spline (SplineBART, gray lines; DSplineBART, black lines). Solid curves and gray bands show the posterior mean and 95\% credible interval over the common support of the target and trial propensity scores; dashed curves over the extrapolation region are left unshaded.}\vspace{-0.3cm}
\begin{center}
\includegraphics[width=\textwidth]{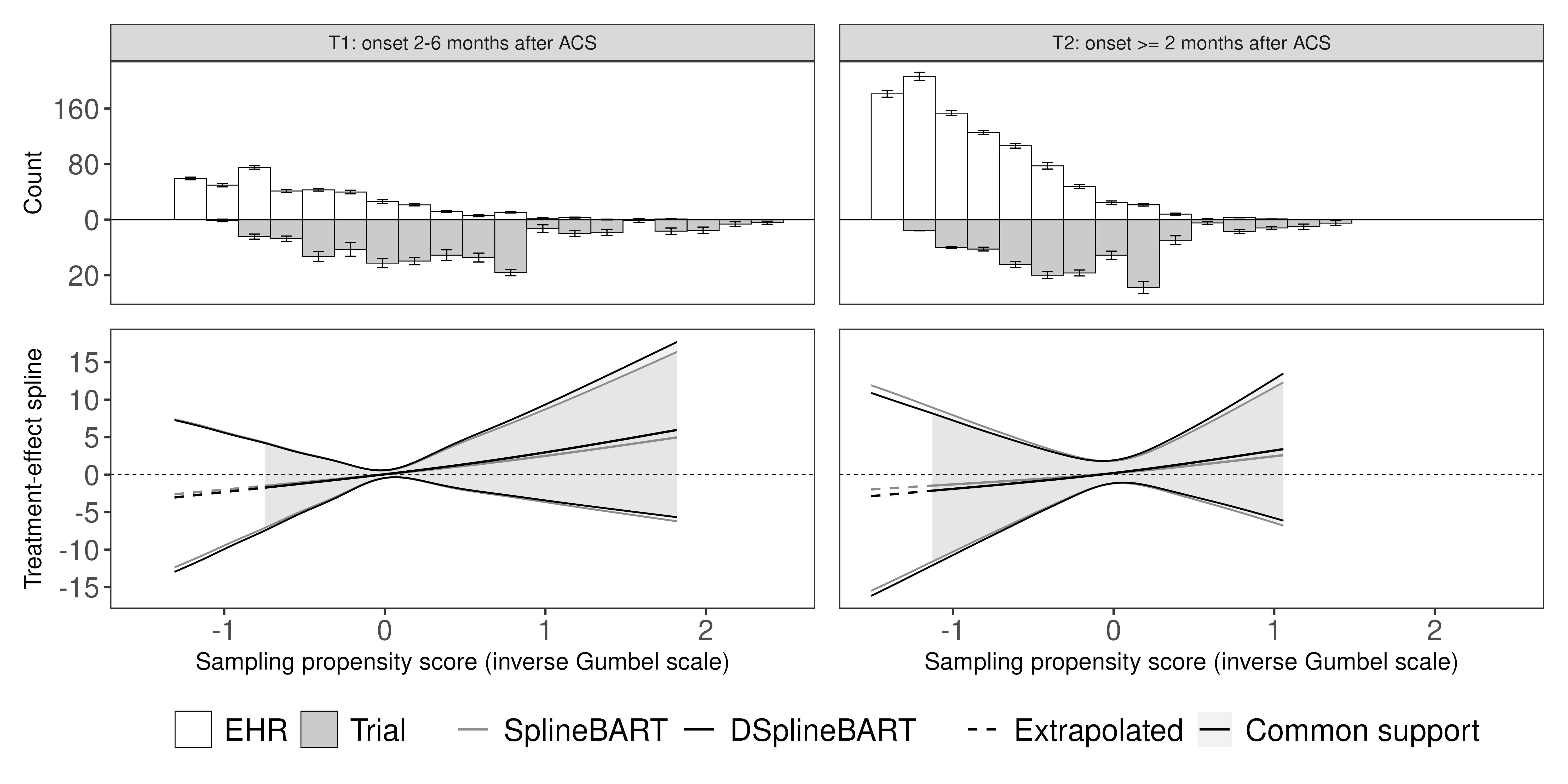}
\end{center}\vspace{-0.5cm}
\label{fig:real_spline}
\end{figure}

Figure~\ref{fig:real_spline} compares the estimated sampling propensity score distributions between the trial sample and each EHR cohort and presents posterior means and 95\% credible intervals of treatment-effect splines over 20 imputations. As expected, trial participants generally have higher estimated sampling propensity scores than individuals in the EHR cohorts. SplineBART and DSplineBART yield similar treatment-effect splines, although DSplineBART generally has a steeper slope, particularly in $T_2$. Both models show an approximately linear, positive relationship between the treatment-effect spline and the inverse-Gumbel transformed sampling propensity score. The extrapolated portions of the splines (dashed lines) extend the fitted boundary trends, predicting greater intervention-related reductions in BDI scores at very low sampling propensity scores.  This pattern is consistent with the larger estimated treatment benefits from SplineBART and, especially, DSplineBART in Figure \ref{fig:gen_results}.

\subsection{Sensitivity Analysis}
\label{sec:sen}
In the analysis above, we assume exchangeability over $S$: $(Y(0), Y(1)) \perp S \cond \bX$ (Assumption A4). Violation of A4 may lead to differences in conditional potential outcome means between the trial and target populations, biasing the transported effect estimates. Because estimates of $\e_T[Y(0)]$ were similar across methods, we focused on the sensitivity analysis on $\e_T[Y(1) - Y(0)]$. We adopted the ``bias function'' approach of \citet{Dahabreh2023}, defining the bias on treatment effect as a user-specified function $\delta(X_i)\equiv\e[Y_i(1) - Y_i(0) \cond X_i, S=1]-\e_T[Y_i(1) - Y_i(0) \cond X_i]$. We specified $\delta(X_i)=\hat{\tau}_i\beta$, where $\hat{\tau}_i=-\log(-\log(\hat{\pi}(X_i)))$. The MATE of PSPI models for $\e_T[Y(1) - Y(0)]$ is therefore
\[\frac{1}{N}\sum_{i} \{\hat{Y}_i(1) - \hat{Y}_i(0)-\hat{\tau}_i\beta\},\]
with varying priors placed on $\beta$. As $\hat{Y}_i(1) - \hat{Y}_i(0)$ is drawn with mean $\text{BART}_s(X_i) + \text{NS}(\hat{\tau}_i\cond \theta_{s})$, the linear term $\hat{\tau}_i\beta$ probes sensitivity to both spline misspecification and linear extrapolation. Since $\beta$ is not identifiable from EHR cohorts, in practice, we sampled $\beta$ from priors and applied the bias $\delta(X_i)$ to posterior draws of $\hat{Y}_i(1) - \hat{Y}_i(0)$. 

\begin{figure}[t]
\caption{Bias-function sensitivity analysis for the DSplineBART estimates of the MATE in the two EHR cohorts.}\vspace{-0.3cm}
\begin{center}
\includegraphics[width=\textwidth]{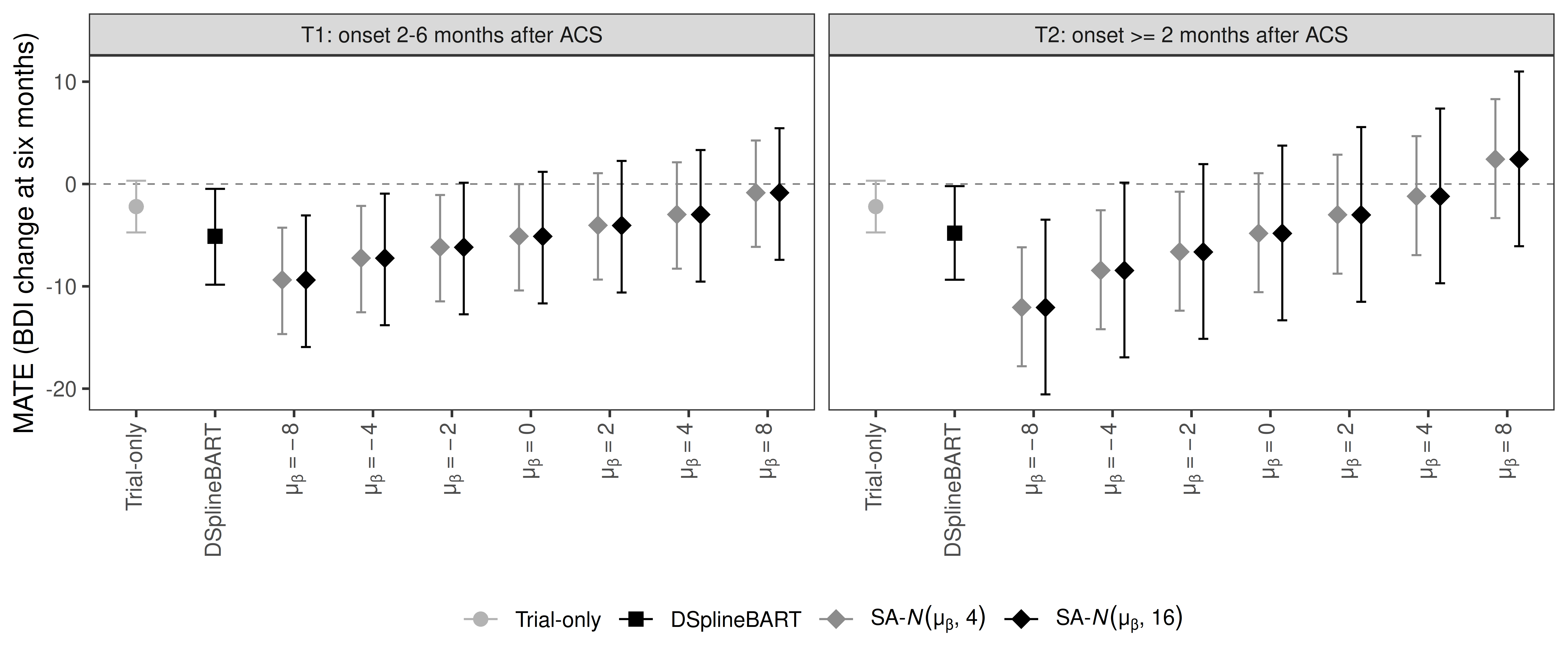}
\end{center}
\label{fig:real_sa}
\end{figure}

Figure~\ref{fig:real_sa} presents the sensitivity analysis for the MATE estimated using DSplineBART. We specified priors $\beta\sim\mathcal{N}(\mu_{\beta}, \sigma^2_{\beta})$ with $\mu_{\beta}\in\{-8,-4,-2,0,2,4,8\}$ and $\sigma^2_{\beta}\in\{2^2,4^2\}$. These choices placed $\beta$ on the scale of the linear slope coefficient $\theta_1$ of the spline $\text{NS}(\hat{\tau}_i\cond\theta_s)$, whose posterior was close to $\mathcal{N}(2, 4^2)$ in both EHR cohorts. The prior means extended to twice the  posterior standard deviation in either direction from zero, while the prior standard deviations were one-half to equal to that posterior standard deviation. The estimated treatment benefit was moderately sensitive to the prior on $\beta$. In both $T_1$ and $T_2$,  increasing $\sigma_{\beta}$ from 2 to 4 widened the credible intervals without changing the point estimates. Smaller values of $\mu_{\beta}$ yielded more negative MATE estimates, indicating greater treatment benefits. The point estimate approached zero at $\mu_{\beta}=8$ in $T_1$ and $\mu_{\beta}=4$ in $T_2$. Credible intervals excluded zero only at sufficiently negative values of $\mu_{\beta}$, and this occurred over a narrower range of $\mu_{\beta}$ under the larger value of $\sigma^2_{\beta}$. Similar patterns were observed for the MATE of SplineBART and for the PATE estimates (see Supplementary Materials C.2).

\section{Discussion}
\label{sec:discussion}

Trial findings may not extend to target populations when the distributions of treatment effect modifiers differ between the trial and target populations. We developed Propensity Score Predictive Inference (PSPI), an interaction-based Bayesian outcome modeling framework for generalizability and transportability analyses with target populations represented by EHR cohorts. Its two variants, SplineBART and DSplineBART, combine BART's flexibility in modeling high-dimensional covariates with separate natural cubic spline components that explicitly model sampling-related variation through sampling propensity score. This separation allows the sampling propensity score to contribute beyond its potentially attenuated role within a regularized BART model and may improve robustness to misspecification of the BART components. The spline components also provide an explicit extrapolation rule when target observations fall outside the trial's observed score range, extending fitted boundary trends linearly on the transformed score scale. To improve extrapolation and reduce variability, we further introduce the inverse Gumbel transformation on the sampling propensity score to expand score values near 0 more gradually and allocate a larger share of the transformed range between the trial and EHR cohorts. Together, these features provide a flexible framework for estimating target treatment effects and mean potential outcomes while making the extrapolation assumptions explicit.

The simulations confirmed the robustness of PSPI models across diverse settings. For estimating the target average treatment effect, SplineBART and DSplineBART consistently achieved lower bias, smaller RMSE, and closer to nominal coverage than alternative methods, especially under substantial covariate shift, where other methods often showed instability and undercoverage. DSplineBART also showed additional gains in estimating potential outcomes, being the only method to consistently attain nominal coverage across most scenarios.

In the application, we used PSPI to transport the treatment effect observed in the CODIACS Vanguard Trial to two target populations of post-ACS patients with depressive symptoms, represented by All of Us EHR cohorts $T_1$ and $T_2$. Compared with the unadjusted trial-only estimate of a 2.2-point greater BDI reduction under intervention than control, SplineBART and DSplineBART estimated larger benefits of 4.9 and 5.1 points in $T_1$ and 4.7 and 4.9 points in $T_2$, respectively. Differences across methods arose primarily from estimates of the treated potential outcome, whereas estimates under control were similar. The interaction-based formulation allowed treatment effect heterogeneity to be modeled explicitly and separately from outcome prognosis. Plots of the posterior treatment-effect splines illustrated the contribution of the transformed sampling propensity score within the trial's observed range and made the extrapolated trends beyond that range explicit. The sensitivity analysis further demonstrated how PSPI can assess the dependence of conclusions on departures from exchangeability over trial participation (Assumption A4).

Although our spline-based models perform well for extrapolation under covariate shift beyond the trial's observed range in the simulations, we are unable to state that our models can avoid model misspecification. The extrapolation relies heavily on the assumptions that the relationship between the propensity score and treatment effects is similar for out-of-sample units and those near the sample boundaries. Sensitivity analyses must be conducted by allowing different functional forms of the propensity score for out-of-sample units, which may further improve extrapolation across diverse generalizability and transportability settings. We recommend the linear term on the propensity score as the ``bias function'' (see Section~\ref{sec:sen}), as the term probes sensitivity to both outcome-model misspecification and linear extrapolation. The proposed methods are implemented in the open-source \texttt{R} package \href{https://cran.r-project.org/web/packages/PSPI/index.html}{\texttt{PSPI}}, available on \texttt{CRAN}. The package provides flexible tools for estimating treatment effects and conducting sensitivity analyses for generalizability and transportability using PSPI models.

Although PSPI shares a similar interaction-based outcome model formula with Bayesian Causal Forests (BCF; \citealp{Hahn2020}), the two methods address fundamentally different problems. BCF models the outcome as $\mu_{A_i}(X_i) = \mathrm{BART}(X_i, \hat\pi_t(X_i)) + A_i \cdot \mathrm{BART}(X_i)$, where $\hat\pi_t(\bX)$ estimates the treatment propensity score $\Pr(A=1 \cond \bX)$. BCF was developed for observational causal inference, where treatment is not randomized, whereas PSPI extends treatment effects from a randomized trial to a target population. The two propensity scores also play different roles. BCF includes $\hat\pi_t(\bX)$ to mitigate regularization-induced confounding: when the covariates are strongly correlated with treatment assignment, a regularization prior that shrinks the prognostic effect tends to reallocate part of that signal to the treatment, biasing treatment effect estimation; including $\hat\pi_t(\bX)$ only in the prognostic term lets this variation be captured directly, removing such bias. PSPI instead addresses differences between a randomized trial and a target population. Its sampling propensity score summarizes trial participation and may capture variation relevant to both outcome prognosis and treatment effect heterogeneity, motivating its inclusion in both components.

Future work may extend the PSPI framework to more complex treatment settings and outcome types. Ordinal or survival outcomes would require tailored outcome models. Treatment-specific interaction terms could accommodate multi-arm trials or continuous interventions. Extending findings from an observational study to a target population would require addressing both treatment confounding and population differences, potentially by incorporating both treatment and sampling propensity scores into the outcome model. Finally, trial recruitment strategies could account for the distribution of treatment effect modifiers in the intended target population, addressing generalizability and transportability at the design stage as well as during analysis.

\section*{Supplementary Material}
\textbf{Derivations, Simulation Studies, and Data Application.}
Part A: Derivations and model details.
Part B: Additional simulation details and results.
Part C: Eligibility criteria and sensitivity analyses for the data application.

\bibliography{reference}

\end{document}


\begin{frontmatter}
\title{Supplementary Materials for ``Bayesian Propensity Score Predictive Inference for Extending Trial Findings to Target Populations''}
\runtitle{Extending Trial Findings to Target Populations}

\end{frontmatter}

\hypersetup{linkcolor=black}
\tableofcontents
\hypersetup{linkcolor=blue}

\suppsection{Derivations and Model Details}

\subsection{Proof of Identifiability}
\label{app:proof}
Throughout, $S=1$ labels trial participants and $S=0$ EHR cohort members outside the trial, $\mathbb{P}_T$ denotes the target population, defined as the population that the EHR cohort represents, and $\e$ and $\e_T$ denote the expectations with respect to the superpopulation $\mathbb{P}$ and $\mathbb{P}_T$, respectively. Under Assumptions A1--A5 of the main text, the potential outcome means and the average treatment effect in the target population are identifiable, and a single argument covers both the nested and the non-nested design. For $a = 0, 1$,
\small
\begin{align*}
\e_T[Y(a)]
& = \e_T\!\left[\e_T[Y(a) \cond \bX]\right]
  & \text{(law of total expectation)} \\
& = \e_T\!\left[\e[Y(a) \cond \bX, S = 1]\right]
  & \text{(definition of $\mathbb{P}_T$ and A4)} \\
& = \e_T\!\left[\e[Y(a) \cond \bX, A = a, S = 1]\right]
  & \text{(A2: exchangeability over $A$)} \\
& = \e_T\!\left[\e[Y \cond \bX, A = a, S = 1]\right],
  & \text{(A1: consistency)}
\end{align*}
\normalsize
where the outer expectation is over the covariate distribution of the target population, supplied by the EHR cohort. The second equality combines the definition of $\mathbb{P}_T$ with A4, and its justification differs slightly between the two designs. In the nested design, the EHR cohort represents $\mathbb{P}_T$ and contains both trial participants and non-participants, so
\small
\begin{align*}
\e_T[Y(a) \cond \bX]
&= \Pr(S=1 \cond \bX)\,\e[Y(a) \cond \bX, S=1] + \Pr(S=0 \cond \bX)\,\e[Y(a) \cond \bX, S=0] \\
&= \Pr(S=1 \cond \bX)\,\e[Y(a) \cond \bX, S=1] + \Pr(S=0 \cond \bX)\,\e[Y(a) \cond \bX, S=1]
  & \text{(A4)} \\
&= \e[Y(a) \cond \bX, S=1].
\end{align*}
\normalsize
In the non-nested design, probability statements involving $S$ refer to the pooled observed data, in which trial draws carry $S=1$ and cohort draws carry $S=0$. The EHR cohort consists of the $S=0$ individuals, so $\e_T[Y(a) \cond \bX] = \e[Y(a) \cond \bX, S=0]$ by definition of $\mathbb{P}_T$, and A4 equates it to $\e[Y(a) \cond \bX, S=1]$. In both designs, A5 guarantees $\Pr(S=1 \cond \bX) > 0$ over the support of the target covariate distribution, so the conditional expectation given $S=1$ is well defined wherever the outer expectation places mass, and A3 guarantees that both treatment arms are represented there, so further conditioning on $A = a$ is also well defined. The average treatment effect follows by the difference:
\small
\begin{align*}
\e_T[Y(1) - Y(0)]
= \e_T\!\left[\e[Y \cond \bX, A = 1, S = 1] - \e[Y \cond \bX, A = 0, S = 1]\right].
\end{align*}
\normalsize
\subsection{Prior Specifications for BART}
For fitting the BART model component, we adopted the default prior specifications as recommended in \cite{Chipman2010}, where regularization priors are independently placed for $p({T}_t)$ and $p({B}_t \cond {T}_t)$. Specifically, for $p({T}_t)$, (i) at each interior node, the splitting variable is chosen uniformly from the set of available predictors; (ii) given the splitting variable, the split point is chosen uniformly from the range of observed values for that variable; (iii) the probability that a node at depth $d$ splits, is given to be proportional to $\alpha (1 + d)^{-\beta}$, with hyperparameters $\alpha \in (0, 1)$ and $\beta>0$. Given the tree structure ${T}_t$, the elements of $p({B}_t \cond {T}_t)$ are modeled as independent normal distributions $\mathcal{N}(0, (4k^2N_T)^{-1})$, with shrinkage hyperparameter $k>0$. \citet{Chipman2010} sets the hyperparameters $\alpha = 0.95$, $\beta = 2$, and $k = 2$ as default automatic choices.

\subsection{Gibbs Samplers for SplineBART and DSplineBART}
We present here the algorithms of Gibbs sampling for updating SplineBART and DSplineBART. The notation $[\cdot]_{\{\cdot\}}$ is used to represent the subset of observations in $[\cdot]$ that satisfy the conditions specified within ${\{\cdot\}}$. For example, $[\bX]_{\{S=1, A=1\}}$ refers to the set of covariates corresponding to individuals who are both part of the trial (S = 1) and belong to the treatment group (A = 1). 
The posterior update of BART uses a backfitting MCMC approach, which iteratively updates each tree's parameters ${T}_t$ and ${B}_t$ conditioning on all other trees and the residuals. To update the posterior distribution of an individual tree, \citet{Pratola2016} proposes an efficient Metropolis-Hastings algorithm that effectively explores the space of possible tree structures, and this approach is commonly implemented in the \texttt{BART} package \citep{Sparapani2021}, upon which our models are built. 

Web Algorithm 1 presents the Gibbs sampler for updating the SplineBART model. The updates of $\text{BART}_s(\bX)$ and $\text{NS}(\hat{\tau}\cond\cdot)$ are based only on the trial samples in the treatment group. Steps 3--5 update the natural cubic spline $\text{NS}(\hat{\tau}\cond\cdot)$. Steps 8--9 generate predictions of the potential outcomes in the target.

Web Algorithm 2 outlines the Gibbs sampler for updating the DSplineBART model. Similarly to the previous algorithm, the updates of $\text{BART}_s(\bX)$ and $\text{NS}_s(\hat{\tau}\cond\cdot)$ are conducted using only the trial samples from the treatment group. Steps 2-4 and 6-8 update the natural cubic splines $\text{NS}_h(\hat{\tau}\cond\cdot)$ and $\text{NS}_s(\hat{\tau}\cond\cdot)$, respectively. Steps 11-12 generate predictions of the potential outcomes in the targets.

\begin{algorithm}[!htb]
\caption{An iteration of the Gibbs sampler for updating PSPI-SplineBART}\label{alg:gibbs_S_est}
\textbf{Input:} Covariates $\bX$, outcome $Y$, transformed estimated propensity scores $\hat{\tau}$, treatment indicator $A$, trial participation indicator $S$, natural cubic spline basis $N_1(\hat{\tau}), \ldots, N_{K-2}(\hat{\tau})$.\\
\textbf{Output:} Posterior samples of $\text{BART}_h(X_i, \hat{\tau}_i)$, $\text{BART}_s(X_i)$, $\sigma^2$, $\theta_1, \ldots ,\theta_{K-1}$, $\gamma^2$, $\hat{Y}_i(1)$ and $\hat{Y}_i({0})$, $i=1,2,\ldots, N$.
\begin{algorithmic}[1]
\State Update $\text{BART}_h(\cdot)$ using $[\bX, \hat{\tau}]_{\{S = 1\}}$ as predictors, and $[Y - A\cdot(\text{BART}_s(\bX) + \text{NS}(\hat{\tau}\cond\cdot))]_{\{S = 1\}}$ as outcomes.
\State Update $\text{BART}_s(\cdot)$ using $[\bX]_{\{S=1,A=1\}}$ as predictors, and $[Y - \text{BART}_h(\bX, \hat{\tau}) - \text{NS}(\hat{\tau}\cond\cdot)]_{\{S=1,A=1\}}$ as outcomes. 
\State Sample $(\theta_1, \theta_2, \ldots, \theta_{K-1})^T$ from $\mathcal{MVN}_{K-1}\!\left(\left(\mathbf{D}^T\mathbf{D} + \mathbf{\Gamma}\right)^{-1}\, \mathbf{D}^TU,\ \sigma^2\left(\mathbf{D}^T\mathbf{D} + \mathbf{\Gamma}\right)^{-1}\right)$, where $\mathbf{\Gamma} = \mathrm{diag}(0,\, \tfrac{\sigma^2}{\gamma^2}, \ldots, \tfrac{\sigma^2}{\gamma^2})$ and $\;U = [Y - \text{BART}_h(\bX, \hat{\tau}) - \text{BART}_s(\bX)]_{\{S=1,A=1\}}$, $\mathbf{D} = [\hat{\tau},\, N_1(\hat{\tau}), \ldots, N_{K-2}(\hat{\tau})]_{\{S=1,A=1\}}$.
\State Sample $\gamma^2 \sim \text{InvGamma}(\frac{K}{2}, 1 + \frac{1}{2}\sum_{k=2}^{K-1}\theta_k^2)$.
\State Calculate $\text{NS}(\hat{\tau}_i\cond\cdot) = \theta_1 \hat{\tau}_i+\sum_{k=1}^{K-2} \theta_{k+1} N_k(\hat{\tau}_i)$, $i=1,2,\ldots,N$.
\State Calculate $SSR=\sum_{i=1}^NS_i\cdot(Y_i - \text{BART}_h(X_i, \hat{\tau}_i) - A_i\cdot [\text{BART}_s(X_i) + \text{NS}(\hat{\tau}_i\cond\cdot)])^2$.
\State Sample $\sigma^2$ from $\text{InvGamma}(\frac{\upsilon + n}{2}, \frac{\upsilon\lambda + SSR}{2})$.
\State Sample $\hat{Y}_i({1})$ from $\mathcal{N}(\text{BART}_h(X_i, \hat{\tau}_i) + \text{BART}_s(X_i) + \text{NS}(\hat{\tau}_i\cond\cdot), \sigma^2)$, for $i=1,2,\ldots,N$.
\State Sample $\hat{Y}_i({0})$ from $\mathcal{N}(\text{BART}_h(X_i, \hat{\tau}_i), \sigma^2)$, for $i=1,2,\ldots,N$.
\end{algorithmic}
\end{algorithm}

\begin{algorithm}[!htb]
\caption{An iteration of the Gibbs sampler for updating PSPI-DSplineBART}\label{alg:gibbs_DS_est}
\textbf{Input:} Covariates $\bX$, outcome $Y$, transformed estimated propensity scores $\hat{\tau}$, treatment indicator $A$, trial participation indicator $S$, natural cubic spline bases $N_1(\hat{\tau}), \ldots, N_{K-2}(\hat{\tau})$ and $N_{h,1}(\hat{\tau}), \ldots, N_{h,H-2}(\hat{\tau})$.\\
\textbf{Output:} Posterior samples of $\text{BART}_h(X_i)$, $\theta_{h,1}, \ldots, \theta_{h, H-1}$, $\gamma^2_h$, $\text{BART}_s(X_i)$, $\theta_1, \ldots ,\theta_{K-1}$, $\gamma^2$, $\sigma^2$, $\hat{Y}_i(1)$ and $\hat{Y}_i(0)$, $i=1,2,\ldots, N$.
\begin{algorithmic}[1]
\State Update $\text{BART}_h(\cdot)$ using $[\bX]_{\{S = 1\}}$ as predictors, and $[Y - \text{NS}_h(\hat{\tau}\cond\cdot) - A\cdot(\text{BART}_s(\bX) + \text{NS}_s(\hat{\tau}\cond\cdot))]_{\{S = 1\}}$ as outcomes.
\State Sample $(\theta_{h,1}, \ldots, \theta_{h,H-1})^T$ from $\mathcal{MVN}_{H-1}\!\left((\mathbf{B}^T\mathbf{B} + \mathbf{\Gamma}_h)^{-1}\mathbf{B}^TV,\ \sigma^2(\mathbf{B}^T\mathbf{B} + \mathbf{\Gamma}_h)^{-1}\right)$, where $\mathbf{\Gamma}_h = \mathrm{diag}(0,\, \tfrac{\sigma^2}{\gamma_h^2}, \ldots, \tfrac{\sigma^2}{\gamma_h^2})$, $\;V = [Y - \text{BART}_h(\bX) - A\cdot(\text{BART}_s(\bX) + \text{NS}_s(\hat{\tau}\cond\cdot))]_{\{S=1\}}$, and $\mathbf{B} = [\hat{\tau},\, N_{h,1}(\hat{\tau}), \ldots, N_{h,H-2}(\hat{\tau})]_{\{S=1\}}$.
\State Sample $\gamma_h^2 \sim \text{InvGamma}(1 + \frac{H-2}{2}, 1 + \frac{1}{2}\sum_{l=2}^{H-1}\theta_{h,l}^2)$.
\State Calculate $\text{NS}_h(\hat{\tau}_i\cond\cdot) = \theta_{h,1} \hat{\tau}_i+\sum_{l=1}^{H-2} \theta_{h, l+1} N_{h,l}(\hat{\tau}_i)$, $i=1,2,\ldots, N$.
\State Update $\text{BART}_s(\cdot)$ using $[\bX]_{\{S=1,A=1\}}$ as predictors, and $[Y - \text{BART}_h(\bX) - \text{NS}_h(\hat{\tau}\cond\cdot) - \text{NS}_s(\hat{\tau}\cond\cdot)]_{\{S=1,A=1\}}$ as outcomes.
\State Sample $(\theta_1, \ldots, \theta_{K-1})^T$ from $\mathcal{MVN}_{K-1}\!\left((\mathbf{D}^T\mathbf{D} + \mathbf{\Gamma})^{-1}\mathbf{D}^TU,\ \sigma^2(\mathbf{D}^T\mathbf{D} + \mathbf{\Gamma})^{-1}\right)$, where $\mathbf{\Gamma} = \mathrm{diag}(0,\, \tfrac{\sigma^2}{\gamma^2}, \ldots, \tfrac{\sigma^2}{\gamma^2})$, $\;U = [Y - \text{BART}_h(\bX) - \text{NS}_h(\hat{\tau}\cond\cdot) - \text{BART}_s(\bX)]_{\{S=1,A=1\}}$, and $\mathbf{D} = [\hat{\tau},\, N_1(\hat{\tau}), \ldots, N_{K-2}(\hat{\tau})]_{\{S=1,A=1\}}$.
\State Sample $\gamma^2 \sim \text{InvGamma}(\frac{K}{2}, 1 + \frac{1}{2}\sum_{k=2}^{K-1}\theta_k^2)$.
\State Calculate $\text{NS}_s(\hat{\tau}_i\cond\cdot) = \theta_1 \hat{\tau}_i+\sum_{k=1}^{K-2} \theta_{k+1} N_k(\hat{\tau}_i)$, $i=1,2,\ldots, N$.
\State Calculate $SSR=\sum_{i=1}^NS_i\cdot(Y_i - \text{BART}_h(X_i) - \text{NS}_h(\hat{\tau}_i\cond\cdot) - A_i\cdot [\text{BART}_s(X_i) + \text{NS}_s(\hat{\tau}_i\cond\cdot)])^2$.
\State Sample $\sigma^2$ from $\text{InvGamma}(\frac{\upsilon + n}{2}, \frac{\upsilon\lambda + SSR}{2})$.
\State Sample $\hat{Y}_i({1})$ from $\mathcal{N}(\text{BART}_h(X_i) + \text{NS}_h(\hat{\tau}_i\cond\cdot) + \text{BART}_s(X_i) + \text{NS}_s(\hat{\tau}_i\cond\cdot), \sigma^2)$, for $i=1,2,\ldots,N$.
\State Sample $\hat{Y}_i({0})$ from $\mathcal{N}(\text{BART}_h(X_i) + \text{NS}_h(\hat{\tau}_i\cond\cdot), \sigma^2)$, for $i=1,2,\ldots,N$.
\end{algorithmic}
\end{algorithm}

\clearpage
\newpage

\subsection{Properties of the Inverse Gumbel Transformation}
Web Figure~\ref{fig:Gumbel} compares the curves of the inverse Gumbel (InvGumbel) function and the Logit transformation. The InvGumbel transformation has three key advantages compared to the Logit function in modeling $\hat{\pi}$: 
\begin{figure}[!h]
\begin{center}
\includegraphics[width=0.7\textwidth]{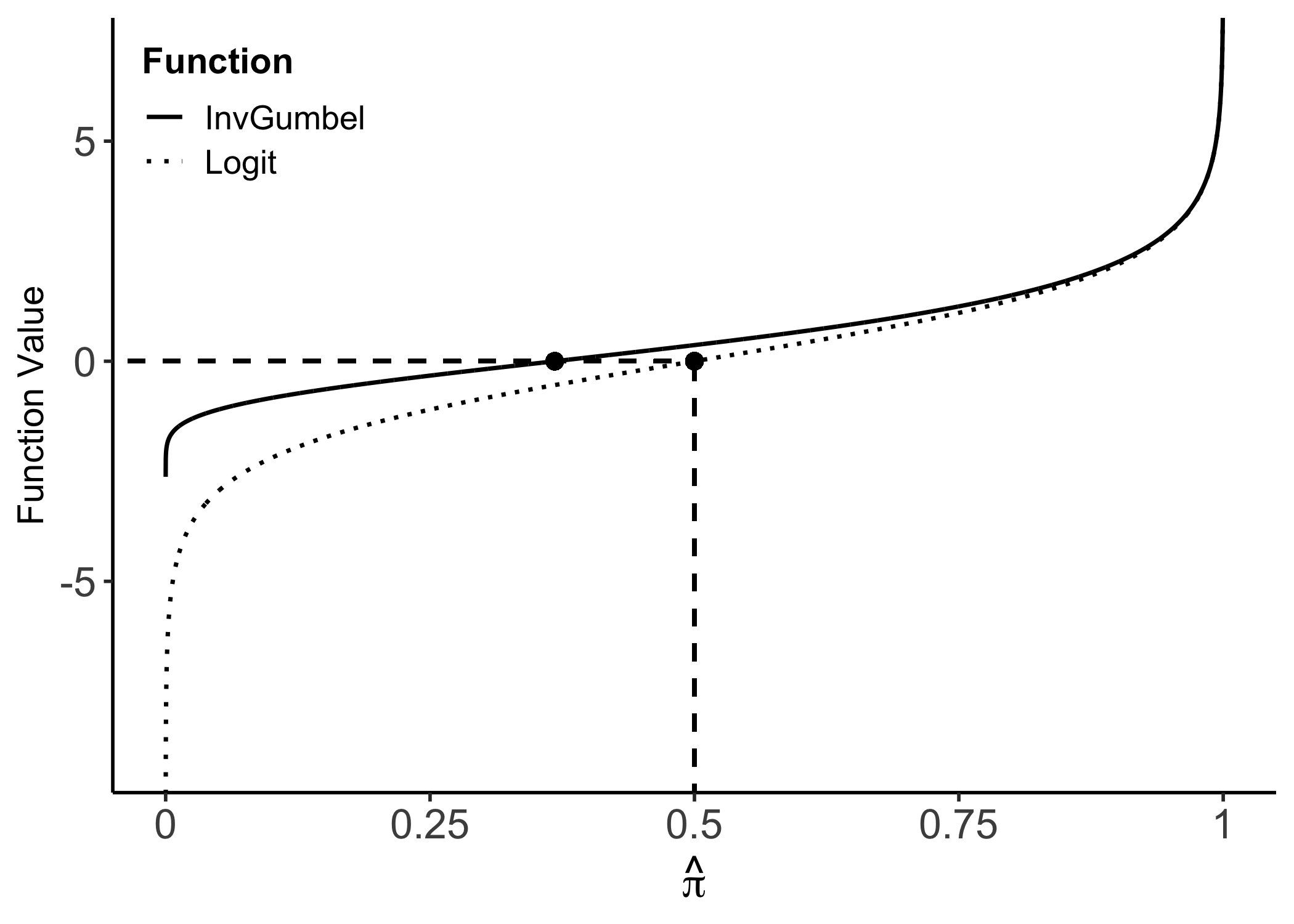}
\end{center}
\caption{Curves of InvGumbel function and Logit function}
\label{fig:Gumbel}
\end{figure}

\begin{enumerate}
\item As $\hat{\pi}_i \to 1$, $\phi(\hat{\pi}_i) = - \log (-\log(\hat{\pi})) \to \text{Logit}(\hat{\pi}_i)$. The InvGumbel function behaves similarly to the Logit function for large values of $\hat{\pi}$.
\item When $\hat{\pi}_i$ is close to 0, the InvGumbel transformation decreases more slowly than the Logit function, resulting in a more gradual expansion of the range. For example, for $\hat{\pi}_i \in(10^{-10}, 10^{-1})$, the Logit function maps to $(-23.02, -2.2)$, whereas the InvGumbel function maps to $(-3.14, -0.83)$. Moreover, compared with the Identity scale $\hat{\pi}$, the InvGumbel transformation still provides a flexible extrapolation region that is not overly narrow, helping to prevent undercoverage.

\item Compared with the Logit transformation, the InvGumbel transformation allocates a larger share of the transformed range to the right end of the distribution, where most trial sample units are located. This allows the natural cubic spline to better capture the association between the treatment effect and $\hat{\pi}$. At the same time, the range for out-of-sample units is narrower under the InvGumbel transformation, leading to more stable extrapolation performance.
\end{enumerate}

\clearpage
\suppsection{Details of Simulation Studies}

\subsection{Comparison Methods}
\label{sec:comparison}
We here only present the details of MATEs. For every comparison method we estimated the potential outcome means and obtained the average treatment effect by their difference. Throughout, $\hat{\pi}_i=\hat{\pi}(X_i)$ is the estimated sampling propensity score, fit by a probit BART model of $S$ on all covariates among the observed individuals and shared across every method that requires it. Sums over $\{i:S_i=1\}$ run over the $n$ trial participants, while $\frac{1}{N}\sum_i$ averages over the $N$ individuals of the EHR cohort. The estimators are listed below.

\begin{enumerate}
\item \textbf{Trial-only}: the mean among trial participants, $\dfrac{\sum_{i:S_i=1}\mathbb{I}(A_i=a)Y_i}{\sum_{i:S_i=1}\mathbb{I}(A_i=a)}$.

\item \textbf{IPPW} \citep{Dahabreh2019}: the inverse probability of participation weighting estimator with normalized weights, $\dfrac{\sum_{i:S_i=1} W_i(a)\mathbb{I}(A_i=a)Y_i}{\sum_{i:S_i=1} W_i(a)\mathbb{I}(A_i=a)}$, with $W_i(a)=\dfrac{1}{\hat{\pi}_i\Pr(A=a\cond S=1)}$ for generalizability and $W_i(a)=\dfrac{1-\hat{\pi}_i}{\hat{\pi}_i\Pr(A=a\cond S=1)}$ for transportability.

\item \textbf{A-IPPW} \citep{Dahabreh2019}: the augmented estimator with normalized weights, $\dfrac{\sum_{i:S_i=1} W_i(a)\mathbb{I}(A_i=a)[Y_i-\hat{Y}_i(a)]}{\sum_{i:S_i=1} W_i(a)\mathbb{I}(A_i=a)}+\dfrac{1}{N}\sum_{i}\hat{Y}_i(a)$, where $W_i(a)$ is as in IPPW for the corresponding setting and the imputed potential outcomes $\hat{Y}_i(a)$ are obtained from two BART models fit on the two trial arms.

\item \textbf{BART} \citep{Kern2017}: an outcome regression with $\mu_{A_i}=\text{BART}(X_i,A_i)$.

\item \textbf{BART-PS}: BART with the propensity score added, $\mu_{A_i}=\text{BART}(X_i,A_i,\hat{\pi}_i)$.

\item \textbf{InterBART}: a PSPI model without $\hat{\pi}$, $\mu_{A_i}=\text{BART}_h(X_i)+A_i\cdot\text{BART}_s(X_i)$.

\item \textbf{InterBART-PS}: InterBART with $\hat{\pi}$ added to both components, $\mu_{A_i}=\text{BART}_h(X_i,\hat{\pi}_i)+A_i\cdot\text{BART}_s(X_i,\hat{\pi}_i)$.

\item \textbf{FullBART}: a PSPI variant of SplineBART replacing the spline with a separate BART, $\mu_{A_i}=\text{BART}_h(X_i,\hat{\pi}_i)+A_i\cdot[\text{BART}_s(X_i)+\text{BART}_{\pi}(\hat{\pi}_i)]$, under the identifiability constraint $\frac{1}{n}\sum_{i:S_i=1}\text{BART}_s(X_i)=0$.

\item \textbf{True model}: a Bayesian regression using the true functional form of the outcomes, implemented in Stan \citep{Carpenter2017}.
\end{enumerate}

For interval estimation, 95\% confidence intervals for the trial-only estimator were constructed using the Welch two-sample \textit{t}-test. For IPPW and A-IPPW, we applied the approach in \citet{Antonelli2022} to construct 95\% intervals, with the number of bootstrap resamples set to $500$. For the fully Bayesian outcome models, posterior means and 95\% equal-tailed intervals were reported directly.

\subsection{Additional Simulation Details and Results}
In this section, we present additional design and results of simulation studies.

\subsubsection{Transportability}
In the main paper, we presented the simulation design of scenarios SS1--SS4 for transportability. Here we additionally include scenario SS5 with imbalanced treatment assignments:

\noindent \textbf{SS5. Linear setting with imbalanced treatment assignment}: We considered the same linear setting as SS1, with EHR cohort size $N=500$. Treatment was randomized among trial participants with $\Pr(A=1)=0.3$ or $\Pr(A=1)=0.7$.

The additional simulation results are provided in Web Tables~\ref{tab:supp_ss1}--\ref{tab:supp_ss5}: SS1 (linear) with EHR cohort sizes $N=1000$ and $N=2000$ (Web Table~\ref{tab:supp_ss1}); SS2 (linear, covariate shift), SS3 (nonlinear), and SS4 (nonlinear, covariate shift) with $N=1000$ (Web Table~\ref{tab:supp_ss234}); and SS5 (linear, imbalanced treatment assignment) with $\Pr(A=1)=0.3$ and $0.7$ (Web Table~\ref{tab:supp_ss5}).

\begin{table}[H]
\caption{Comparison of bias ($\times100$), RMSE ($\times100$), coverage rate ($\times100$), and average width ($\times100$) of 95\% confidence/credible intervals in simulation scenario SS1 (linear) with EHR cohort sizes $N=1000$ and $N=2000$.}
\label{tab:supp_ss1}
\resizebox{\textwidth}{!}{%
\begin{tabular}{lcccccccccccccc}
\toprule
\multicolumn{15}{c}{\textbf{SS1. Linear, $N = 1000$, $n=150$}} \\
\hline
\multirow{2}{*}{\textbf{Estimators}} & \multicolumn{4}{c}{\textbf{$\e_T[Y(1)-Y(0)]$}} & & \multicolumn{4}{c}{\textbf{$\e_T[Y(1)]$}} & & \multicolumn{4}{c}{\textbf{$\e_T[Y(0)]$}} \\ \cline{2-5} \cline{7-10} \cline{12-15}
  & \textbf{Bias} & \textbf{RMSE} & \textbf{Coverage} & \textbf{Width} & & \textbf{Bias} & \textbf{RMSE}  & \textbf{Coverage} & \textbf{Width} & & \textbf{Bias} & \textbf{RMSE}  & \textbf{Coverage} & \textbf{Width} \\
 \hline
\textbf{Existing methods} & & & & & & & & & & & & & &\\
\ \ Trial-only & 216 & 228 & 16 & 289 & & 447 & 452 & 0 & 265 & & 231 & 233 & 0 & 116\\
\ \ IPPW & 66 & 152 & \bf 95 & 657 & & 170 & 205 & 77 & 613 & & 104 & 119 & 54 & 269\\
\ \ A-IPPW & 37 & \bf 67 & 82 & 194 & & 102 & 111 & 31 & 163 & & 65 & 73 & 39 & 114\\
\ \ BART & 92 & 103 & 45 & 172 & & 98 & 107 & 21 & 134 & & 6 & 35 & 95 & 131\\
\ \ BART-PS & 94 & 106 & 49 & 181 & & 111 & 121 & 24 & 154 & & 16 & 42 & 92 & 153\\
\textbf{Interaction-based outcome models} & & & & & & & & & & & & & &\\
\ \ InterBART & 91 & 99 & 46 & 169 & & 101 & 108 & 22 & 142 & & 10 & 33 & 95 & 129\\
\ \ InterBART-PS & 101 & 108 & 37 & 171 & & 116 & 124 & 20 & 162 & & 14 & 40 & 94 & 151\\
\ \ FullBART & 81 & 92 & 61 & 186 & & 108 & 117 & 30 & 167 & & 27 & 46 & 90 & 154\\
\ \ SplineBART & \bf -1 & \bf 64 & \bf 95 & 255 & & 45 & 71 & 86 & 211 & & 46 & 60 & 80 & 161\\
\ \ DSplineBART & \bf 11 & \bf 64 & \bf 96 & 255 & & \bf 22 & \bf 57 & \bf 92 & 203 & & 11 & 45 & \bf 95 & 177\\
\textbf{True model} & -2 & 28 & 93 & 105 & & -1 & 23 & 94 & 87 & & 1 & 15 & 94 & 59\\
\midrule
\textbf{With Bayesian bootstrap} & & & & & & & & & & & & & &\\
\ \ SplineBART & \bf -2 & \bf 66 & \bf 96 & 259 & & 45 & 72 & 88 & 222 & & 47 & 61 & 82 & 165\\
\ \ DSplineBART & \bf 9 & \bf 66 & \bf 96 & 259 & & \bf 22 & \bf 59 & \bf 94 & 214 & & 12 & 46 & \bf 95 & 182\\
\ \ True model & -3 & 31 & 94 & 121 & & -1 & 28 & 97 & 117 & & 2 & 17 & 95 & 69\\
\midrule
\midrule
\multicolumn{15}{c}{\textbf{SS1. Linear, $N = 2000$, $n=150$}} \\
\hline
\multirow{2}{*}{\textbf{Estimators}} & \multicolumn{4}{c}{\textbf{$\e_T[Y(1)-Y(0)]$}} & & \multicolumn{4}{c}{\textbf{$\e_T[Y(1)]$}} & & \multicolumn{4}{c}{\textbf{$\e_T[Y(0)]$}} \\ \cline{2-5} \cline{7-10} \cline{12-15}
  & \textbf{Bias} & \textbf{RMSE} & \textbf{Coverage} & \textbf{Width} & & \textbf{Bias} & \textbf{RMSE}  & \textbf{Coverage} & \textbf{Width} & & \textbf{Bias} & \textbf{RMSE}  & \textbf{Coverage} & \textbf{Width} \\
 \hline
\textbf{Existing methods} & & & & & & & & & & & & & &\\
\ \ Trial-only & 216 & 228 & 16 & 289 & & 447 & 452 & 0 & 265 & & 231 & 233 & 0 & 116\\
\ \ IPPW & 65 & 154 & \bf 95 & 660 & & 165 & 203 & 74 & 614 & & 100 & 116 & 58 & 272\\
\ \ A-IPPW & 38 & 67 & 81 & 190 & & 102 & 111 & 27 & 156 & & 64 & 73 & 38 & 113\\
\ \ BART & 91 & 102 & 45 & 172 & & 98 & 107 & 22 & 134 & & 8 & 35 & 94 & 130\\
\ \ BART-PS & 95 & 107 & 46 & 180 & & 112 & 122 & 22 & 152 & & 17 & 42 & 92 & 150\\
\textbf{Interaction-based outcome models} & & & & & & & & & & & & & &\\
\ \ InterBART & 91 & 99 & 44 & 169 & & 102 & 109 & 22 & 141 & & 10 & 33 & 94 & 129\\
\ \ InterBART-PS & 101 & 108 & 34 & 170 & & 117 & 125 & 19 & 160 & & 16 & 39 & 93 & 148\\
\ \ FullBART & 81 & 91 & 59 & 185 & & 109 & 118 & 29 & 165 & & 28 & 46 & 89 & 151\\
\ \ SplineBART & \bf 6 & \bf 61 & \bf 95 & 237 & & 49 & 72 & 86 & 199 & & 43 & 57 & 84 & 158\\
\ \ DSplineBART & \bf 18 & \bf 63 & \bf 91 & 234 & & \bf 27 & \bf 57 & 89 & 190 & & 9 & 42 & \bf 94 & 166\\
\textbf{True model} & -2 & 28 & 93 & 105 & & -1 & 23 & 94 & 87 & & 1 & 15 & 94 & 58\\
\midrule
\textbf{With Bayesian bootstrap} & & & & & & & & & & & & & &\\
\ \ SplineBART & \bf 5 & \bf 61 & \bf 96 & 239 & & 49 & 72 & 84 & 205 & & 44 & 58 & 85 & 161\\
\ \ DSplineBART & \bf 17 & \bf 63 & \bf 94 & 236 & & \bf 26 & \bf 57 & \bf 90 & 196 & & 9 & 43 & \bf 94 & 168\\
\ \ True model & -3 & 29 & 94 & 113 & & -1 & 25 & 97 & 103 & & 1 & 16 & 95 & 63\\
\bottomrule
\end{tabular}
}
\end{table}

\begin{table}[H]
\caption{Comparison of bias ($\times100$), RMSE ($\times100$), coverage rate ($\times100$), and average width ($\times100$) of 95\% confidence/credible intervals in simulation scenarios SS2 (linear, covariate shift), SS3 (nonlinear), and SS4 (nonlinear, covariate shift) with EHR cohort size $N=1000$.}
\label{tab:supp_ss234}
\resizebox{\textwidth}{!}{%
\begin{tabular}{lcccccccccccccc}
\toprule
\multicolumn{15}{c}{\textbf{SS2. Linear, strong covariate shift, $N = 1000$, $n=150$}} \\
\hline
\multirow{2}{*}{\textbf{Estimators}} & \multicolumn{4}{c}{\textbf{$\e_T[Y(1)-Y(0)]$}} & & \multicolumn{4}{c}{\textbf{$\e_T[Y(1)]$}} & & \multicolumn{4}{c}{\textbf{$\e_T[Y(0)]$}} \\ \cline{2-5} \cline{7-10} \cline{12-15}
  & \textbf{Bias} & \textbf{RMSE} & \textbf{Coverage} & \textbf{Width} & & \textbf{Bias} & \textbf{RMSE}  & \textbf{Coverage} & \textbf{Width} & & \textbf{Bias} & \textbf{RMSE}  & \textbf{Coverage} & \textbf{Width} \\
 \hline
\textbf{Existing methods} & & & & & & & & & & & & & &\\
\ \ Trial-only & 500 & 505 & 0 & 247 & & 804 & 807 & 0 & 228 & & 304 & 305 & 0 & 98\\
\ \ IPPW & 236 & 280 & 63 & 631 & & 381 & 401 & 30 & 587 & & 145 & 157 & 37 & 259\\
\ \ A-IPPW & 146 & 166 & 38 & 230 & & 254 & 262 & 1 & 187 & & 108 & 115 & 18 & 137\\
\ \ BART & 232 & 238 & 1 & 184 & & 250 & 257 & 0 & 172 & & 18 & 45 & 95 & 168\\
\ \ BART-PS & 214 & 222 & 7 & 211 & & 243 & 252 & 2 & 209 & & 29 & 57 & 92 & 208\\
\textbf{Interaction-based outcome models} & & & & & & & & & & & & & &\\
\ \ InterBART & 237 & 243 & 0 & 194 & & 263 & 270 & 0 & 191 & & 26 & 47 & 94 & 176\\
\ \ InterBART-PS & 250 & 255 & 0 & 196 & & 269 & 277 & 1 & 227 & & 19 & 51 & 97 & 214\\
\ \ FullBART & 209 & 216 & 6 & 225 & & 251 & 260 & 2 & 235 & & 43 & 62 & 92 & 220\\
\ \ SplineBART & \bf -11 & \bf 114 & \bf 96 & 481 & & 72 & 129 & 90 & 420 & & 84 & 95 & 74 & 230\\
\ \ DSplineBART & \bf -2 & \bf 118 & \bf 96 & 493 & & \bf 47 & \bf 118 & \bf 92 & 417 & & 50 & 87 & \bf 96 & 364\\
\textbf{True model} & 2 & 33 & 92 & 124 & & 1 & 31 & 94 & 115 & & -1 & 11 & 95 & 46\\
\midrule
\textbf{With Bayesian bootstrap} & & & & & & & & & & & & & &\\
\ \ SplineBART & \bf -13 & \bf 113 & \bf 96 & 483 & & 72 & 127 & 91 & 424 & & 84 & 96 & 72 & 232\\
\ \ DSplineBART & \bf -4 & \bf 117 & \bf 97 & 495 & & \bf 47 & \bf 116 & \bf 92 & 421 & & 50 & 87 & \bf 96 & 366\\
\ \ True model & 1 & 34 & 95 & 138 & & 0 & 34 & 95 & 139 & & 0 & 14 & 94 & 59\\
\midrule
\midrule
\multicolumn{15}{c}{\textbf{SS3. Nonlinear, $N = 1000$, $n=150$}} \\
\hline
\multirow{2}{*}{\textbf{Estimators}} & \multicolumn{4}{c}{\textbf{$\e_T[Y(1)-Y(0)]$}} & & \multicolumn{4}{c}{\textbf{$\e_T[Y(1)]$}} & & \multicolumn{4}{c}{\textbf{$\e_T[Y(0)]$}} \\ \cline{2-5} \cline{7-10} \cline{12-15}
  & \textbf{Bias} & \textbf{RMSE} & \textbf{Coverage} & \textbf{Width} & & \textbf{Bias} & \textbf{RMSE}  & \textbf{Coverage} & \textbf{Width} & & \textbf{Bias} & \textbf{RMSE}  & \textbf{Coverage} & \textbf{Width} \\
 \hline
\textbf{Existing methods} & & & & & & & & & & & & & &\\
\ \ Trial-only & 340 & 362 & 22 & 479 & & 826 & 833 & 0 & 388 & & 486 & 492 & 0 & 284\\
\ \ IPPW & 165 & 286 & 90 & 980 & & 428 & 461 & 42 & 816 & & 263 & 295 & 46 & 588\\
\ \ A-IPPW & 100 & 145 & 87 & 433 & & 238 & 253 & 24 & 360 & & 138 & 148 & 42 & 256\\
\ \ BART & 158 & 178 & 52 & 308 & & 182 & 202 & 34 & 284 & & 24 & 72 & 94 & 279\\
\ \ BART-PS & 158 & 176 & 56 & 324 & & 233 & 250 & 27 & 330 & & 75 & 103 & 84 & 326\\
\textbf{Interaction-based outcome models} & & & & & & & & & & & & & &\\
\ \ InterBART & 186 & 195 & 24 & 277 & & 201 & 216 & 28 & 297 & & 15 & 66 & 96 & 286\\
\ \ InterBART-PS & 203 & 211 & 15 & 277 & & 256 & 268 & 20 & 351 & & 52 & 87 & 93 & 338\\
\ \ FullBART & 153 & 168 & 60 & 343 & & 241 & 255 & 28 & 366 & & 88 & 110 & 86 & 354\\
\ \ SplineBART & \bf -23 & \bf 142 & \bf 98 & 663 & & 99 & 166 & 92 & 576 & & 122 & 139 & 77 & 366\\
\ \ DSplineBART & \bf -2 & \bf 140 & \bf 97 & 658 & & \bf 42 & \bf 137 & \bf 96 & 547 & & 44 & 103 & \bf 99 & 488\\
\textbf{True model} & 0 & 58 & 92 & 212 & & -2 & 61 & 96 & 236 & & -2 & 53 & 96 & 209\\
\midrule
\textbf{With Bayesian bootstrap} & & & & & & & & & & & & & &\\
\ \ SplineBART & \bf -24 & \bf 143 & \bf 99 & 665 & & 98 & 166 & 92 & 584 & & 122 & 140 & 79 & 374\\
\ \ DSplineBART & \bf -3 & \bf 141 & \bf 98 & 660 & & \bf 42 & \bf 138 & \bf 96 & 555 & & 44 & 104 & \bf 99 & 495\\
\ \ True model & -1 & 60 & 93 & 222 & & -3 & 64 & 95 & 255 & & -2 & 56 & 95 & 220\\
\midrule
\midrule
\multicolumn{15}{c}{\textbf{SS4. Nonlinear, strong covariate shift, $N = 1000$, $n=150$}} \\
\hline
\multirow{2}{*}{\textbf{Estimators}} & \multicolumn{4}{c}{\textbf{$\e_T[Y(1)-Y(0)]$}} & & \multicolumn{4}{c}{\textbf{$\e_T[Y(1)]$}} & & \multicolumn{4}{c}{\textbf{$\e_T[Y(0)]$}} \\ \cline{2-5} \cline{7-10} \cline{12-15}
  & \textbf{Bias} & \textbf{RMSE} & \textbf{Coverage} & \textbf{Width} & & \textbf{Bias} & \textbf{RMSE}  & \textbf{Coverage} & \textbf{Width} & & \textbf{Bias} & \textbf{RMSE}  & \textbf{Coverage} & \textbf{Width} \\
 \hline
\textbf{Existing methods} & & & & & & & & & & & & & &\\
\ \ Trial-only & 403 & 417 & 4 & 419 & & 873 & 877 & 0 & 336 & & 470 & 474 & 0 & 253\\
\ \ IPPW & 263 & 312 & 66 & 728 & & 610 & 623 & 7 & 630 & & 347 & 357 & 12 & 424\\
\ \ A-IPPW & 168 & 186 & 62 & 382 & & 333 & 340 & 0 & 309 & & 166 & 173 & 20 & 240\\
\ \ BART & 230 & 238 & 6 & 243 & & 305 & 315 & 1 & 250 & & 75 & 99 & 77 & 252\\
\ \ BART-PS & 218 & 226 & 12 & 268 & & 316 & 326 & 3 & 315 & & 98 & 120 & 80 & 314\\
\textbf{Interaction-based outcome models} & & & & & & & & & & & & & &\\
\ \ InterBART & 241 & 245 & 1 & 238 & & 314 & 321 & 0 & 277 & & 73 & 94 & 84 & 270\\
\ \ InterBART-PS & 251 & 255 & 0 & 239 & & 335 & 343 & 2 & 344 & & 84 & 108 & 87 & 335\\
\ \ FullBART & 203 & 210 & 19 & 316 & & 316 & 325 & 4 & 361 & & 114 & 132 & 81 & 352\\
\ \ SplineBART & \bf -5 & \bf 149 & \bf 99 & 804 & & 141 & 200 & \bf 94 & 717 & & 146 & 159 & 70 & 359\\
\ \ DSplineBART & \bf 8 & \bf 151 & \bf 99 & 814 & & \bf 100 & \bf 179 & \bf 96 & 708 & & 92 & 148 & \bf 97 & 617\\
\textbf{True model} & 4 & 54 & 95 & 216 & & -2 & 70 & 95 & 275 & & -6 & 64 & 96 & 250\\
\midrule
\textbf{With Bayesian bootstrap} & & & & & & & & & & & & & &\\
\ \ SplineBART & \bf -5 & \bf 149 & \bf 99 & 805 & & 141 & 200 & \bf 94 & 723 & & 146 & 160 & 68 & 366\\
\ \ DSplineBART & \bf 8 & \bf 151 & \bf 99 & 815 & & \bf 100 & \bf 180 & \bf 95 & 714 & & 92 & 149 & \bf 97 & 621\\
\ \ True model & 4 & 56 & 96 & 226 & & -1 & 71 & 97 & 292 & & -5 & 65 & 95 & 260\\
\bottomrule
\end{tabular}
}
\end{table}

\begin{table}[H]
\caption{Comparison of bias ($\times100$), RMSE ($\times100$), coverage rate ($\times100$), and average width ($\times100$) of 95\% confidence/credible intervals in simulation scenario SS5 (linear with imbalanced treatment assignment) with EHR cohort size $N=500$ and $\Pr(A=1)=0.3$ or $0.7$.}
\label{tab:supp_ss5}
\resizebox{\textwidth}{!}{%
\begin{tabular}{lcccccccccccccc}
\toprule
\multicolumn{15}{c}{\textbf{SS5. Linear, $\Pr(A=1)=0.3$, $N = 500$, $n=150$}} \\
\hline
\multirow{2}{*}{\textbf{Estimators}} & \multicolumn{4}{c}{\textbf{$\e_T[Y(1)-Y(0)]$}} & & \multicolumn{4}{c}{\textbf{$\e_T[Y(1)]$}} & & \multicolumn{4}{c}{\textbf{$\e_T[Y(0)]$}} \\ \cline{2-5} \cline{7-10} \cline{12-15}
  & \textbf{Bias} & \textbf{RMSE} & \textbf{Coverage} & \textbf{Width} & & \textbf{Bias} & \textbf{RMSE}  & \textbf{Coverage} & \textbf{Width} & & \textbf{Bias} & \textbf{RMSE}  & \textbf{Coverage} & \textbf{Width} \\
 \hline
\textbf{Existing methods} & & & & & & & & & & & & & &\\
\ \ Trial-only & 219 & 238 & 33 & 356 & & 450 & 460 & 0 & 344 & & 231 & 233 & 0 & 98\\
\ \ IPPW & 87 & 178 & \bf 91 & 717 & & 192 & 238 & 76 & 691 & & 106 & 115 & 49 & 241\\
\ \ A-IPPW & 99 & 124 & 62 & 235 & & 149 & 163 & 28 & 222 & & 49 & 56 & 48 & 101\\
\ \ BART & 111 & 126 & 34 & 181 & & 129 & 142 & 20 & 167 & & 18 & 33 & 91 & 108\\
\ \ BART-PS & 108 & 122 & 42 & 193 & & 134 & 148 & 25 & 187 & & 26 & 43 & 86 & 132\\
\textbf{Interaction-based outcome models} & & & & & & & & & & & & & &\\
\ \ InterBART & 105 & 117 & 39 & 184 & & 126 & 137 & 22 & 174 & & 22 & 34 & 88 & 108\\
\ \ InterBART-PS & 113 & 124 & 34 & 185 & & 140 & 151 & 22 & 190 & & 26 & 43 & 88 & 132\\
\ \ FullBART & 97 & 111 & 52 & 196 & & 129 & 142 & 30 & 196 & & 32 & 46 & 83 & 133\\
\ \ SplineBART & \bf -6 & \bf 83 & \bf 93 & 304 & & 41 & 89 & 88 & 282 & & 47 & 58 & 72 & 135\\
\ \ DSplineBART & \bf 3 & \bf 83 & \bf 94 & 301 & & \bf 13 & \bf 78 & \bf 92 & 275 & & 11 & 36 & \bf 96 & 151\\
\textbf{True model} & 0 & 34 & 94 & 128 & & 0 & 30 & 96 & 117 & & 0 & 13 & 94 & 51\\
\midrule
\textbf{With Bayesian bootstrap} & & & & & & & & & & & & & &\\
\ \ SplineBART & \bf -8 & \bf 83 & \bf 94 & 312 & & 39 & 89 & 92 & 297 & & 47 & 59 & 75 & 144\\
\ \ DSplineBART & \bf 0 & \bf 83 & \bf 94 & 308 & & \bf 11 & \bf 79 & \bf 93 & 291 & & 11 & 38 & \bf 96 & 159\\
\ \ True model & -2 & 37 & 97 & 153 & & -2 & 38 & 97 & 162 & & 0 & 18 & 97 & 71\\
\midrule
\midrule
\multicolumn{15}{c}{\textbf{SS5. Linear, $\Pr(A=1)=0.7$, $N = 500$, $n=150$}} \\
\hline
\multirow{2}{*}{\textbf{Estimators}} & \multicolumn{4}{c}{\textbf{$\e_T[Y(1)-Y(0)]$}} & & \multicolumn{4}{c}{\textbf{$\e_T[Y(1)]$}} & & \multicolumn{4}{c}{\textbf{$\e_T[Y(0)]$}} \\ \cline{2-5} \cline{7-10} \cline{12-15}
  & \textbf{Bias} & \textbf{RMSE} & \textbf{Coverage} & \textbf{Width} & & \textbf{Bias} & \textbf{RMSE}  & \textbf{Coverage} & \textbf{Width} & & \textbf{Bias} & \textbf{RMSE}  & \textbf{Coverage} & \textbf{Width} \\
 \hline
\textbf{Existing methods} & & & & & & & & & & & & & &\\
\ \ Trial-only & 222 & 235 & 13 & 270 & & 452 & 457 & 0 & 227 & & 230 & 234 & 0 & 149\\
\ \ IPPW & 73 & 142 & \bf 97 & 611 & & 182 & 202 & 72 & 547 & & 109 & 128 & 59 & 298\\
\ \ A-IPPW & -11 & \bf 63 & 88 & 207 & & 76 & 86 & 48 & 153 & & 87 & 98 & 37 & 150\\
\ \ BART & 98 & 111 & 47 & 194 & & 84 & 93 & 25 & 119 & & -14 & 52 & 92 & 174\\
\ \ BART-PS & 101 & 114 & 50 & 205 & & 96 & 105 & 26 & 145 & & -5 & 52 & 94 & 197\\
\textbf{Interaction-based outcome models} & & & & & & & & & & & & & &\\
\ \ InterBART & 117 & 126 & 28 & 183 & & 93 & 101 & 21 & 127 & & -24 & 53 & 91 & 168\\
\ \ InterBART-PS & 132 & 140 & 20 & 184 & & 108 & 116 & 21 & 155 & & -24 & 56 & 92 & 190\\
\ \ FullBART & 90 & 101 & 64 & 214 & & 100 & 109 & 30 & 159 & & 10 & 51 & 96 & 204\\
\ \ SplineBART & \bf -19 & \bf 68 & \bf 96 & 296 & & 37 & 60 & \bf 93 & 197 & & 57 & 77 & 84 & 216\\
\ \ DSplineBART & \bf 0 & \bf 71 & \bf 96 & 302 & & \bf 21 & \bf 49 & \bf 96 & 185 & & 21 & 70 & \bf 95 & 257\\
\textbf{True model} & 4 & 27 & 96 & 109 & & 2 & 18 & 96 & 75 & & -2 & 19 & 96 & 78\\
\midrule
\textbf{With Bayesian bootstrap} & & & & & & & & & & & & & &\\
\ \ SplineBART & \bf -22 & \bf 72 & \bf 95 & 301 & & 35 & 65 & \bf 92 & 220 & & 57 & 77 & 84 & 224\\
\ \ DSplineBART & \bf -2 & \bf 74 & \bf 96 & 309 & & \bf 19 & \bf 56 & \bf 93 & 210 & & 21 & 70 & \bf 95 & 264\\
\ \ True model & 2 & 33 & 96 & 137 & & 0 & 33 & 96 & 134 & & -2 & 21 & 96 & 93\\
\bottomrule
\end{tabular}
}
\end{table}

\subsubsection{Generalizability}
We present the designs and results of the generalizability simulations here. For generalizability, we first drew an EHR cohort from the superpopulation using the \texttt{sample()} function in \texttt{R} for simple random sampling without replacement. We then drew a trial sample of size $n=150$ from the cohort using the same \texttt{sample()} function without replacement, with the \texttt{prob} argument given by the setting-specific participation models of SS1--SS5.

We reuse the same scenarios SS1--SS5 as in the transportability simulations: the potential-outcome models and the trial-participation (\texttt{prob}) mechanisms are identical. The design differs only in the sampling structure, matching the nested design: the EHR cohort is a simple random sample from the superpopulation, and the trial sample is drawn from within the cohort, so no separate target population is formed. The estimands are $\e_T[Y(1)-Y(0)]$, $\e_T[Y(1)]$, and $\e_T[Y(0)]$, identical to the transportability simulations. EHR cohort sizes and treatment-assignment probabilities are as in the corresponding scenarios.

The additional simulation results are provided in Web Tables~\ref{tab:supp_gen_ss1}--\ref{tab:supp_gen_ss45}. 

\begin{table}[H]
\caption{\textbf{Generalizability.} Comparison of bias ($\times100$), RMSE ($\times100$), coverage rate ($\times100$), and average width ($\times100$) of 95\% confidence/credible intervals for scenario SS1 (linear) with EHR cohort sizes $N=500$, $1000$, and $2000$.}
\label{tab:supp_gen_ss1}
\resizebox{\textwidth}{!}{%
\begin{tabular}{lcccccccccccccc}
\toprule
\multicolumn{15}{c}{\textbf{Generalizability --- SS1. Linear, $N = 500$, $n=150$}} \\
\hline
\multirow{2}{*}{\textbf{Estimators}} & \multicolumn{4}{c}{\textbf{$\e_T[Y(1)-Y(0)]$}} & & \multicolumn{4}{c}{\textbf{$\e_T[Y(1)]$}} & & \multicolumn{4}{c}{\textbf{$\e_T[Y(0)]$}} \\ \cline{2-5} \cline{7-10} \cline{12-15}
  & \textbf{Bias} & \textbf{RMSE} & \textbf{Coverage} & \textbf{Width} & & \textbf{Bias} & \textbf{RMSE}  & \textbf{Coverage} & \textbf{Width} & & \textbf{Bias} & \textbf{RMSE}  & \textbf{Coverage} & \textbf{Width} \\
 \hline
\textbf{Existing methods} & & & & & & & & & & & & & &\\
\ \ Trial-only & 132 & 150 & 54 & 293 & & 297 & 304 & 1 & 270 & & 165 & 167 & 0 & 116\\
\ \ IPPW & 48 & 136 & \bf 93 & 561 & & 126 & 164 & 77 & 518 & & 78 & 91 & 61 & 224\\
\ \ A-IPPW & 18 & \bf 44 & \bf 95 & 172 & & 65 & 71 & 67 & 157 & & 46 & 52 & 57 & 99\\
\ \ BART & 57 & 66 & 68 & 145 & & 62 & 69 & 40 & 110 & & 5 & 26 & 95 & 106\\
\ \ BART-PS & 59 & 69 & 73 & 154 & & 69 & 78 & 43 & 128 & & 10 & 32 & 94 & 127\\
\textbf{Interaction-based outcome models} & & & & & & & & & & & & & &\\
\ \ InterBART & 53 & 61 & 72 & 146 & & 62 & 68 & 44 & 116 & & 8 & 25 & 95 & 105\\
\ \ InterBART-PS & 63 & 70 & 64 & 148 & & 72 & 80 & 41 & 134 & & 9 & 31 & 96 & 125\\
\ \ FullBART & 52 & 62 & 80 & 159 & & 68 & 77 & 51 & 138 & & 16 & 34 & 93 & 128\\
\ \ SplineBART & \bf -10 & \bf 51 & \bf 96 & 211 & & 26 & 53 & 90 & 171 & & 36 & 46 & 85 & 130\\
\ \ DSplineBART & \bf 0 & \bf 50 & \bf 98 & 214 & & \bf 9 & \bf 43 & \bf 95 & 166 & & 10 & 34 & \bf 94 & 147\\
\textbf{True model} & 0 & 20 & 99 & 92 & & 1 & 16 & 99 & 75 & & 1 & 13 & 97 & 54\\
\midrule
\textbf{With Bayesian bootstrap} & & & & & & & & & & & & & &\\
\ \ SplineBART & \bf -11 & \bf 56 & \bf 96 & 222 & & 23 & 60 & 90 & 198 & & 34 & 46 & 88 & 141\\
\ \ DSplineBART & \bf -2 & \bf 54 & \bf 97 & 225 & & \bf 6 & \bf 52 & \bf 95 & 195 & & 8 & 35 & \bf 97 & 157\\
\ \ True model & -1 & 27 & 98 & 125 & & -2 & 32 & 96 & 134 & & -1 & 17 & 96 & 74\\
\midrule
\midrule
\multicolumn{15}{c}{\textbf{Generalizability --- SS1. Linear, $N = 1000$, $n=150$}} \\
\hline
\multirow{2}{*}{\textbf{Estimators}} & \multicolumn{4}{c}{\textbf{$\e_T[Y(1)-Y(0)]$}} & & \multicolumn{4}{c}{\textbf{$\e_T[Y(1)]$}} & & \multicolumn{4}{c}{\textbf{$\e_T[Y(0)]$}} \\ \cline{2-5} \cline{7-10} \cline{12-15}
  & \textbf{Bias} & \textbf{RMSE} & \textbf{Coverage} & \textbf{Width} & & \textbf{Bias} & \textbf{RMSE}  & \textbf{Coverage} & \textbf{Width} & & \textbf{Bias} & \textbf{RMSE}  & \textbf{Coverage} & \textbf{Width} \\
 \hline
\textbf{Existing methods} & & & & & & & & & & & & & &\\
\ \ Trial-only & 159 & 171 & 40 & 292 & & 347 & 352 & 0 & 269 & & 188 & 190 & 0 & 116\\
\ \ IPPW & 51 & 143 & \bf 95 & 615 & & 134 & 173 & 78 & 572 & & 83 & 100 & 59 & 241\\
\ \ A-IPPW & 25 & \bf 55 & 90 & 178 & & 78 & 86 & 49 & 151 & & 52 & 59 & 50 & 105\\
\ \ BART & 71 & 81 & 53 & 157 & & 76 & 84 & 35 & 121 & & 5 & 28 & 97 & 115\\
\ \ BART-PS & 71 & 81 & 59 & 165 & & 86 & 96 & 38 & 139 & & 15 & 35 & 96 & 135\\
\textbf{Interaction-based outcome models} & & & & & & & & & & & & & &\\
\ \ InterBART & 68 & 76 & 60 & 155 & & 76 & 84 & 36 & 127 & & 9 & 28 & 95 & 114\\
\ \ InterBART-PS & 75 & 83 & 52 & 158 & & 90 & 98 & 34 & 146 & & 14 & 34 & 94 & 134\\
\ \ FullBART & 61 & 71 & 74 & 170 & & 84 & 93 & 43 & 151 & & 23 & 38 & 93 & 136\\
\ \ SplineBART & \bf -2 & \bf 51 & \bf 98 & 231 & & 36 & 59 & 88 & 190 & & 38 & 50 & 88 & 142\\
\ \ DSplineBART & \bf 8 & \bf 50 & \bf 98 & 233 & & \bf 17 & \bf 47 & \bf 93 & 183 & & 9 & 35 & \bf 97 & 158\\
\textbf{True model} & 0 & 24 & 95 & 97 & & 0 & 19 & 96 & 80 & & 0 & 14 & 96 & 55\\
\midrule
\textbf{With Bayesian bootstrap} & & & & & & & & & & & & & &\\
\ \ SplineBART & \bf -2 & \bf 52 & \bf 98 & 236 & & 35 & 60 & 92 & 203 & & 37 & 50 & 87 & 147\\
\ \ DSplineBART & \bf 7 & \bf 51 & \bf 98 & 238 & & \bf 16 & \bf 50 & \bf 94 & 196 & & 8 & 37 & \bf 97 & 162\\
\ \ True model & -1 & 27 & 96 & 114 & & -1 & 26 & 97 & 112 & & -1 & 17 & 96 & 65\\
\midrule
\midrule
\multicolumn{15}{c}{\textbf{Generalizability --- SS1. Linear, $N = 2000$, $n=150$}} \\
\hline
\multirow{2}{*}{\textbf{Estimators}} & \multicolumn{4}{c}{\textbf{$\e_T[Y(1)-Y(0)]$}} & & \multicolumn{4}{c}{\textbf{$\e_T[Y(1)]$}} & & \multicolumn{4}{c}{\textbf{$\e_T[Y(0)]$}} \\ \cline{2-5} \cline{7-10} \cline{12-15}
  & \textbf{Bias} & \textbf{RMSE} & \textbf{Coverage} & \textbf{Width} & & \textbf{Bias} & \textbf{RMSE}  & \textbf{Coverage} & \textbf{Width} & & \textbf{Bias} & \textbf{RMSE}  & \textbf{Coverage} & \textbf{Width} \\
 \hline
\textbf{Existing methods} & & & & & & & & & & & & & &\\
\ \ Trial-only & 168 & 182 & 38 & 289 & & 365 & 372 & 0 & 266 & & 197 & 199 & 0 & 116\\
\ \ IPPW & 69 & 158 & \bf 92 & 629 & & 151 & 191 & 79 & 578 & & 82 & 102 & 63 & 262\\
\ \ A-IPPW & 28 & \bf 59 & 90 & 179 & & 81 & 92 & 42 & 146 & & 53 & 60 & 52 & 108\\
\ \ BART & 72 & 84 & 57 & 163 & & 79 & 89 & 35 & 125 & & 7 & 30 & 95 & 121\\
\ \ BART-PS & 72 & 85 & 58 & 170 & & 87 & 99 & 37 & 142 & & 15 & 37 & 94 & 138\\
\textbf{Interaction-based outcome models} & & & & & & & & & & & & & &\\
\ \ InterBART & 68 & 78 & 61 & 160 & & 79 & 88 & 35 & 131 & & 11 & 30 & 94 & 118\\
\ \ InterBART-PS & 76 & 86 & 53 & 162 & & 90 & 100 & 33 & 147 & & 15 & 36 & 95 & 137\\
\ \ FullBART & 60 & 74 & 73 & 175 & & 84 & 95 & 45 & 153 & & 24 & 40 & 91 & 139\\
\ \ SplineBART & \bf 7 & \bf 57 & \bf 96 & 223 & & 41 & 66 & 83 & 183 & & 34 & 47 & 86 & 147\\
\ \ DSplineBART & \bf 16 & \bf 57 & \bf 96 & 221 & & \bf 21 & \bf 54 & \bf 92 & 176 & & 5 & 34 & \bf 97 & 154\\
\textbf{True model} & -2 & 26 & 93 & 99 & & -1 & 21 & 94 & 82 & & 0 & 14 & 95 & 56\\
\midrule
\textbf{With Bayesian bootstrap} & & & & & & & & & & & & & &\\
\ \ SplineBART & \bf 7 & \bf 56 & \bf 95 & 225 & & 41 & 66 & 86 & 190 & & 33 & 47 & 88 & 150\\
\ \ DSplineBART & \bf 16 & \bf 57 & \bf 96 & 223 & & \bf 20 & \bf 55 & \bf 92 & 183 & & 4 & 35 & \bf 98 & 156\\
\ \ True model & -2 & 28 & 94 & 107 & & -2 & 25 & 94 & 99 & & 0 & 15 & 96 & 61\\
\bottomrule
\end{tabular}
}
\end{table}

\begin{table}[H]
\caption{\textbf{Generalizability.} Comparison of bias ($\times100$), RMSE ($\times100$), coverage rate ($\times100$), and average width ($\times100$) of 95\% confidence/credible intervals for scenarios SS2 (linear, covariate shift) and SS3 (nonlinear) with EHR cohort sizes $N=500$ and $1000$.}
\label{tab:supp_gen_ss23}
\resizebox{0.95\textwidth}{!}{%
\begin{tabular}{lcccccccccccccc}
\toprule
\multicolumn{15}{c}{\textbf{Generalizability --- SS2. Linear, strong covariate shift, $N = 500$, $n=150$}} \\
\hline
\multirow{2}{*}{\textbf{Estimators}} & \multicolumn{4}{c}{\textbf{$\e_T[Y(1)-Y(0)]$}} & & \multicolumn{4}{c}{\textbf{$\e_T[Y(1)]$}} & & \multicolumn{4}{c}{\textbf{$\e_T[Y(0)]$}} \\ \cline{2-5} \cline{7-10} \cline{12-15}
  & \textbf{Bias} & \textbf{RMSE} & \textbf{Coverage} & \textbf{Width} & & \textbf{Bias} & \textbf{RMSE}  & \textbf{Coverage} & \textbf{Width} & & \textbf{Bias} & \textbf{RMSE}  & \textbf{Coverage} & \textbf{Width} \\
 \hline
\textbf{Existing methods} & & & & & & & & & & & & & &\\
\ \ Trial-only & 214 & 223 & 8 & 262 & & 349 & 354 & 0 & 241 & & 135 & 137 & 0 & 104\\
\ \ IPPW & 152 & 178 & 68 & 435 & & 251 & 263 & 24 & 407 & & 99 & 105 & 26 & 167\\
\ \ A-IPPW & 51 & \bf 65 & 76 & 162 & & 102 & 108 & 22 & 150 & & 52 & 56 & 40 & 90\\
\ \ BART & 97 & 102 & 10 & 120 & & 107 & 111 & 6 & 113 & & 9 & 28 & 95 & 111\\
\ \ BART-PS & 83 & 90 & 44 & 150 & & 100 & 109 & 30 & 156 & & 17 & 42 & 94 & 154\\
\textbf{Interaction-based outcome models} & & & & & & & & & & & & & &\\
\ \ InterBART & 97 & 101 & 13 & 131 & & 109 & 113 & 8 & 125 & & 12 & 26 & 96 & 116\\
\ \ InterBART-PS & 103 & 107 & 10 & 133 & & 113 & 120 & 19 & 164 & & 10 & 39 & 96 & 157\\
\ \ FullBART & 74 & 82 & 62 & 169 & & 100 & 109 & 38 & 175 & & 25 & 44 & 94 & 164\\
\ \ SplineBART & \bf -21 & 99 & \bf 93 & 364 & & 23 & 95 & 88 & 319 & & 44 & 55 & 90 & 165\\
\ \ DSplineBART & \bf -15 & 101 & \bf 95 & 378 & & \bf 13 & \bf 93 & 88 & 326 & & 29 & 63 & \bf 97 & 280\\
\textbf{True model} & -1 & 19 & 97 & 87 & & 0 & 17 & 96 & 76 & & 1 & 10 & 97 & 43\\
\midrule
\textbf{With Bayesian bootstrap} & & & & & & & & & & & & & &\\
\ \ SplineBART & \bf -23 & 102 & \bf 93 & 370 & & 19 & 100 & 86 & 333 & & 42 & 55 & 88 & 173\\
\ \ DSplineBART & \bf -16 & 104 & \bf 94 & 384 & & \bf 10 & \bf 98 & 89 & 339 & & 27 & 63 & \bf 98 & 285\\
\ \ True model & -2 & 27 & 96 & 121 & & -3 & 33 & 95 & 135 & & -1 & 15 & 97 & 67\\
\midrule
\midrule
\multicolumn{15}{c}{\textbf{Generalizability --- SS2. Linear, strong covariate shift, $N = 1000$, $n=150$}} \\
\hline
\multirow{2}{*}{\textbf{Estimators}} & \multicolumn{4}{c}{\textbf{$\e_T[Y(1)-Y(0)]$}} & & \multicolumn{4}{c}{\textbf{$\e_T[Y(1)]$}} & & \multicolumn{4}{c}{\textbf{$\e_T[Y(0)]$}} \\ \cline{2-5} \cline{7-10} \cline{12-15}
  & \textbf{Bias} & \textbf{RMSE} & \textbf{Coverage} & \textbf{Width} & & \textbf{Bias} & \textbf{RMSE}  & \textbf{Coverage} & \textbf{Width} & & \textbf{Bias} & \textbf{RMSE}  & \textbf{Coverage} & \textbf{Width} \\
 \hline
\textbf{Existing methods} & & & & & & & & & & & & & &\\
\ \ Trial-only & 314 & 319 & 0 & 248 & & 508 & 511 & 0 & 229 & & 195 & 196 & 0 & 99\\
\ \ IPPW & 166 & 213 & 69 & 573 & & 279 & 302 & 38 & 533 & & 113 & 123 & 38 & 216\\
\ \ A-IPPW & 79 & \bf 95 & 62 & 189 & & 152 & 158 & 5 & 157 & & 73 & 79 & 31 & 112\\
\ \ BART & 144 & 148 & 3 & 142 & & 154 & 160 & 3 & 135 & & 11 & 35 & 90 & 131\\
\ \ BART-PS & 127 & 134 & 24 & 174 & & 142 & 151 & 16 & 178 & & 15 & 45 & 94 & 175\\
\textbf{Interaction-based outcome models} & & & & & & & & & & & & & &\\
\ \ InterBART & 142 & 147 & 4 & 152 & & 158 & 164 & 3 & 148 & & 16 & 36 & 93 & 137\\
\ \ InterBART-PS & 151 & 156 & 3 & 154 & & 159 & 166 & 10 & 187 & & 8 & 41 & 96 & 178\\
\ \ FullBART & 113 & 120 & 36 & 192 & & 140 & 149 & 20 & 198 & & 28 & 48 & 94 & 185\\
\ \ SplineBART & \bf -32 & 109 & \bf 94 & 416 & & 23 & 99 & 93 & 363 & & 55 & 68 & 84 & 191\\
\ \ DSplineBART & \bf -30 & 116 & \bf 95 & 430 & & \bf 9 & 100 & \bf 95 & 365 & & 39 & 76 & \bf 97 & 316\\
\textbf{True model} & 0 & 26 & 95 & 99 & & 0 & 24 & 95 & 89 & & 1 & 11 & 96 & 43\\
\midrule
\textbf{With Bayesian bootstrap} & & & & & & & & & & & & & &\\
\ \ SplineBART & \bf -33 & 110 & \bf 95 & 419 & & 21 & 100 & 93 & 368 & & 54 & 68 & 84 & 194\\
\ \ DSplineBART & \bf -31 & 117 & \bf 95 & 432 & & \bf 7 & 103 & \bf 93 & 370 & & 38 & 77 & \bf 96 & 318\\
\ \ True model & -1 & 31 & 96 & 116 & & -1 & 32 & 94 & 119 & & 0 & 14 & 96 & 56\\
\midrule
\midrule
\multicolumn{15}{c}{\textbf{Generalizability --- SS3. Nonlinear, $N = 500$, $n=150$}} \\
\hline
\multirow{2}{*}{\textbf{Estimators}} & \multicolumn{4}{c}{\textbf{$\e_T[Y(1)-Y(0)]$}} & & \multicolumn{4}{c}{\textbf{$\e_T[Y(1)]$}} & & \multicolumn{4}{c}{\textbf{$\e_T[Y(0)]$}} \\ \cline{2-5} \cline{7-10} \cline{12-15}
  & \textbf{Bias} & \textbf{RMSE} & \textbf{Coverage} & \textbf{Width} & & \textbf{Bias} & \textbf{RMSE}  & \textbf{Coverage} & \textbf{Width} & & \textbf{Bias} & \textbf{RMSE}  & \textbf{Coverage} & \textbf{Width} \\
 \hline
\textbf{Existing methods} & & & & & & & & & & & & & &\\
\ \ Trial-only & 212 & 242 & 55 & 459 & & 517 & 524 & 0 & 369 & & 304 & 310 & 0 & 275\\
\ \ IPPW & 133 & 222 & \bf 92 & 779 & & 297 & 319 & 45 & 616 & & 165 & 194 & 60 & 482\\
\ \ A-IPPW & 62 & \bf 104 & 87 & 318 & & 132 & 151 & 53 & 277 & & 70 & 81 & 79 & 206\\
\ \ BART & 90 & 108 & 66 & 239 & & 99 & 120 & 48 & 207 & & 9 & 51 & 96 & 205\\
\ \ BART-PS & 91 & 109 & 73 & 257 & & 132 & 150 & 46 & 252 & & 41 & 73 & 91 & 250\\
\textbf{Interaction-based outcome models} & & & & & & & & & & & & & &\\
\ \ InterBART & 112 & 120 & 48 & 221 & & 110 & 127 & 47 & 215 & & -1 & 50 & 97 & 209\\
\ \ InterBART-PS & 123 & 130 & 42 & 222 & & 147 & 162 & 38 & 265 & & 25 & 63 & 95 & 258\\
\ \ FullBART & 97 & 112 & 74 & 272 & & 140 & 157 & 45 & 277 & & 43 & 70 & 94 & 271\\
\ \ SplineBART & \bf -2 & \bf 102 & \bf 93 & 426 & & 76 & 124 & 86 & 373 & & 78 & 93 & 84 & 264\\
\ \ DSplineBART & \bf 16 & \bf 109 & \bf 94 & 437 & & \bf 54 & \bf 111 & 89 & 370 & & 38 & 80 & \bf 96 & 331\\
\textbf{True model} & 0 & 37 & 95 & 147 & & 0 & 41 & 94 & 147 & & -1 & 34 & 94 & 132\\
\midrule
\textbf{With Bayesian bootstrap} & & & & & & & & & & & & & &\\
\ \ SplineBART & \bf -3 & \bf 102 & \bf 94 & 431 & & 73 & 124 & 90 & 396 & & 76 & 94 & 88 & 285\\
\ \ DSplineBART & \bf 15 & \bf 109 & \bf 95 & 442 & & \bf 51 & \bf 110 & \bf 90 & 394 & & 36 & 81 & \bf 96 & 349\\
\ \ True model & 0 & 42 & 96 & 173 & & -3 & 48 & 95 & 201 & & -3 & 40 & 96 & 165\\
\midrule
\midrule
\multicolumn{15}{c}{\textbf{Generalizability --- SS3. Nonlinear, $N = 1000$, $n=150$}} \\
\hline
\multirow{2}{*}{\textbf{Estimators}} & \multicolumn{4}{c}{\textbf{$\e_T[Y(1)-Y(0)]$}} & & \multicolumn{4}{c}{\textbf{$\e_T[Y(1)]$}} & & \multicolumn{4}{c}{\textbf{$\e_T[Y(0)]$}} \\ \cline{2-5} \cline{7-10} \cline{12-15}
  & \textbf{Bias} & \textbf{RMSE} & \textbf{Coverage} & \textbf{Width} & & \textbf{Bias} & \textbf{RMSE}  & \textbf{Coverage} & \textbf{Width} & & \textbf{Bias} & \textbf{RMSE}  & \textbf{Coverage} & \textbf{Width} \\
 \hline
\textbf{Existing methods} & & & & & & & & & & & & & &\\
\ \ Trial-only & 260 & 283 & 42 & 470 & & 671 & 677 & 0 & 379 & & 411 & 416 & 0 & 280\\
\ \ IPPW & 143 & 254 & \bf 93 & 927 & & 356 & 380 & 45 & 742 & & 213 & 253 & 50 & 564\\
\ \ A-IPPW & 84 & \bf 127 & 86 & 383 & & 183 & 200 & 40 & 318 & & 99 & 112 & 60 & 231\\
\ \ BART & 117 & 136 & 60 & 270 & & 129 & 149 & 48 & 248 & & 13 & 61 & 93 & 240\\
\ \ BART-PS & 120 & 140 & 60 & 285 & & 189 & 205 & 30 & 295 & & 69 & 94 & 87 & 287\\
\textbf{Interaction-based outcome models} & & & & & & & & & & & & & &\\
\ \ InterBART & 142 & 151 & 37 & 251 & & 147 & 162 & 40 & 257 & & 5 & 58 & 95 & 248\\
\ \ InterBART-PS & 159 & 166 & 26 & 248 & & 207 & 219 & 23 & 310 & & 49 & 79 & 93 & 296\\
\ \ FullBART & 121 & 136 & 70 & 309 & & 196 & 210 & 34 & 324 & & 75 & 95 & 90 & 310\\
\ \ SplineBART & \bf 2 & \bf 123 & \bf 96 & 547 & & 103 & 156 & 88 & 479 & & 101 & 118 & 80 & 315\\
\ \ DSplineBART & \bf 21 & \bf 121 & \bf 97 & 547 & & \bf 60 & \bf 126 & \bf 95 & 449 & & 39 & 88 & \bf 96 & 399\\
\textbf{True model} & 2 & 48 & 95 & 179 & & -1 & 45 & 97 & 189 & & -3 & 44 & 93 & 167\\
\midrule
\textbf{With Bayesian bootstrap} & & & & & & & & & & & & & &\\
\ \ SplineBART & \bf 2 & \bf 122 & \bf 97 & 549 & & 102 & 155 & 88 & 488 & & 100 & 118 & 83 & 325\\
\ \ DSplineBART & \bf 21 & \bf 120 & \bf 97 & 549 & & \bf 60 & \bf 125 & \bf 94 & 460 & & 39 & 90 & \bf 97 & 408\\
\ \ True model & 2 & 50 & 95 & 191 & & -2 & 49 & 97 & 213 & & -4 & 45 & 96 & 181\\
\bottomrule
\end{tabular}
}
\end{table}

\begin{table}[H]
\caption{\textbf{Generalizability.} Comparison of bias ($\times100$), RMSE ($\times100$), coverage rate ($\times100$), and average width ($\times100$) of 95\% confidence/credible intervals for scenario SS4 (nonlinear, covariate shift) with $N=500$ and $1000$, and scenario SS5 (linear with imbalanced treatment assignment) with $N=500$ and $\Pr(A=1)=0.3$ or $0.7$.}
\label{tab:supp_gen_ss45}
\resizebox{0.95\textwidth}{!}{%
\begin{tabular}{lcccccccccccccc}
\toprule
\multicolumn{15}{c}{\textbf{Generalizability --- SS4. Nonlinear, strong covariate shift, $N = 500$, $n=150$}} \\
\hline
\multirow{2}{*}{\textbf{Estimators}} & \multicolumn{4}{c}{\textbf{$\e_T[Y(1)-Y(0)]$}} & & \multicolumn{4}{c}{\textbf{$\e_T[Y(1)]$}} & & \multicolumn{4}{c}{\textbf{$\e_T[Y(0)]$}} \\ \cline{2-5} \cline{7-10} \cline{12-15}
  & \textbf{Bias} & \textbf{RMSE} & \textbf{Coverage} & \textbf{Width} & & \textbf{Bias} & \textbf{RMSE}  & \textbf{Coverage} & \textbf{Width} & & \textbf{Bias} & \textbf{RMSE}  & \textbf{Coverage} & \textbf{Width} \\
 \hline
\textbf{Existing methods} & & & & & & & & & & & & & &\\
\ \ Trial-only & 229 & 249 & 40 & 406 & & 529 & 535 & 0 & 330 & & 301 & 306 & 0 & 239\\
\ \ IPPW & 167 & 201 & 76 & 526 & & 406 & 415 & 8 & 449 & & 240 & 246 & 8 & 286\\
\ \ A-IPPW & 58 & \bf 88 & 87 & 279 & & 130 & 142 & 49 & 246 & & 73 & 82 & 67 & 183\\
\ \ BART & 82 & 94 & 60 & 192 & & 109 & 122 & 38 & 184 & & 27 & 54 & 89 & 182\\
\ \ BART-PS & 75 & 89 & 76 & 225 & & 114 & 135 & 56 & 256 & & 40 & 72 & 89 & 256\\
\textbf{Interaction-based outcome models} & & & & & & & & & & & & & &\\
\ \ InterBART & 102 & 109 & 42 & 191 & & 121 & 132 & 33 & 198 & & 19 & 48 & 94 & 192\\
\ \ InterBART-PS & 111 & 117 & 33 & 194 & & 132 & 148 & 51 & 276 & & 21 & 63 & 96 & 267\\
\ \ FullBART & 74 & \bf 88 & 85 & 255 & & 120 & 138 & 62 & 290 & & 45 & 74 & 95 & 282\\
\ \ SplineBART & \bf -63 & 134 & \bf 92 & 499 & & 15 & 116 & \bf 96 & 460 & & 78 & 95 & 80 & 271\\
\ \ DSplineBART & \bf -51 & 134 & \bf 94 & 518 & & \bf -2 & \bf 114 & \bf 96 & 469 & & 49 & 99 & \bf 96 & 414\\
\textbf{True model} & 0 & 31 & 97 & 138 & & 0 & 36 & 94 & 142 & & 0 & 35 & 94 & 125\\
\midrule
\textbf{With Bayesian bootstrap} & & & & & & & & & & & & & &\\
\ \ SplineBART & \bf -63 & 134 & \bf 91 & 503 & & 12 & 118 & \bf 96 & 479 & & 76 & 96 & 84 & 289\\
\ \ DSplineBART & \bf -52 & 135 & \bf 94 & 522 & & \bf -5 & \bf 117 & \bf 97 & 487 & & 47 & 100 & \bf 97 & 427\\
\ \ True model & -1 & 37 & 96 & 167 & & -3 & 44 & 97 & 199 & & -2 & 41 & 95 & 160\\
\midrule
\midrule
\multicolumn{15}{c}{\textbf{Generalizability --- SS4. Nonlinear, strong covariate shift, $N = 1000$, $n=150$}} \\
\hline
\multirow{2}{*}{\textbf{Estimators}} & \multicolumn{4}{c}{\textbf{$\e_T[Y(1)-Y(0)]$}} & & \multicolumn{4}{c}{\textbf{$\e_T[Y(1)]$}} & & \multicolumn{4}{c}{\textbf{$\e_T[Y(0)]$}} \\ \cline{2-5} \cline{7-10} \cline{12-15}
  & \textbf{Bias} & \textbf{RMSE} & \textbf{Coverage} & \textbf{Width} & & \textbf{Bias} & \textbf{RMSE}  & \textbf{Coverage} & \textbf{Width} & & \textbf{Bias} & \textbf{RMSE}  & \textbf{Coverage} & \textbf{Width} \\
 \hline
\textbf{Existing methods} & & & & & & & & & & & & & &\\
\ \ Trial-only & 328 & 342 & 11 & 414 & & 743 & 747 & 0 & 331 & & 415 & 419 & 0 & 251\\
\ \ IPPW & 237 & 279 & 64 & 631 & & 561 & 572 & 5 & 531 & & 324 & 333 & 8 & 356\\
\ \ A-IPPW & 124 & \bf 152 & 70 & 347 & & 252 & 260 & 7 & 281 & & 128 & 137 & 41 & 219\\
\ \ BART & 167 & 175 & 20 & 224 & & 220 & 229 & 5 & 229 & & 53 & 77 & 86 & 227\\
\ \ BART-PS & 159 & 168 & 35 & 258 & & 235 & 247 & 16 & 306 & & 76 & 102 & 86 & 303\\
\textbf{Interaction-based outcome models} & & & & & & & & & & & & & &\\
\ \ InterBART & 182 & 187 & 7 & 221 & & 230 & 238 & 6 & 253 & & 49 & 73 & 91 & 243\\
\ \ InterBART-PS & 193 & 198 & 4 & 225 & & 251 & 262 & 16 & 330 & & 58 & 87 & 93 & 317\\
\ \ FullBART & 151 & 161 & 45 & 301 & & 236 & 249 & 22 & 347 & & 85 & 107 & 89 & 334\\
\ \ SplineBART & \bf -5 & \bf 147 & \bf 98 & 678 & & 112 & 180 & 90 & 613 & & 117 & 132 & 74 & 335\\
\ \ DSplineBART & \bf 7 & \bf 149 & \bf 98 & 692 & & \bf 84 & \bf 155 & \bf 95 & 601 & & 77 & 128 & \bf 96 & 540\\
\textbf{True model} & 5 & 46 & 94 & 186 & & 2 & 59 & 95 & 224 & & -3 & 48 & 96 & 198\\
\midrule
\textbf{With Bayesian bootstrap} & & & & & & & & & & & & & &\\
\ \ SplineBART & \bf -5 & \bf 148 & \bf 98 & 679 & & 112 & 180 & \bf 94 & 619 & & 116 & 132 & 78 & 343\\
\ \ DSplineBART & \bf 7 & \bf 149 & \bf 98 & 694 & & \bf 84 & \bf 156 & \bf 94 & 608 & & 76 & 129 & \bf 95 & 545\\
\ \ True model & 5 & 50 & 92 & 197 & & 2 & 62 & 96 & 245 & & -3 & 50 & 96 & 211\\
\midrule
\midrule
\multicolumn{15}{c}{\textbf{Generalizability --- SS5. Linear, $\Pr(A=1)=0.3$, $N = 500$, $n=150$}} \\
\hline
\multirow{2}{*}{\textbf{Estimators}} & \multicolumn{4}{c}{\textbf{$\e_T[Y(1)-Y(0)]$}} & & \multicolumn{4}{c}{\textbf{$\e_T[Y(1)]$}} & & \multicolumn{4}{c}{\textbf{$\e_T[Y(0)]$}} \\ \cline{2-5} \cline{7-10} \cline{12-15}
  & \textbf{Bias} & \textbf{RMSE} & \textbf{Coverage} & \textbf{Width} & & \textbf{Bias} & \textbf{RMSE}  & \textbf{Coverage} & \textbf{Width} & & \textbf{Bias} & \textbf{RMSE}  & \textbf{Coverage} & \textbf{Width} \\
 \hline
\textbf{Existing methods} & & & & & & & & & & & & & &\\
\ \ Trial-only & 129 & 157 & 74 & 364 & & 293 & 306 & 10 & 351 & & 165 & 166 & 0 & 98\\
\ \ IPPW & 47 & 156 & \bf 95 & 641 & & 124 & 184 & 82 & 609 & & 76 & 86 & 60 & 208\\
\ \ A-IPPW & 62 & 84 & 76 & 203 & & 97 & 111 & 51 & 194 & & 36 & 41 & 62 & 89\\
\ \ BART & 68 & 80 & 59 & 154 & & 81 & 92 & 40 & 138 & & 13 & 24 & 90 & 86\\
\ \ BART-PS & 63 & 77 & 66 & 162 & & 83 & 96 & 44 & 153 & & 20 & 32 & 88 & 106\\
\textbf{Interaction-based outcome models} & & & & & & & & & & & & & &\\
\ \ InterBART & 60 & 73 & 68 & 159 & & 77 & 88 & 47 & 146 & & 17 & 25 & 90 & 85\\
\ \ InterBART-PS & 66 & 78 & 64 & 160 & & 86 & 98 & 44 & 157 & & 20 & 32 & 90 & 106\\
\ \ FullBART & 57 & 72 & 71 & 167 & & 80 & 93 & 50 & 162 & & 23 & 34 & 88 & 106\\
\ \ SplineBART & \bf -13 & \bf 69 & \bf 92 & 239 & & 22 & 70 & 91 & 218 & & 35 & 42 & 76 & 106\\
\ \ DSplineBART & \bf -6 & \bf 66 & \bf 94 & 238 & & \bf 4 & \bf 63 & \bf 94 & 215 & & 10 & 27 & \bf 96 & 118\\
\textbf{True model} & 0 & 26 & 97 & 109 & & 1 & 23 & 97 & 98 & & 1 & 11 & 97 & 47\\
\midrule
\textbf{With Bayesian bootstrap} & & & & & & & & & & & & & &\\
\ \ SplineBART & \bf -15 & \bf 74 & \bf 91 & 248 & & 19 & 78 & 89 & 240 & & 34 & 43 & 80 & 118\\
\ \ DSplineBART & \bf -7 & \bf 72 & \bf 93 & 248 & & \bf 1 & \bf 72 & \bf 91 & 237 & & 8 & 29 & \bf 97 & 129\\
\ \ True model & -1 & 32 & 98 & 138 & & -2 & 35 & 98 & 149 & & -1 & 16 & 97 & 69\\
\midrule
\midrule
\multicolumn{15}{c}{\textbf{Generalizability --- SS5. Linear, $\Pr(A=1)=0.7$, $N = 500$, $n=150$}} \\
\hline
\multirow{2}{*}{\textbf{Estimators}} & \multicolumn{4}{c}{\textbf{$\e_T[Y(1)-Y(0)]$}} & & \multicolumn{4}{c}{\textbf{$\e_T[Y(1)]$}} & & \multicolumn{4}{c}{\textbf{$\e_T[Y(0)]$}} \\ \cline{2-5} \cline{7-10} \cline{12-15}
  & \textbf{Bias} & \textbf{RMSE} & \textbf{Coverage} & \textbf{Width} & & \textbf{Bias} & \textbf{RMSE}  & \textbf{Coverage} & \textbf{Width} & & \textbf{Bias} & \textbf{RMSE}  & \textbf{Coverage} & \textbf{Width} \\
 \hline
\textbf{Existing methods} & & & & & & & & & & & & & &\\
\ \ Trial-only & 129 & 146 & 55 & 272 & & 293 & 298 & 0 & 227 & & 164 & 169 & 0 & 151\\
\ \ IPPW & 53 & 123 & \bf 96 & 531 & & 126 & 145 & 80 & 459 & & 73 & 101 & 65 & 269\\
\ \ A-IPPW & -18 & \bf 51 & \bf 92 & 178 & & 46 & 54 & 78 & 138 & & 64 & 74 & 46 & 130\\
\ \ BART & 58 & 72 & 70 & 162 & & 50 & 57 & 41 & 92 & & -7 & 40 & 94 & 143\\
\ \ BART-PS & 61 & 74 & 71 & 172 & & 60 & 69 & 43 & 114 & & -1 & 43 & 94 & 161\\
\textbf{Interaction-based outcome models} & & & & & & & & & & & & & &\\
\ \ InterBART & 69 & 78 & 60 & 156 & & 57 & 63 & 39 & 99 & & -13 & 38 & 93 & 138\\
\ \ InterBART-PS & 81 & 89 & 48 & 159 & & 67 & 74 & 43 & 123 & & -14 & 43 & 93 & 155\\
\ \ FullBART & 52 & 66 & 82 & 180 & & 62 & 70 & 54 & 126 & & 10 & 44 & 93 & 165\\
\ \ SplineBART & \bf -22 & 61 & \bf 94 & 229 & & 24 & 43 & 92 & 149 & & 46 & 67 & 79 & 171\\
\ \ DSplineBART & \bf -7 & 66 & \bf 94 & 239 & & \bf 12 & \bf 36 & \bf 94 & 143 & & 19 & 65 & 90 & 200\\
\textbf{True model} & 0 & 24 & 94 & 95 & & 0 & 14 & 98 & 63 & & 0 & 18 & 95 & 71\\
\midrule
\textbf{With Bayesian bootstrap} & & & & & & & & & & & & & &\\
\ \ SplineBART & \bf -23 & 67 & \bf 92 & 237 & & 21 & 50 & 92 & 181 & & 44 & 67 & 84 & 181\\
\ \ DSplineBART & \bf -8 & 72 & \bf 94 & 247 & & \bf 9 & \bf 44 & \bf 93 & 176 & & 17 & 65 & 90 & 210\\
\ \ True model & -2 & 31 & 95 & 127 & & -3 & 32 & 96 & 128 & & -2 & 21 & 98 & 87\\
\bottomrule
\end{tabular}
}
\end{table}
\subsection{Comparison of Transformations of $\hat{\pi}$}
We compared the performance of DSplineBART under four transformations $\phi$ of the estimated participation score $\hat{\pi}$: Identity, Logit, Cloglog, and InvGumbel, where InvGumbel is the transformation used throughout the main analysis; its column reproduces the corresponding DSplineBART results from the main simulations. Results for the transportability design are summarized in Web Tables~\ref{tab:transform_trans_lin} and~\ref{tab:transform_trans_nonlin} for the linear and nonlinear scenarios, respectively; the results for the generalizability design are similar and are therefore omitted. All entries use only the EHR-cohort estimand (MATE).

\begin{table}[H]
\caption{\textbf{Transportability.} Transformations comparison of $\phi(\hat{\pi})$ in DSplineBART: bias ($\times100$), RMSE ($\times100$), coverage rate ($\times100$), and average width ($\times100$) of 95\% credible intervals for the linear scenarios SS1, SS2 (with covariate shift), and SS5 (with imbalanced treatment assignment).}
\label{tab:transform_trans_lin}
\resizebox{\textwidth}{!}{%
\begin{tabular}{lcccccccccccccc}
\toprule
\multicolumn{15}{c}{\textbf{Transportability --- SS1. Linear, $N = 500$, $n=150$}} \\
\hline
\multirow{2}{*}{\textbf{Transformations}} & \multicolumn{4}{c}{\textbf{$\e_T[Y(1)-Y(0)]$}} & & \multicolumn{4}{c}{\textbf{$\e_T[Y(1)]$}} & & \multicolumn{4}{c}{\textbf{$\e_T[Y(0)]$}} \\ \cline{2-5} \cline{7-10} \cline{12-15}
  & \textbf{Bias} & \textbf{RMSE} & \textbf{Coverage} & \textbf{Width} & & \textbf{Bias} & \textbf{RMSE}  & \textbf{Coverage} & \textbf{Width} & & \textbf{Bias} & \textbf{RMSE}  & \textbf{Coverage} & \textbf{Width} \\
 \hline
\ \ Identity & 34 & 60 & 89 & 190 & & 77 & 88 & 54 & 156 & & 42 & 55 & 80 & 140\\
\ \ Logit & -14 & 105 & 92 & 368 & & -32 & 88 & 94 & 303 & & -18 & 73 & 95 & 268\\
\ \ Cloglog & -6 & 115 & 90 & 372 & & -24 & 93 & 92 & 309 & & -19 & 82 & 92 & 268\\
\ \ \textbf{InvGumbel} & 4 & 65 & 95 & 271 & & 18 & 56 & 93 & 214 & & 13 & 46 & 94 & 187\\
\midrule
\midrule
\multicolumn{15}{c}{\textbf{Transportability --- SS1. Linear, $N = 1000$, $n=150$}} \\
\hline
\multirow{2}{*}{\textbf{Transformations}} & \multicolumn{4}{c}{\textbf{$\e_T[Y(1)-Y(0)]$}} & & \multicolumn{4}{c}{\textbf{$\e_T[Y(1)]$}} & & \multicolumn{4}{c}{\textbf{$\e_T[Y(0)]$}} \\ \cline{2-5} \cline{7-10} \cline{12-15}
  & \textbf{Bias} & \textbf{RMSE} & \textbf{Coverage} & \textbf{Width} & & \textbf{Bias} & \textbf{RMSE}  & \textbf{Coverage} & \textbf{Width} & & \textbf{Bias} & \textbf{RMSE}  & \textbf{Coverage} & \textbf{Width} \\
 \hline
\ \ Identity & 56 & 72 & 77 & 179 & & 103 & 111 & 26 & 149 & & 47 & 59 & 72 & 135\\
\ \ Logit & 4 & 108 & 90 & 343 & & -13 & 87 & 92 & 286 & & -17 & 78 & 91 & 249\\
\ \ Cloglog & 9 & 116 & 87 & 340 & & -6 & 90 & 90 & 289 & & -15 & 86 & 85 & 243\\
\ \ \textbf{InvGumbel} & 11 & 64 & 96 & 255 & & 22 & 57 & 92 & 203 & & 11 & 45 & 95 & 177\\
\midrule
\midrule
\multicolumn{15}{c}{\textbf{Transportability --- SS1. Linear, $N = 2000$, $n=150$}} \\
\hline
\multirow{2}{*}{\textbf{Transformations}} & \multicolumn{4}{c}{\textbf{$\e_T[Y(1)-Y(0)]$}} & & \multicolumn{4}{c}{\textbf{$\e_T[Y(1)]$}} & & \multicolumn{4}{c}{\textbf{$\e_T[Y(0)]$}} \\ \cline{2-5} \cline{7-10} \cline{12-15}
  & \textbf{Bias} & \textbf{RMSE} & \textbf{Coverage} & \textbf{Width} & & \textbf{Bias} & \textbf{RMSE}  & \textbf{Coverage} & \textbf{Width} & & \textbf{Bias} & \textbf{RMSE}  & \textbf{Coverage} & \textbf{Width} \\
 \hline
\ \ Identity & 70 & 82 & 65 & 173 & & 115 & 123 & 13 & 147 & & 46 & 57 & 72 & 135\\
\ \ Logit & 17 & 113 & 85 & 317 & & -1 & 89 & 90 & 272 & & -18 & 76 & 86 & 228\\
\ \ Cloglog & 24 & 125 & 80 & 317 & & 3 & 95 & 86 & 270 & & -21 & 81 & 86 & 229\\
\ \ \textbf{InvGumbel} & 18 & 63 & 91 & 234 & & 27 & 57 & 89 & 190 & & 9 & 42 & 94 & 166\\
\midrule
\midrule
\multicolumn{15}{c}{\textbf{Transportability --- SS2. Linear, strong covariate shift, $N = 500$, $n=150$}} \\
\hline
\multirow{2}{*}{\textbf{Transformations}} & \multicolumn{4}{c}{\textbf{$\e_T[Y(1)-Y(0)]$}} & & \multicolumn{4}{c}{\textbf{$\e_T[Y(1)]$}} & & \multicolumn{4}{c}{\textbf{$\e_T[Y(0)]$}} \\ \cline{2-5} \cline{7-10} \cline{12-15}
  & \textbf{Bias} & \textbf{RMSE} & \textbf{Coverage} & \textbf{Width} & & \textbf{Bias} & \textbf{RMSE}  & \textbf{Coverage} & \textbf{Width} & & \textbf{Bias} & \textbf{RMSE}  & \textbf{Coverage} & \textbf{Width} \\
 \hline
\ \ Identity & 60 & 101 & 86 & 292 & & 138 & 157 & 45 & 260 & & 78 & 92 & 73 & 222\\
\ \ Logit & -191 & 318 & 89 & 956 & & -194 & 304 & 85 & 839 & & -2 & 176 & 98 & 775\\
\ \ Cloglog & -202 & 374 & 84 & 1044 & & -212 & 348 & 83 & 936 & & -9 & 213 & 96 & 843\\
\ \ \textbf{InvGumbel} & -14 & 126 & 96 & 503 & & 33 & 120 & 93 & 432 & & 47 & 90 & 96 & 374\\
\midrule
\midrule
\multicolumn{15}{c}{\textbf{Transportability --- SS2. Linear, strong covariate shift, $N = 1000$, $n=150$}} \\
\hline
\multirow{2}{*}{\textbf{Transformations}} & \multicolumn{4}{c}{\textbf{$\e_T[Y(1)-Y(0)]$}} & & \multicolumn{4}{c}{\textbf{$\e_T[Y(1)]$}} & & \multicolumn{4}{c}{\textbf{$\e_T[Y(0)]$}} \\ \cline{2-5} \cline{7-10} \cline{12-15}
  & \textbf{Bias} & \textbf{RMSE} & \textbf{Coverage} & \textbf{Width} & & \textbf{Bias} & \textbf{RMSE}  & \textbf{Coverage} & \textbf{Width} & & \textbf{Bias} & \textbf{RMSE}  & \textbf{Coverage} & \textbf{Width} \\
 \hline
\ \ Identity & 114 & 132 & 58 & 256 & & 194 & 205 & 9 & 229 & & 80 & 91 & 69 & 201\\
\ \ Logit & -186 & 328 & 85 & 952 & & -180 & 293 & 87 & 860 & & 6 & 191 & 94 & 784\\
\ \ Cloglog & -181 & 368 & 80 & 998 & & -186 & 323 & 86 & 912 & & -4 & 221 & 93 & 824\\
\ \ \textbf{InvGumbel} & -2 & 118 & 96 & 493 & & 47 & 118 & 92 & 417 & & 50 & 87 & 96 & 364\\
\bottomrule
\end{tabular}
}
\end{table}

\begin{table}[H]
\caption{\textbf{Transportability.} Transformations comparison of $\phi(\hat{\pi})$ in DSplineBART: bias ($\times100$), RMSE ($\times100$), coverage rate ($\times100$), and average width ($\times100$) of 95\% credible intervals for the nonlinear scenarios SS3 and SS4 (with covariate shift).}
\label{tab:transform_trans_nonlin}
\resizebox{\textwidth}{!}{%
\begin{tabular}{lcccccccccccccc}
\toprule
\multicolumn{15}{c}{\textbf{Transportability --- SS3. Nonlinear, $N = 500$, $n=150$}} \\
\hline
\multirow{2}{*}{\textbf{Transformations}} & \multicolumn{4}{c}{\textbf{$\e_T[Y(1)-Y(0)]$}} & & \multicolumn{4}{c}{\textbf{$\e_T[Y(1)]$}} & & \multicolumn{4}{c}{\textbf{$\e_T[Y(0)]$}} \\ \cline{2-5} \cline{7-10} \cline{12-15}
  & \textbf{Bias} & \textbf{RMSE} & \textbf{Coverage} & \textbf{Width} & & \textbf{Bias} & \textbf{RMSE}  & \textbf{Coverage} & \textbf{Width} & & \textbf{Bias} & \textbf{RMSE}  & \textbf{Coverage} & \textbf{Width} \\
 \hline
\ \ Identity & 11 & 105 & 93 & 381 & & 106 & 137 & 76 & 346 & & 95 & 117 & 80 & 319\\
\ \ Logit & -81 & 332 & 93 & 1257 & & -70 & 291 & 92 & 1066 & & 11 & 220 & 97 & 1005\\
\ \ Cloglog & -32 & 399 & 89 & 1316 & & -13 & 345 & 89 & 1149 & & 20 & 283 & 94 & 1056\\
\ \ \textbf{InvGumbel} & -20 & 146 & 98 & 683 & & 32 & 136 & 96 & 560 & & 52 & 112 & 98 & 507\\
\midrule
\midrule
\multicolumn{15}{c}{\textbf{Transportability --- SS3. Nonlinear, $N = 1000$, $n=150$}} \\
\hline
\multirow{2}{*}{\textbf{Transformations}} & \multicolumn{4}{c}{\textbf{$\e_T[Y(1)-Y(0)]$}} & & \multicolumn{4}{c}{\textbf{$\e_T[Y(1)]$}} & & \multicolumn{4}{c}{\textbf{$\e_T[Y(0)]$}} \\ \cline{2-5} \cline{7-10} \cline{12-15}
  & \textbf{Bias} & \textbf{RMSE} & \textbf{Coverage} & \textbf{Width} & & \textbf{Bias} & \textbf{RMSE}  & \textbf{Coverage} & \textbf{Width} & & \textbf{Bias} & \textbf{RMSE}  & \textbf{Coverage} & \textbf{Width} \\
 \hline
\ \ Identity & 60 & 105 & 90 & 341 & & 168 & 185 & 44 & 318 & & 108 & 124 & 72 & 303\\
\ \ Logit & -7 & 360 & 89 & 1175 & & -19 & 316 & 91 & 1050 & & -12 & 256 & 92 & 928\\
\ \ Cloglog & 35 & 393 & 86 & 1214 & & 35 & 351 & 89 & 1084 & & 0 & 272 & 92 & 944\\
\ \ \textbf{InvGumbel} & -2 & 140 & 97 & 658 & & 42 & 137 & 96 & 547 & & 44 & 103 & 99 & 488\\
\midrule
\midrule
\multicolumn{15}{c}{\textbf{Transportability --- SS4. Nonlinear, strong covariate shift, $N = 500$, $n=150$}} \\
\hline
\multirow{2}{*}{\textbf{Transformations}} & \multicolumn{4}{c}{\textbf{$\e_T[Y(1)-Y(0)]$}} & & \multicolumn{4}{c}{\textbf{$\e_T[Y(1)]$}} & & \multicolumn{4}{c}{\textbf{$\e_T[Y(0)]$}} \\ \cline{2-5} \cline{7-10} \cline{12-15}
  & \textbf{Bias} & \textbf{RMSE} & \textbf{Coverage} & \textbf{Width} & & \textbf{Bias} & \textbf{RMSE}  & \textbf{Coverage} & \textbf{Width} & & \textbf{Bias} & \textbf{RMSE}  & \textbf{Coverage} & \textbf{Width} \\
 \hline
\ \ Identity & 35 & 114 & 92 & 419 & & 162 & 187 & 60 & 377 & & 126 & 148 & 70 & 347\\
\ \ Logit & -114 & 386 & 98 & 1700 & & -82 & 348 & 97 & 1471 & & 32 & 280 & 98 & 1421\\
\ \ Cloglog & -120 & 461 & 95 & 1773 & & -84 & 415 & 95 & 1640 & & 36 & 348 & 96 & 1522\\
\ \ \textbf{InvGumbel} & 1 & 163 & 99 & 833 & & 88 & 180 & 96 & 721 & & 88 & 147 & 98 & 653\\
\midrule
\midrule
\multicolumn{15}{c}{\textbf{Transportability --- SS4. Nonlinear, strong covariate shift, $N = 1000$, $n=150$}} \\
\hline
\multirow{2}{*}{\textbf{Transformations}} & \multicolumn{4}{c}{\textbf{$\e_T[Y(1)-Y(0)]$}} & & \multicolumn{4}{c}{\textbf{$\e_T[Y(1)]$}} & & \multicolumn{4}{c}{\textbf{$\e_T[Y(0)]$}} \\ \cline{2-5} \cline{7-10} \cline{12-15}
  & \textbf{Bias} & \textbf{RMSE} & \textbf{Coverage} & \textbf{Width} & & \textbf{Bias} & \textbf{RMSE}  & \textbf{Coverage} & \textbf{Width} & & \textbf{Bias} & \textbf{RMSE}  & \textbf{Coverage} & \textbf{Width} \\
 \hline
\ \ Identity & 88 & 124 & 84 & 367 & & 226 & 239 & 24 & 328 & & 138 & 153 & 62 & 308\\
\ \ Logit & -51 & 426 & 94 & 1587 & & -50 & 375 & 95 & 1455 & & 0 & 334 & 94 & 1344\\
\ \ Cloglog & -41 & 491 & 88 & 1622 & & -38 & 426 & 92 & 1557 & & 3 & 379 & 93 & 1362\\
\ \ \textbf{InvGumbel} & 8 & 151 & 99 & 814 & & 100 & 179 & 96 & 708 & & 92 & 148 & 97 & 617\\
\midrule
\midrule
\multicolumn{15}{c}{\textbf{Transportability --- SS5. Linear, $\Pr(A=1)=0.3$, $N = 500$, $n=150$}} \\
\hline
\multirow{2}{*}{\textbf{Transformations}} & \multicolumn{4}{c}{\textbf{$\e_T[Y(1)-Y(0)]$}} & & \multicolumn{4}{c}{\textbf{$\e_T[Y(1)]$}} & & \multicolumn{4}{c}{\textbf{$\e_T[Y(0)]$}} \\ \cline{2-5} \cline{7-10} \cline{12-15}
  & \textbf{Bias} & \textbf{RMSE} & \textbf{Coverage} & \textbf{Width} & & \textbf{Bias} & \textbf{RMSE}  & \textbf{Coverage} & \textbf{Width} & & \textbf{Bias} & \textbf{RMSE}  & \textbf{Coverage} & \textbf{Width} \\
 \hline
\ \ Identity & 31 & 71 & 87 & 209 & & 73 & 95 & 67 & 195 & & 42 & 51 & 73 & 113\\
\ \ Logit & -21 & 126 & 92 & 410 & & -38 & 122 & 90 & 386 & & -17 & 58 & 95 & 219\\
\ \ Cloglog & -16 & 139 & 87 & 414 & & -32 & 127 & 88 & 389 & & -16 & 68 & 92 & 223\\
\ \ \textbf{InvGumbel} & 3 & 83 & 94 & 301 & & 13 & 78 & 92 & 275 & & 11 & 36 & 96 & 151\\
\midrule
\midrule
\multicolumn{15}{c}{\textbf{Transportability --- SS5. Linear, $\Pr(A=1)=0.7$, $N = 500$, $n=150$}} \\
\hline
\multirow{2}{*}{\textbf{Transformations}} & \multicolumn{4}{c}{\textbf{$\e_T[Y(1)-Y(0)]$}} & & \multicolumn{4}{c}{\textbf{$\e_T[Y(1)]$}} & & \multicolumn{4}{c}{\textbf{$\e_T[Y(0)]$}} \\ \cline{2-5} \cline{7-10} \cline{12-15}
 & \textbf{Bias} & \textbf{RMSE} & \textbf{Coverage} & \textbf{Width} & & \textbf{Bias} & \textbf{RMSE}  & \textbf{Coverage} & \textbf{Width} & & \textbf{Bias} & \textbf{RMSE}  & \textbf{Coverage} & \textbf{Width} \\
 \hline
\ \ Identity & 27 & 64 & 87 & 208 & & 76 & 86 & 42 & 137 & & 49 & 73 & 76 & 185\\
\ \ Logit & -1 & 123 & 92 & 411 & & -19 & 72 & 93 & 268 & & -17 & 109 & 91 & 358\\
\ \ Cloglog & 23 & 145 & 85 & 402 & & -12 & 77 & 92 & 271 & & -35 & 131 & 85 & 351\\
\ \ \textbf{InvGumbel} & 0 & 71 & 96 & 302 & & 21 & 49 & 96 & 185 & & 21 & 70 & 95 & 257\\
\bottomrule
\end{tabular}
}
\end{table}

\clearpage
\newpage
\suppsection{Details of Data Application}
\subsection{Inclusion and Exclusion Criteria}

Web Table~\ref{tab:eligibility_codiacs} summarizes the eligibility criteria of the CODIACS Vanguard trial. We constructed the target EHR cohorts to approximate these criteria using information available in the All of Us EHR data. The target cohorts were defined using criteria in Web Table~\ref{tab:eligibility_aou}.

\begin{table}[H]
\centering
\begin{threeparttable}
\caption{Inclusion and Exclusion Criteria for the CODIACS Vanguard Trial}
\label{tab:eligibility_codiacs}
\renewcommand{\arraystretch}{0.8}
\begin{tabular}{p{0.9\linewidth}}
\toprule
\textbf{Inclusion Criteria:}
\begin{itemize}[leftmargin=*,noitemsep,nolistsep]
    \item Age 35 years or older.
    \item English- or Spanish-speaking.
    \item 2 to 6 months after hospitalization for an ACS.
    \item Elevated depressive symptoms defined using the Beck Depression Inventory I (BDI): BDI score $\geq 10$ on 2 screening occasions or $>15$ on 1 occasion.
\end{itemize} \\
\midrule
\textbf{Exclusion Criteria:}
\begin{itemize}[leftmargin=*,noitemsep,nolistsep]
    \item Excluded for safety, retention, or intervention futility reasons (see \citet{Whang2012} for details).
\end{itemize}
\\
\bottomrule
\end{tabular}
\end{threeparttable}
\end{table}

\begin{table}[H]
\centering
\begin{threeparttable}
\caption{Inclusion and Exclusion Criteria for the All of Us EHR Cohorts. Because BDI screening results were not available in the All of Us EHR data, an EHR depression diagnosis was used as a proxy for elevated depressive symptoms. Trial eligibility criteria that could not be reliably applied to EHR data, including language and study-specific retention criteria, were not additionally imposed.}
\label{tab:eligibility_aou}
\renewcommand{\arraystretch}{0.8}
\begin{tabular}{p{0.9\linewidth}}
\toprule
\textbf{Inclusion Criteria:}
\begin{itemize}[leftmargin=*,noitemsep,nolistsep]
    \item Age 35 years or older on 01/29/2012.
    \item An EHR diagnosis of acute coronary syndrome, unstable angina, or myocardial infarction.
    \item An EHR diagnosis of depressive disorder or a related depressive condition. Depression diagnoses with psychotic features, postpartum depression, and other diagnoses not corresponding to the trial depression criterion were excluded from the depression definition.
    \item For $T_1$, the index depression diagnosis occurred 2--6 months after an ACS diagnosis.
    \item For $T_2$, the index depression diagnosis occurred at least 2 months after an ACS diagnosis.
\end{itemize} \\
\midrule
\textbf{Exclusion Criteria:}
\begin{itemize}[leftmargin=*,noitemsep,nolistsep]
    \item EHR diagnoses indicating bipolar disorder, psychotic disorders, cognitive impairment or dementia, suicidal or self-injurious behavior, substance abuse or drug dependence, or selected severe personality disorders.
    \item EHR diagnoses indicating severe liver disease, including cirrhosis, esophageal varices, portal hypertension, or hepatic encephalopathy.
    \item Serum creatinine $\geq 2.0$ mg/dL based on the most recent interpretable serum or plasma creatinine measurement within 1 year before the index depression diagnosis. Individuals without an interpretable measurement during this period were retained.
\end{itemize} \\
\bottomrule
\end{tabular}
\end{threeparttable}
\end{table}

\subsection{Additional Sensitivity Analysis}
\begin{figure}[tbp]
\centering
\begin{subfigure}{\textwidth}
\centering
\caption{MATE of SplineBART}
\label{fig:sa_mate_spline}
\includegraphics[width=0.9\textwidth]{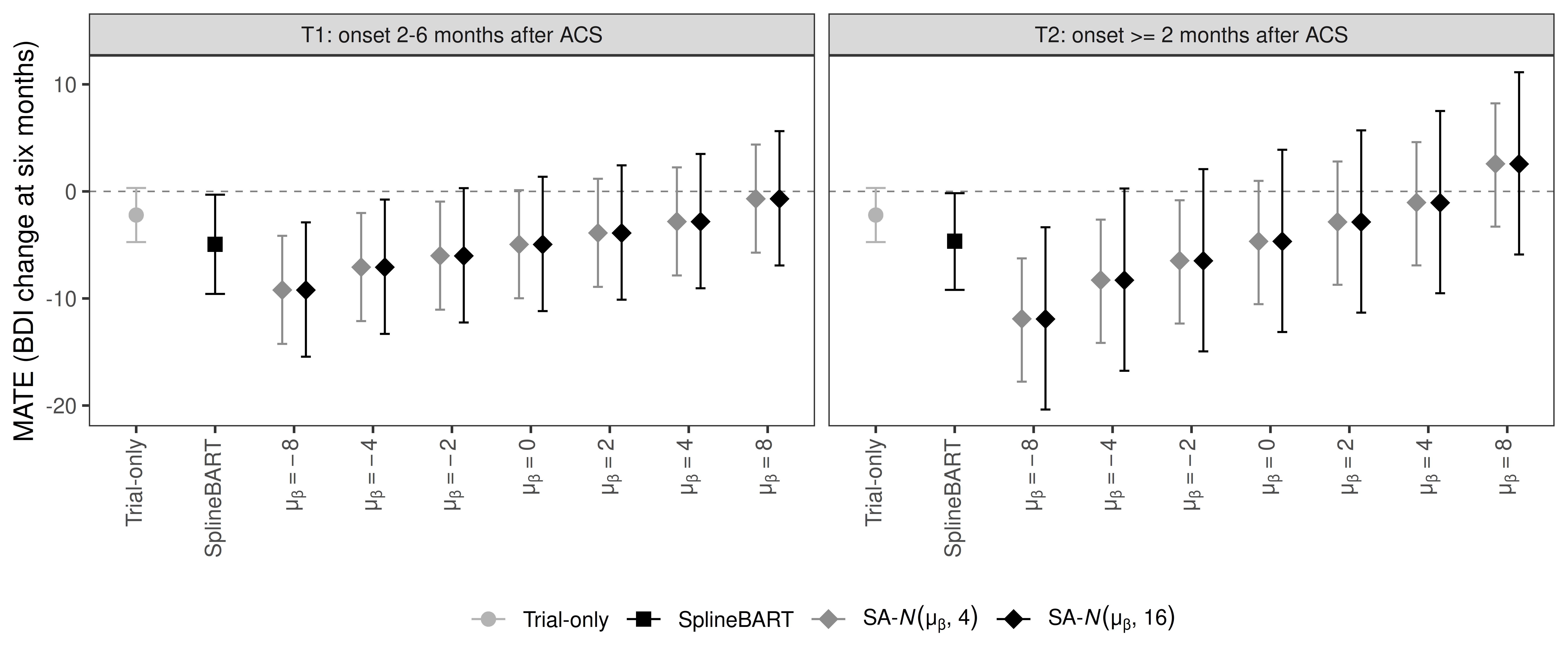}

\end{subfigure}

\vspace{0.2cm}

\begin{subfigure}{\textwidth}
\centering
\caption{PATE of SplineBART}
\label{fig:sa_pate_spline}
\includegraphics[width=0.9\textwidth]{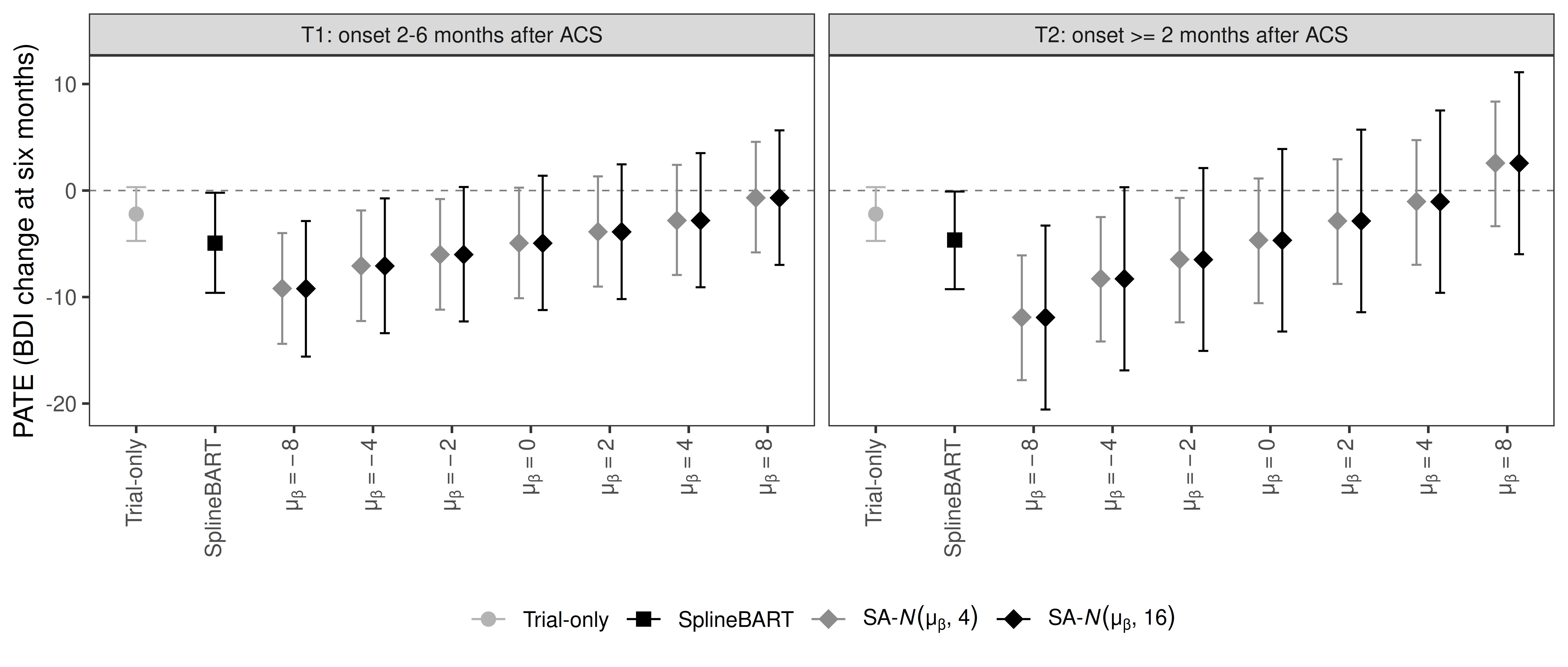}

\end{subfigure}

\vspace{0.2cm}

\begin{subfigure}{\textwidth}
\centering
\caption{PATE of DSplineBART}
\label{fig:sa_pate_dspline}
\includegraphics[width=0.9\textwidth]{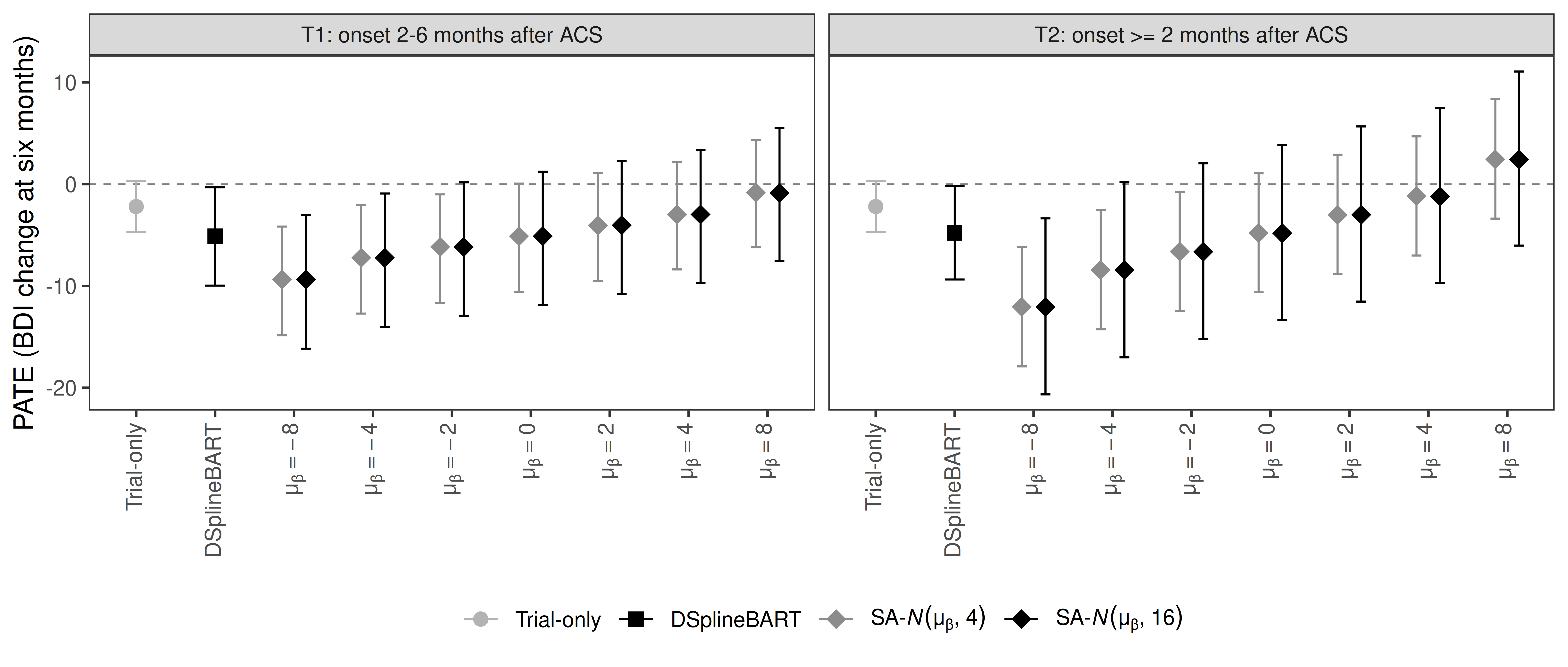}
\end{subfigure}

\caption{Bias-function sensitivity analyses in target samples $T_1$ and $T_2$, complementing the DSplineBART MATE analysis in the main text: (a) the MATE of SplineBART, (b) the PATE of SplineBART, and (c) the PATE of DSplineBART. Each applies the bias function $\delta(X_i)=\hat{\tau}_i\beta$ with $\beta\sim\mathcal{N}(\mu_\beta,\sigma^2_\beta)$ over the same grid of $\mu_\beta$ and $\sigma^2_\beta$, on the scale of the 6-month BDI-score change.}
\label{fig:sa_supp}
\end{figure}

The additional bias-function sensitivity analyses, for the MATE of SplineBART and the PATEs of SplineBART and DSplineBART, are presented in Web Figure~\ref{fig:sa_supp}.

\clearpage
\newpage


\bibliographystyle{imsart-nameyear} 
\bibliography{reference}